\documentclass{emulateapj}
\usepackage{natbib}
\newcommand{\hi}{\mbox{\rm \ion{H}{1}}}
\newcommand{\hii}{\mbox{\rm \ion{H}{2}}}
\newcommand{\htwo}{\mbox{\rm H$_2$}}

\shorttitle{Star Formation Rates in Nearby Galaxies}
\shortauthors{Leroy et al.}

\begin{document}

\slugcomment{Accepted for Publication in the Astronomical Journal} 

\title{Estimating the Star Formation Rate at 1~kpc Scales in Nearby
  Galaxies}

\author{Adam K. Leroy\altaffilmark{1},
  Frank Bigiel\altaffilmark{2},
  W.J.G. de Blok\altaffilmark{3},
  Samuel Boissier\altaffilmark{4},
  Alberto Bolatto \altaffilmark{5},
  Elias Brinks\altaffilmark{6},
  Barry Madore\altaffilmark{7},
  Juan-Carlos Munoz-Mateos\altaffilmark{1},
  Eric Murphy\altaffilmark{7},
  Karin Sandstrom\altaffilmark{8},
  Andreas Schruba\altaffilmark{8},
  Fabian Walter\altaffilmark{8}}
\altaffiltext{1}{National Radio Astronomy Observtory, 520 Edgemont Road, Charlottesville, VA 22903, USA}
\altaffiltext{2}{Institut f\"ur Theoretische Astrophysik, Universit\"at Heidelberg, Albert-Ueberle-Str. 2, 69120, Heidelberg, Germany}
\altaffiltext{3}{Astrophysics, Cosmology and Gravity Centre, Department of Astronomy, University of Cape Town, Private Bag X3, Rondebosch 7701, South Africa}
\altaffiltext{4}{Laboratoire d'Astrophysique de Marseille, Universit\'e de Provence, CNRS (UMR6110), 38 rue Fr\'ed\'eric Joliot Curie, 13388 Marseille Cedex 13}
\altaffiltext{5}{Department of Astronomy, University of Maryland, College Park, MD, USA}
\altaffiltext{6}{Centre for Astrophysics Research, University of Hertfordshire, Hatfield AL10 9AB, United Kingdom}
\altaffiltext{7}{Observatories of the Carnegie Institution for Science, 813 Santa Barbara Street, Pasadena, CA 91101, USA}
\altaffiltext{8}{Max Planck Institute f\"ur Astronomie, K\"onigstuhl 17, 69117, Heidelberg, Germany}

\begin{abstract}
Using combinations of H$\alpha$, ultraviolet (UV), and infrared (IR)
emission, we estimate the star formation rate (SFR) surface density,
$\Sigma_{\rm SFR}$, at 1~kpc resolution for $30$ disk galaxies that
are targets of the IRAM HERACLES CO survey. We present a new
physically-motivated IR spectral energy distribution-based approach to
account for possible contributions to 24$\mu$m emission not associated
with recent star formation. Considering a variety of ``reference''
SFRs from the literature, we revisit the calibration of the 24$\mu$m
term in hybrid (UV+IR or H$\alpha$+IR) tracers. We show that the
overall calibration of this term remains uncertain at the factor of
two level because of the lack of wide-field, robust reference SFR
estimates. Within this uncertainty, published calibrations represent a
reasonable starting point for 1~kpc-wide areas of star-forming disk
galaxies but we re-derive and refine the calibration of the IR term in these tracers to match our resolution and approach to 24$\mu$m emission. We compare a large suite of $\Sigma_{\rm SFR}$ estimates and find that above $\Sigma_{\rm SFR}
\sim 10^{-3}$~M$_\odot$~yr$^{-1}$~kpc$^{-2}$ the systematic
differences among tracers are less than a factor of two across two
orders of magnitude dynamic range. We caution that methodology and
data both become serious issues below this level. We note from simple
model considerations that focusing on a part of a galaxy dominated by
a single stellar population the intrinsic uncertainty in H$\alpha$ and
FUV-based SFRs are $\sim 0.3$ and $\sim 0.5$~dex.
\end{abstract}


%--------------------------------
\section{Introduction}
\label{sec:intro}
%---------------------------------

Comparing the distributions of recent star formation and interstellar
gas can illuminate the preconditions for star formation, the impact of
local conditions on the conversion of gas to stars, and the process of
galaxy assembly. Making these comparisons quantitative requires local
estimates of the recent star formation rate (SFR) that span whole
galaxies. Such estimates can be reconstructed from the observational
signatures of massive stars, which live only a short time after they
are born. These signatures include recombination line and thermal radio emission, 
both produced by ionizing photons emitted by the most massive stars. In regions of 
active star formation, direct photospheric emission from massive stars also dominates 
ultraviolet (UV) emission and provides the bulk of starlight reprocessed by dust into 
infrared (IR) emission.

This paper investigates the use of such observations to estimate the surface density of recent star
formation, $\Sigma_{\rm SFR}$, at 1~kpc resolution in nearby galaxies. We focus on 30 galaxies
that are targets of the IRAM\footnote{IRAM is supported by CNRS/INSU
  (France), the MPG (Germany) and the IGN (Spain).} HERACLES Large Program \citep{LEROY09}. HERACLES surveyed CO $J=2\rightarrow1$
emission and we have \hi\ maps for each
target, mostly from the VLA THINGS survey \citep{WALTER08}. As a result, we know
the distributions of both the molecular (H$_2$) and atomic (\hi )
interstellar medium (ISM). THINGS and HERACLES aim to
constrain theories of star formation via comparisons of \hi , H$_2$,
SFR, and environment at matched resolution across large
areas. $\Sigma_{\rm SFR}$ estimates matched to the gas maps play a key
role in such tests a main goal of this paper is to motivate our approach to such comparisons.

There is no definitive approach to estimate $\Sigma_{\rm SFR}$ at kpc
resolutions. Indeed, less attention has been paid to estimating $\Sigma_{\rm
  SFR}$ for large parts of galaxies than to SFRs for individual
regions or whole galaxies. We therefore begin at a low level and focus on issues related to this intermediate scale as we work towards  $\Sigma_{\rm SFR}$ estimates. These include the scatter in SFR estimates that arises from considering individual stellar populations, the fraction of IR emission not associated with recent star formation, and the appropriate calibration for "hybrid" UV+IR or H$\alpha$+IR tracers.

We begin by defining our approach, which follows other recent work by hybridizing UV or H$\alpha$ emission with IR emission (Sections \ref{sec:uvha} and \ref{sec:hybrid}). We consider how UV and H$\alpha$-based estimates will be affected by isolating individual stellar populations rather than considering an integrated, continuously star-forming population (Section \ref{sec:uvha}). This will be an increasing issue moving from integrated galaxies to sub-galactic scales and we examine Starburst99 \citep{LEITHERER99} simulations to estimate the magnitude of the issue. We discuss basic trends among observed intensities of UV, H$\alpha$, and IR emission at 1~kpc resolution, including their correlations and relative magnitudes (Section \ref{sec:hybrid}). We then explore the effects and potential magnitude of IR emission not directly associated with star formation (Section \ref{sec:cirrus}). We adopt a physically motivated approach to this topic, interpreting 24$\mu$m emission through the lens of dust models fit to the whole IR spectral energy distribution (SED) and test its effect (Section \ref{sec:cirrus}). Based on these calculations, we explore the calibration of empirical ``hybrid'' (IR+H$\alpha$ and IR+UV) tracers for our regime (Section \ref{sec:w}) and discuss previous work on this topic. We emphasize the substantial uncertainty that still remains in the absolute calibration of such tracers despite much recent attention. We conclude by comparing a large suite of $\Sigma_{\rm SFR}$ estimates (Section \ref{sec:compare}) and then discussing uncertainties in SFR estimates, outlining our recommended approach to estimate $\Sigma_{\rm SFR}$, and noting several implications of our work for studying star-formation in galaxies (Section \ref{sec:disc}). We present our conclusions in Section \ref{sec:conc}.

We focus on IR emission measured by the {\em Spitzer} Space Telescope, emphasizing the use of 24$\mu$m to trace embedded star formation. The 24$\mu$m maps produced by {\em Spitzer} have very good sensitivity and resolution well matched to our ISM data. Many studies of nearby galaxies have revealed a strong empirical correlation between 24$\mu$m emission and sites of active star formation \citep[e.g.,][]{CALZETTI05,ALONSOHERRERO06,PEREZGONZALEZ06,PRESCOTT07} and it has been used extensively to investigate the correlations between gas and star formation \citep[e.g.,][]{KENNICUTT07,BIGIEL08,LEROY08,WILSON09,RAHMAN10}. Critically, as part of the SINGS and LVL legacy programs, {\em Spitzer} produced well-calibrated maps, publicly available maps for each of our targets. Based on these considerations, we focus on {\em Spitzer} data here but expect that similar studies using the short wavelength {\em Herschel} bands will yield complementary results in the near future.

%-----------------------------------------
\section{Data}
\label{sec:data}

We focus on $\Sigma_{\rm SFR}$ to complement our
database of \hi\ and CO maps. We therefore present measurements for
galaxies meeting the following criteria: 1) a HERACLES CO map
containing a robust detection, 2) {\em Spitzer} data from
3.6--160$\mu$m, and 3) inclination $\lesssim 75\degr$. This yields the
sample of $30$ disk galaxies listed in Table \ref{tab:sample}. This
table also gives the distance, the physical resolution corresponding to the (limiting) 13$\arcsec$ angular resolution of HERACLES
at that distance, inclination, position
angle, optical radius, and the source of the H$\alpha$ data that we
will use for that galaxy. We adopt redshift-independent distances and
orientation from the careful literature compilation in \citet{KENNICUTT11} wherever possible, \citet{WALTER08} elsewhere, and from LEDA
\citep{PRUGNIEL98} and NED when neither are available.

We will estimate $\Sigma_{\rm SFR}$ from combinations of broadband
infrared (IR), UV, and H$\alpha$ emission with references to other
estimates from the literature. By construction HERACLES and THINGS
overlap SINGS \citep{KENNICUTT03}, LVL \citet{DALE09}, and the {\em
  GALEX} NGS \citep{GILDEPAZ07} surveys, so H$\alpha$, UV, and IR are
readily available for each target.

We aim to compare tracers of the ISM and star formation. The maps of the ISM limit the resolution of such comparison; the HERACLES CO maps have $13\arcsec$ resolution and the naturally-weighted THINGS \hi\ data are often comparable. For most of our targets, this angular resolution corresponds to $\lesssim 1$~kpc and we adopt $1$~kpc as our working resolution. We convolve each map
to have a symmetric gaussian beam with FWHM$=1$kpc. For the {\em
  Spitzer} 24$\mu$m maps we first convert from the MIPS PSF to a
$13\arcsec$ gaussian beam using a kernel provided by K. Gordon
(priv. comm.) then we convolve to $1$~kpc. This effectively places our
targets at a common distance but does not account for foreshortening
along the minor axis. Five galaxies are too distant to convolve the
ISM maps to 1~kpc. We mark these in Table \ref{tab:sample} and include
them in our analysis at the 13$\arcsec$ angular resolution of the CO
data.

From these convolved maps, we generate a database of intensity
measurements. We sample each map using an hexagonal grid
spaced by 0.5\,kpc, i.e., one half-resolution element. At each
sampling point we measure the CO $J=2\rightarrow1$, 24$\mu$m, FUV,
NUV, {\sc Hi}, and H$\alpha$ intensities. We also note dust
properties, estimated at coarser resolution, for each point. Table
\ref{tab:dataset} summarizes our data set. The rest of this section
explains how we derive the measurements for each point.

\begin{deluxetable*}{lccccccccc}
\tablecaption{Sample Properties} 
\tabletypesize{\footnotesize}
\tablehead{ \colhead{Galaxy} & \colhead{$D$\tablenotemark{a}} & \colhead{res.\tablenotemark{b}} &
  \colhead{$i$} & \colhead{$PA$} & \colhead{$r_{25}$} &
  \colhead{H$\alpha$ Map} & \colhead{Adopted $\log_{10} f_{\rm H\alpha+NII}$} & \colhead{Adopted \ion{N}{2}/H$\alpha$} \\
\colhead{} & \colhead{[Mpc]} & \colhead{[kpc]} & \colhead{[$\arcdeg$]}
& \colhead{[$\arcdeg$]} & \colhead{[$\arcmin$]} & \colhead{} & \colhead{(erg~s$^{-1}$~cm$^{-2}$)}}

\startdata
  NGC\,0337 & 19.3\tablenotemark{c} & 1.24 & 51 & 90 & 1.5 & SINGS \citep{KENNICUTT03} & $-11.43$ & $0.23$ \\
  NGC\,0628 & 7.2 & 0.46 & 7 & 20 & 4.9 & Palomar Las Campanas Atlas & $-10.84$ & $0.35$ \\
  NGC\,0925 & 9.1 & 0.59 & 66 & 287 & 5.4 & \citet{BOSELLI02B} & $-11.10$ & $0.20$ \\
  NGC\,2403 & 3.2 & 0.21 & 63 & 124 & 7.9 & \citet{BOSELLI02B} & $-11.76$ & $0.36$ \\
  NGC\,2841 & 14.1 & 0.91 & 74 & 153 & 3.5 & Palomar Las Campanas Atlas & $-11.53$ & $0.61$ \\
  NGC\,2903 & 8.9 & 0.57 & 65 & 204 & 5.9 & \citet{HOOPES01} & $-10.71$ & $0.56$ \\
  NGC\,2976 & 3.6 & 0.23 & 65 & 335 & 3.6 & LVL \citep{DALE09} & \nodata\tablenotemark{d} & $0.13$ \\
  NGC\,3049 & 19.2\tablenotemark{c} & 1.24 & 58 & 28 & 1.0 & SINGS \citep{KENNICUTT03} & $-11.93$ & $0.40$ \\
  NGC\,3184 & 11.8 & 0.76 & 16 & 179 & 3.7 & SINGS \citep{KENNICUTT03} & $-11.12$ & $0.52$ \\
  NGC\,3198 & 14.1 & 0.91 & 72 & 215 & 3.2 & Palomar Las Campanas Atlas & $-11.4$ & $0.30$ \\
  NGC\,3351 & 9.3 & 0.60 & 41 & 192 & 3.6 & LVL \citep{DALE09} & \nodata\tablenotemark{d} & 0.62 \\
  NGC\,3521 & 11.2 & 0.72 & 73 & 340 & 4.2 & SINGS \citep{KENNICUTT03} & $-10.81$ & $0.57$ \\
  NGC\,3627 & 9.4 & 0.61 & 62 & 173 & 5.1 & SINGS \citep{KENNICUTT03} & $-10.75$ & $0.54$ \\
  NGC\,3938 & 17.9\tablenotemark{c} & 1.15 & 14 & 15 & 1.8 & SINGS \citep{KENNICUTT03} & $-11.25$ & $0.42$ \\
  NGC\,4214 & 2.90 & 0.19 & 44 & 65 & 3.4 & \citet{HUNTER04} & $-10.77$ & $0.16$ \\
  NGC\,4254 & 14.4 & 0.93 & 32 & 55 & 2.5 & GoldMine \citep{GAVAZZI03} & $-10.89$ & $0.45$ \\
  NGC\,4321 & 14.3 & 0.92 & 30 & 153 & 3.0 & GoldMine \citep{GAVAZZI03} & $-11.08$ & $0.43$ \\
  NGC\,4536 & 14.5 & 0.94 & 59 & 299 & 3.5 & GoldMine \citep{GAVAZZI03} & $-11.36$ & $0.45$ \\
  NGC\,4559 & 7.0 & 0.45 & 65 & 328 & 5.2 & SINGS \citep{KENNICUTT03} & $-10.97$ & $0.28$ \\
  NGC\,4569 & 9.86 & 0.64 & 66 & 23 & 4.6 & GoldMine \citep{GAVAZZI03} & $-11.39$ & $0.99$ \\
  NGC\,4579 & 16.4\tablenotemark{c} & 1.06 & 39 & 100 &  2.5 & GoldMine \citep{GAVAZZI03} &  $-11.49$ & $0.62$ \\
  NGC\,4625 & 9.3 & 0.60 & 47 & 330 & 0.7 & LVL \citep{DALE09} & \nodata\tablenotemark{d} & $0.55$ \\
  NGC\,4725 & 11.9 & 0.77 & 54 & 36 & 4.9 & \citet{KNAPEN04} & $-11.42$ & $0.38$ \\
  NGC\,4736 & 4.7 & 0.30 & 41 & 296 & 3.9 & \citet{KNAPEN04} & $-10.72$ & $0.71$ \\
  NGC\,5055 & 7.9 & 0.51 & 59 & 102 & 5.9 & LVL \citep{DALE09} & \nodata\tablenotemark{d} & 0.50 \\
  NGC\,5194 & 7.9 & 0.52 & 20 & 172 & 3.9 & \citet{BOSELLI02B} & -10.42 & 0.60 \\
  NGC\,5457 & 6.7 & 4.3 & 18 & 39 & 12.0 & \citet{HOOPES01} & $-10.22$ & $0.54$ \\
  NGC\,5713 & 21.4\tablenotemark{c} & 1.38 & 48 & 11 & 1.2 & SINGS \citep{KENNICUTT03} & $-11.63$ & $0.55$ \\
  NGC\,6946 & 6.8 & 0.44 & 33 & 243 & 5.7 & SINGS \citep{KENNICUTT03} & $-10.42$ & $0.54$\\
  NGC\,7331 & 14.5 & 0.94 & 76 & 168 & 4.6 & SINGS \citep{KENNICUTT03} & $-11.09$ & 0.61
\enddata
\label{tab:sample}
\tablenotetext{a}{{\rm D}istances adopted, in order of preference, from compilations by \citet{KENNICUTT11}, \citet{WALTER08}, and the NED/LEDA databases.}
\tablenotetext{b}{{\rm Linear} resolution corresponding to $13.3\arcsec$ angular resolution at the distance of the target.}
\tablenotetext{c}{Too distant to
    convolve to 1~kpc resolution. Included in analysis at native
    ($13\arcsec$) resolution.}
\tablenotetext{d}{Flux calibration of LVL H$\alpha$ maps taken to be correct.}
\end{deluxetable*}

\subsection{HERACLES CO}

The HERA CO Line Extragalactic Survey \citep[HERACLES,][]{LEROY09}
used the Heterodyne Receiver Array \citep[HERA,][]{SCHUSTER04} on the
IRAM 30m telescope to map CO $J=2\rightarrow1$ emission from $48$
nearby galaxies. The HERACLES cubes cover out to $r_{25}$ with angular
resolution $13\arcsec$ and typical $1\sigma$ sensitivity $20$~mK per
5~km~s$^{-1}$ channel. Leroy et al. (2012, in prep.) present the full data
set and more details\footnote{All HERACLES data are publicly available from the IRAM and NRAO web pages. Current URL \url{www.cv.nrao.edu/$\sim$aleroy/HERACLES}}.

We estimate \htwo\ mass surface density, $\Sigma_{\rm H2}$, from CO
$J=2\rightarrow1$ intensity via
\begin{equation}
\label{eq:xco}
\Sigma_{\rm H2} [{\rm
    M_{\sun}\,pc^{-2}}]=6.3~\left(\frac{0.7}{R_{21}}\right)~\left(\frac{\alpha_{\rm
    CO}}{4.4}\right)~I_{CO}~[{\rm K\,km\,s^{-1}}],
\end{equation}
where $R_{21}$ is the CO(2$\rightarrow$1)-to-CO(1$\rightarrow$0) line
ratio and $\alpha_{\rm CO}$ is the
CO(1$\rightarrow$0)-to-\htwo\ conversion factor.    The
formula includes a factor of 1.36 to account for helium. We adopt
a line ratio of 0.7\footnote{This line ratio is slightly lower than
  the 0.8 used by \citet{LEROY09}, reflecting the revised efficiency
  used in the data reduction. This value, $0.7$, is the mean ratio of
  integrated CO $J=2\rightarrow1$ HERACLES flux divided by the CO
  $J=1\rightarrow0$ flux measured by \citet{YOUNG95},
  \citet{HELFER03}, or \citet{KUNO07}.}, and a Galactic conversion factor,
$\alpha_{\rm CO} = 4.4$~M$_\odot$~pc$^{-2}$~$\left( {\rm K~km~s}^{-1}
\right)^{-1}$ equivalent to $X_{\rm CO} = 2\times10^{20}\,{\rm
  cm^{-2}\,(K\,km\,s^{-1})^{-1}}$. This value is intermediate among recent determinations of the Milky Way $\alpha_{\rm CO}$  \citep[e.g.,][]{STRONG96,DAME01,HEYER09,ABDOXCO} and represents a commonly adopted "best single value" for work on nearby disk galaxies \citep[e.g.,][]{WONG02,LEROY08}. HERACLES does span a range of metallicities \citep[e.g.,][]{MOUSTAKAS10} and observational evidence suggests that the CO-to-H$_2$ conversion factor varies as a function of metallicity \citep[e.g., see summary in][]{LEROY11}. However, most low-metallicity, high-$\alpha_{\rm CO}$ systems also tend to be dominated by atomic gas and in this paper only the total (\hi + H$_2$) gas supply will be relevant. We therefore neglect metallicity variations in $\alpha_{\rm CO}$ for purposes of estimating the dust-to-gas ratio and $\Sigma_{\rm SFR}$, though we consider these in our subsequent comparison of $\Sigma_{\rm H2}$ and $\Sigma_{\rm SFR}$ (Leroy et al., in prep.).

\subsection{THINGS and Supplemental \hi }

We assemble \hi\ maps for all targets, which we use to mask the CO,
estimate the dust-to-gas ratio, and explore 24$\mu$m cirrus
corrections. These come from THINGS \citep{WALTER08} and a collection
of new and archival VLA\footnote{The National Radio Astronomy
  Observatory is a facility of the National Science Foundation
  operated under cooperative agreement by Associated Universities,
  Inc.} data (including our programs AL731 and AL735). These
supplemental \hi\ are C+D configuration maps with resolutions
$13\arcsec$--$25\arcsec$. We reduced and imaged these a standard way
using the CASA package (see Leroy et al. in prep.). In a few cases the native angular resolution of the \hi\ maps corresponds to a spatial resolution coarser than 1~kpc. In these cases we assume the \hi\ to be smooth at scales smaller than the
resolution. Given the low dynamic range in \hi\ column densities observed at these resolutions \citep{WALTER08} this approximation should have minimal impact on our results. We translate 21-cm intensity into \hi\ surface density
assuming optically thin emission \citep[see references in][]{WALTER08}
and include helium when quoting atomic gas surface density,
$\Sigma_{\rm HI}$. Thus

\begin{equation}
\label{eq:hi}
\Sigma_{\rm HI}~\left[ {\rm M}_\odot~{\rm pc}^{-2} \right] = 0.020~I_{\rm HI}~\left[ {\rm K~km~s}^{-1} \right]
\end{equation}

\subsection{{\em GALEX} UV}

For 24 galaxies, we use NUV and FUV maps from the Nearby Galaxy Survey
\citep[NGS,][]{GILDEPAZ07}. For one galaxy, we use a map from the
Medium Imaging Survey (MIS) and we take maps for 5 targets from the
All-sky Imaging Survey \citep[AIS][]{MARTIN05}. We subtract a small
background from the UV maps, determined after blanking the bright (SNR$>2$), extended emission in the map. The magnitude of this background is typically $\sim 5 \times 10^{-4}$~MJy~sr$^{-1}$. This corresponds to $\Sigma_{\rm SFR} \sim 4 \times 10^{-5}$~M$_\odot$~yr$^{-1}$~kpc$^{-2}$, much lower than the typical $\Sigma_{\rm SFR}$ considered in this paper.
We identify foreground stars via their UV
color, by-eye inspection, and the color-based masks of
\citet{MUNOZMATEOS09A}, which also blank background galaxies. We
correct the FUV maps for the effects of Galactic extinction following
\citet{SCHLEGEL98} and \citet{WYDER07} --- see \citet{LEROY08}.

\subsection{SINGS \& LVL IR}

We use maps of IR emission from 3.6--160$\mu$m from the {\em Spitzer}
Infrared Nearby Galaxies Survey \citep[SINGS,][]{KENNICUTT03} and the
Local Volume Legacy survey \citep[LVL,][]{DALE09}. We mask foreground
stars based on UV color, by-eye inspection, and the color-based masks
from \citet{MUNOZMATEOS09A}.

\subsection{Literature H$\alpha$}

We draw H$\alpha$ maps from the literature for all of our targets. Whenever possible, we draw maps from the LVL survey. These have clean backgrounds and agree very well with previous flux measurements. For the remaining targets we assembled H$\alpha$ maps from a variety of literature sources, including the SINGS data release, the GOLDMine database, and surveys with the Palomar-Las Campanas Observatories. The photometric calibrations for these maps were not always available. Even when such calibrations were present, e.g., for the SINGS H$\alpha$ maps, the integrated fluxes of the maps often scattered substantially about previous measurements \citep[a conclusion verified independently by several of us and also implicitly present in][who mix spectroscopic, literature, and narrow-band imaging measurements for estimates of SINGS galaxies]{KENNICUTT09}. Therefore for non-LVL maps, we set the overall flux scale of each map by searching the literature for  pinning the integrated flux of the H$\alpha$ map to match the average literature value. Table \ref{tab:sample} lists the source and adopted H$\alpha$+\ion{N}{2} flux for each H$\alpha$. When several H$\alpha$ maps were available for a single target, we chose among them based on a by-eye evaluation of the quality of continuum subtraction and overall flat fielding. Because our coarse working resolution, $13\arcsec$, involves heavily smoothing the maps before any analysis these considerations are more important than seeing or sensitivity.

We process the H$\alpha$ maps as follows. Whenever available, we apply the masks of \citet{MUNOZMATEOS09A} to blank bright foreground stars and background galaxies, which often leave residual artifacts in the continuum subtracted image and are in any case not associated with our galaxies. We also mask foreground stars identified by eye. During convolution to our working resolution, we fill in these blank regions with interpolated values.

For non-LVL images we found it necessary to subtract a background determined away from the galaxy. Using the UV and IR 24$\mu$m images, we define a mask that encompasses all bright star formation in each galaxy and then verify by eye that this encloses bright H$\alpha$ emission. We subtracted the median of the map determined outside this region from the whole map. In a few cases that displayed obvious horizontal or vertical striping, we also used this external region to determine and subtract a row- or column-wise median away from the galaxy. For some literature images in which the field of view was fairly small it was not possible to determine a robust median far away from all IR or UV emission. In these cases we subtracted the mode of a histogram of the H$\alpha$ map after carefully blanking all bright H$\alpha$ emission. After this processing but before any correction for Galactic extinction or \ion{N}{2} contamination we renormalized the H$\alpha$ maps so that their integrated flux matches values adopted from the literature.

The H$\alpha$ filter includes both H$\alpha$ and [\ion{N}{2}]. We correct for this effect using the spectroscopic ratios from \citet{KENNICUTT08} and \citet{KENNICUTT09}. If the [\ion{N}{2}]-to-H$\alpha$ ratio is not available, we use the galaxy's $B$-band magnitude with the scaling relation from \citet{KENNICUTT08} to estimate a ratio. We correct the maps for the effect of Galactic extinction following \citet{SCHLEGEL98}.

\subsection{Dust Property Fits}
\label{sec:dustfits}

Of the {\em Spitzer} far-infrared data, only the 24$\mu$m data reach our working resolution but the
whole IR SED provides valuable information. We
derive a variety of information at the coarser ($\approx 40\arcsec$)
resolution of the 160$\mu$m data. Most of this information comes from fitting the IR SED
using the \citet{DRAINE07A} models. This fitting resembles that in \citet{MUNOZMATEOS09B}
but we include several galaxies that they did not and derive dust-to-gas
ratios using our new CO and \hi\ maps.

We carry out the dust fits using the following approach. First, we mask the 8$\mu$m and 24$\mu$m images using the masks of \citet{MUNOZMATEOS09A}. We then subtract a stellar contribution from the 8$\mu$m image using the 3.6$\mu$m image as a template. We use a scaling factor of 0.269,
  adopted from \citet{MUNOZMATEOS09B}; see also \citet{HELOU04}. We convolve the 8, 24, 70, and 160$\mu$m {\em Spitzer} images and the \hi\ and CO maps to all share the PSF of {\em Spitzer} 160$\mu$m. We then construct radial profiles of each IR image treating our targets as thin disks with the orientation parameters in Table \ref{tab:sample}. Before adopting this approach, we experimented with fitting SEDs to each line-of-sight but the lower signal-to-noise ratio and the correlation among uncertainties the fitted dust parameters (e.g., DGR and $U_{\rm min}$) led to instability in the analysis. We found the loss of information from assuming azimuthal symmetry and working in profile to be offset by the improved S/N. 
  
We consider only rings with S/N$>4$ at each band and for each IR SED we calculate $\chi^2$ across a grid of dust emission models following \citet{DRAINE07B} and \citet{MUNOZMATEOS09B}. The free parameters in the grid are: the intensity of the ambient radiation field $U_{\rm min}$ in units of the local interstellar radiation field; the fraction of dust mass illuminated by a distribution of more intense radiation fields (the rest is illuminated by $U_{\rm min}$); $q_{\rm PAH}$,  the mass fraction of dust in PAHs; and the total dust mass. To calculate $\chi^2$ we use uncertainties estimated from the scatter in the convolved maps away from the galaxy  or 10\% of the observed intensity, whichever is higher \citep[this approach follows][]{DRAINE07B}. We identify best-fit parameters from the minimum $\chi^2$ in our grid search.

Based on these calculations we derive the following quantities for
each line of sight: 1) the dust-to-gas ratio, $DGR$, from comparison to the
convolved \hi\ and CO maps; 2) the 24$\mu$m-to-total infrared (TIR)
luminosity ratio --- defined as $\nu L_{\nu} / L_{\rm TIR}$ for $\nu$
at $\lambda = 24\mu$m and $L_{\rm TIR}$ calculated using the
prescription of \citet{DRAINE07A}; 3) the 24$\mu$m emission per unit
dust mass given $U_{\rm min}$ and $q_{\rm PAH}$ for the best-fit
model. Along with $U_{\rm min}$, these quantities will allow us to
explore contamination of the 24$\mu$m band and the impact of dust
abundance on star formation and map 24$\mu$m emission to total IR
emission.

We derive these IR-based quantities from radial profiles of maps with
the $\sim 40\arcsec$ resolution of the 160$\mu$m data. For this
reason, we focus on ratios like the dust-to-gas ratio or
24$\mu$m-to-TIR ratio that we expect to vary weakly within a
resolution element. Under these assumptions, these can be extrapolated to higher
resolution if part of the ratio is known at high resolution.

\begin{deluxetable*}{ll}
\tablecaption{Measurements For Each Sampling Point} 
\tablehead{ \colhead{Quantity} & \colhead{Origin}}
\startdata
FUV & {\em GALEX} \citep[][]{MARTIN05,GILDEPAZ07} \\
NUV & {\em GALEX} \citep[][]{MARTIN05,GILDEPAZ07} \\
H$\alpha$ & Literature (Table \ref{tab:sample}) \\
$I_{\rm 24}$ & {\em Spitzer} SINGS \& LVL\citep{KENNICUTT03,DALE09} \\
$\Sigma_{\rm H2}$ & IRAM 30-m HERACLES \citep[][Leroy et al. in prep.]{LEROY09} \\
$\Sigma_{\rm HI}$ & VLA THINGS \citep{WALTER08} + supplemental \hi \\
$\Sigma_{\rm Dust}$ & {\em Spitzer} IR + \citet{DRAINE07A} \\
Dust-to-Gas Ratio & {\em Spitzer} IR + \citet{DRAINE07A} \\
24$\mu$m-to-TIR ratio & {\em Spitzer} IR + \citet{DRAINE07A} \\
$U_{\rm min}$ & {\em Spitzer} IR + \citet{DRAINE07A} \\
24$\mu$m/$\Sigma_{\rm Dust}$ for $U_{\rm min}$ & {\em Spitzer} IR + \citet{DRAINE07A}
\enddata
\label{tab:dataset}
\end{deluxetable*}

\section{UV and H$\alpha$ Emission}
\label{sec:uvha}

We start with the assumption that the time-averaged rate of recent
star formation can be inferred from the rate of ionizing photon
production or the UV luminosity over part of a galaxy. Ionizing photon
production is driven by the most massive stars, which live only a very
short time. Assuming case B recombination, the distribution of
ionizing photons can be traced by H$\alpha$ emission after correcting
for extinction. UV emission traces mainly photospheric direct emission
from O and B stars.

Both approaches have a long pedigree with well established tradeoffs
and biases. Thanks to its short age sensitivity, H$\alpha$ emission
directly traces the most recent generation of star formation. However,
ionizing photons may leak from their parent regions or be absorbed by
dust, confusing the mapping of recombination line emission to local
star formation. UV emission probes down to lower stellar masses,
rendering it less sensitive to stochasticity and variations in the
initial mass function and more sensitive to star formation at
intermediate ages (a few tens of Myr). Horizontal branch stars and
scattered light may contaminate UV emission, and the longer time
window to see UV emission may bias the SFR estimate to reflect
slightly older populations.

\subsection{Adopted Conversions of UV and H$\alpha$ to SFR}

We take the relation between SFR to H$\alpha$ emission from \citet{CALZETTI07},

\begin{equation}
\label{eq:sfr_ha_lum}
{\rm SFR}~\left[ {\rm M}_\odot~{\rm yr}^{-1} \right] = 5.3 \times 10^{-42}~L_{\rm H\alpha}~\left[ {\rm erg~s}^{-1} \right]~.
\end{equation}

\noindent A similar, more recent calculation by \citet{MURPHY11} adjusts the coefficient to $5.37 \times 10^{-42}$. Both conversions derive from population synthesis modeling. Equation \ref{eq:sfr_ha_lum} adopts the default (in 2007) Starburst99 IMF, which resembles a \citet{KROUPA01} IMF truncated at 120~M$_\odot$. In surface brightness units Equation \ref{eq:sfr_ha_lum} is

\begin{equation}
\label{eq:sfr_ha}
\Sigma_{\rm SFR}~\left[ {\rm M}_\odot~{\rm yr}^{-1}~{\rm kpc}^{-2} \right] = 634~I_{\rm H\alpha}~\left[ {\rm erg~s}^{-1}~{\rm sr}^{-1}\right].
\end{equation}

\noindent This calibration coefficient in Equation \ref{eq:sfr_ha_lum} is about $0.66$ times that suggested by \citet{KENNICUTT98B} with the differences mainly due to the adopted IMF. Given our current understanding of the IMF \citep[e.g.,][]{BASTIAN10}, the conversion in Equation \ref{eq:sfr_ha} should yield more realistic $\Sigma_{\rm SFR}$ than the \citet{KENNICUTT98B} calibration.

We also adopt the relation between SFR and FUV emission from \citet{SALIM07},

\begin{equation}
{\rm SFR}~\left[ {\rm M}_\odot~{\rm yr}^{-1} \right] = 0.68 \times 10^{-28} L_{\rm FUV}~\left[ {\rm erg~s}^{-1}~{\rm Hz}^{-1} \right]~.
\end{equation}

\noindent In surface brightness units this is

\begin{equation}
\label{eq:sfr_fuv}
\Sigma_{\rm SFR}~\left[ {\rm M}_\odot~{\rm yr}^{-1}~{\rm kpc}^{-2} \right] = 8.1 \times 10^{-2}~I_{\rm FUV}~\left[ {\rm MJy~sr}^{-1} \right]~.
\end{equation}

\noindent  Equation \ref{eq:sfr_fuv} comes from comparison of SED modeling to UV flux for
large set of multiband observations. It adopts a \citet{CHABRIER03}
IMF. The coefficient in Equation \ref{eq:sfr_fuv} is $\sim 30\%$ lower than the
\citet{KENNICUTT98B} value, even after accounting for IMF
differences. It is $\approx 20\%$ lower than the more recent calculation by \citet{MURPHY11}, which like \citet{KENNICUTT98B} considers a theoretical population that has been continuously star-forming for $\sim 100$~Myr. \citet{SALIM07} discuss this difference. Given the large
calibration data set and observational grounding, we take the
\citet{SALIM07} FUV calibration to be correct. From the original
works, the uncertainty on the coefficients due to metallicity, IMF
truncation, and star formation history appears to be $\sim
10$--$30\%$.

\subsection{Effects of Discrete Star Formation Events}

\begin{deluxetable}{lcc}
\tablecaption{Calculations for a Simple Burst Model} 
\tablehead{ \colhead{Quantity} & \colhead{H$\alpha$} & \colhead{FUV}}
\startdata
Time & & \\
... of 50\% intensity at 1 Myr & 3.4 Myr & 5.5 Myr \\
... of  5\% intensity at 1 Myr & 5.7 Myr & 27 Myr \\
... of 50\% cumulative emission & 1.7~Myr & 4.8 Myr \\
... of 95\% cumulative emission & 4.7~Myr & 65 Myr \\
... luminosity-weighted $\left< \tau \right>$ & 2 Myr & 14 Myr \\
Intrinsic scatter\tablenotemark{a} & & \\
... out to 5\% peak emission & 0.36 dex & 0.44 dex \\
... out to 95\% cumulative emission & 0.26 dex & 0.49 dex \\
... luminosity weighted & 0.34 dex & 0.63 dex
\enddata
\label{tab:burstcalc}
\tablenotetext{a}{RMS scatter in $\log$ of the H$\alpha$ or FUV intensity for a fixed-mass burst as it evolves to the indicated point. This will roughly correspond to the scatter in SFR estimates as one isolates a single stellar population.}
\end{deluxetable}

\begin{figure*}
\plotone{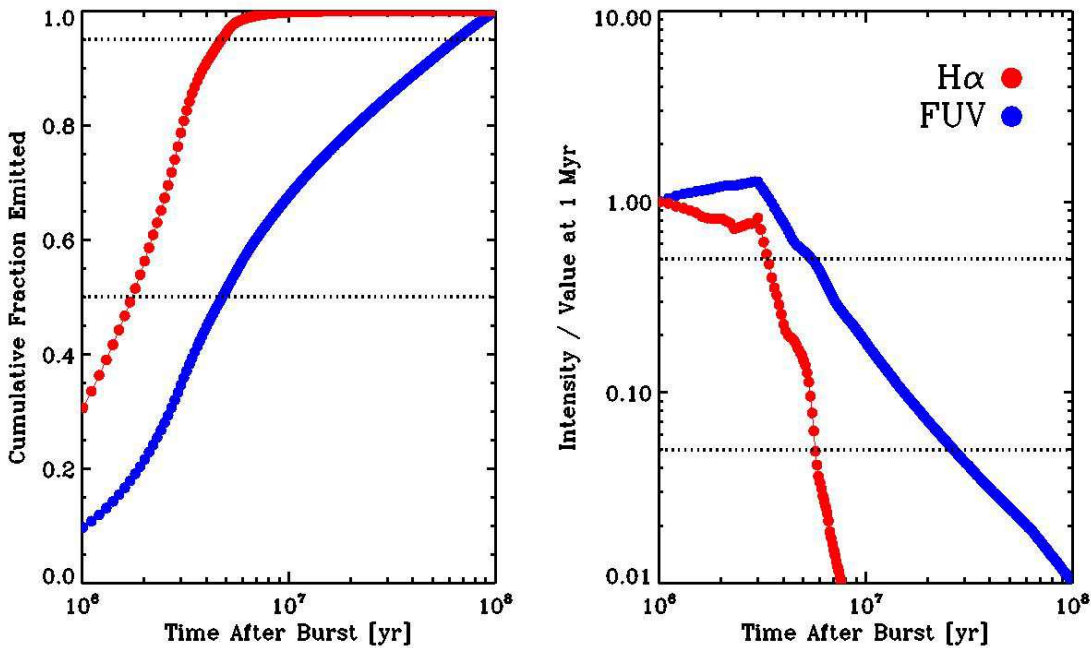}
\caption{{\em Behavior of a Burst of Star Formation.} Starburst99
  calculations for UV (blue) and H$\alpha$ (red) emission from an
  instantaneous burst of star formation. The left panel shows
  cumulative emission as a function of time after the burst, so that one corresponds to all emission over the calculation. The right
  panel shows intensity of the population relative to intensity at 1
  Myr as a function of time. Dashed lines in both panels indicate the
  50\% and 95\% (left) / 5\% (right) levels.}
\label{fig:hauv_time}
\end{figure*}

Equations \ref{eq:sfr_ha} and \ref{eq:sfr_fuv} assume continuous star
formation. This assumption will break down as one considers systems
with low integrated SFR, either very small galaxies or parts of a
galaxy. For such systems, a single ``star formation rate'' becomes an
inadequate approximation and one eventually approaches the case of a
single stellar population with a discrete age. We expect a square kiloparsec portion of an actively star-forming galaxy to contain more than a single single stellar population but investigating this limiting case yields insight into the effect of resolution on SFR estimates.

Figure \ref{fig:hauv_time} illustrates aspects of this breakdown
\citep[see also Figure 6 in][]{GENZEL10}. We plot H$\alpha$ and FUV
emission (at 1500~\AA) as a function of time after an instantaneous
burst of star formation. The calculation uses the Starburst99 code
\citep[][]{LEITHERER99}, runs out to 100 Myr, and adopts the default evolutionary tracks and
IMF.

{\em Age Sensitivity:} We use this simple simulation to quantify the
age sensitivity of the two tracers, which we report in Table
\ref{tab:burstcalc}. We measure the falloff from peak intensity as a
function of time, noting the time that it takes to reach 50\% and 5\%
of the peak. We also derive the time needed to emit 50\% and 95\%
fraction of the cumulative emission given off over the whole
simulation. Finally we calculate the luminosity-weighted average time
after the burst at which a photon is emitted, defined as

\begin{equation} 
\left< \tau \right> = \int \tau L d \tau / \int L d \tau~.
\end{equation}

Both FUV and H$\alpha$ emission emit most of their light within a few
Myr after the burst. FUV emission then has a long ``tail'' (in time)
over which it continues to emit at a low but significant
level. Meanwhile H$\alpha$ drops precipitously before the burst is 10
Myr old. We recover the expected time sensitivity of a few Myr for
H$\alpha$ \citep{VACCA96,MCKEE97} while FUV covers a wide range of
times with a characteristic value of $\sim 10$--$30$ Myr and most
emission gone by 65 Myr.

{\em Intrinsic Scatter:} Luminosity varies with time after the burst,
but the same stellar population produces the luminosity at all times.
Considering ${\rm SFR} \sim M/\delta t$ over a relatively long $\delta
t \sim 100$~Myr, the time-average SFR is the same for all times in the
calculation. In this light, the varying luminosity implies scatter in
the ability to map luminosity to SFR. We calculate this scatter as an
estimate of the intrinsic uncertainty in estimating an SFR in a regime
better described by discrete events.

We assume that a tracer is visible out to time $t_{\rm visible}$ and
that we view it at some random time $t < t_{\rm visible}$. At all $t$
we wish to recover the same SFR. If we use a linear conversion of
luminosity to SFR then the scatter in luminosity for $t < t_{\rm
  visible}$ is the minimum uncertainty in the accuracy of this linear
conversion. We try two values $t_{\rm visible}$: the time at which the
intensity has fallen to 5\% of its value at 1~Myr and the time by
which 95\% of the total emission has occured. We also report the
results of the luminosity-weighted scatter, the second moment in
$\log_{10}$ luminosity weighting by $L d \tau$, which is a more
natural but less intuitive quantity.

Table \ref{tab:burstcalc} reports these estimates. In the limiting
case of discrete bursts we expect a factor of $\sim 2$ uncertainty
($1\sigma$) inferring SFR from H$\alpha$ and a factor of $3$--$4$
uncertainty inferring SFR from FUV. Over a square kpc, we expect to
average several populations so that the scatter in intensity should be
lower than in Table \ref{tab:burstcalc} by $\sim \sqrt{N}$, where $N$
is the number of independent populations that we average \citep[e.g.,
  see][]{SCHRUBA10}.

\section{Hybrid H$\alpha$+IR and FUV+IR SFR Tracers}
\label{sec:hybrid}

\begin{figure*}
\plotone{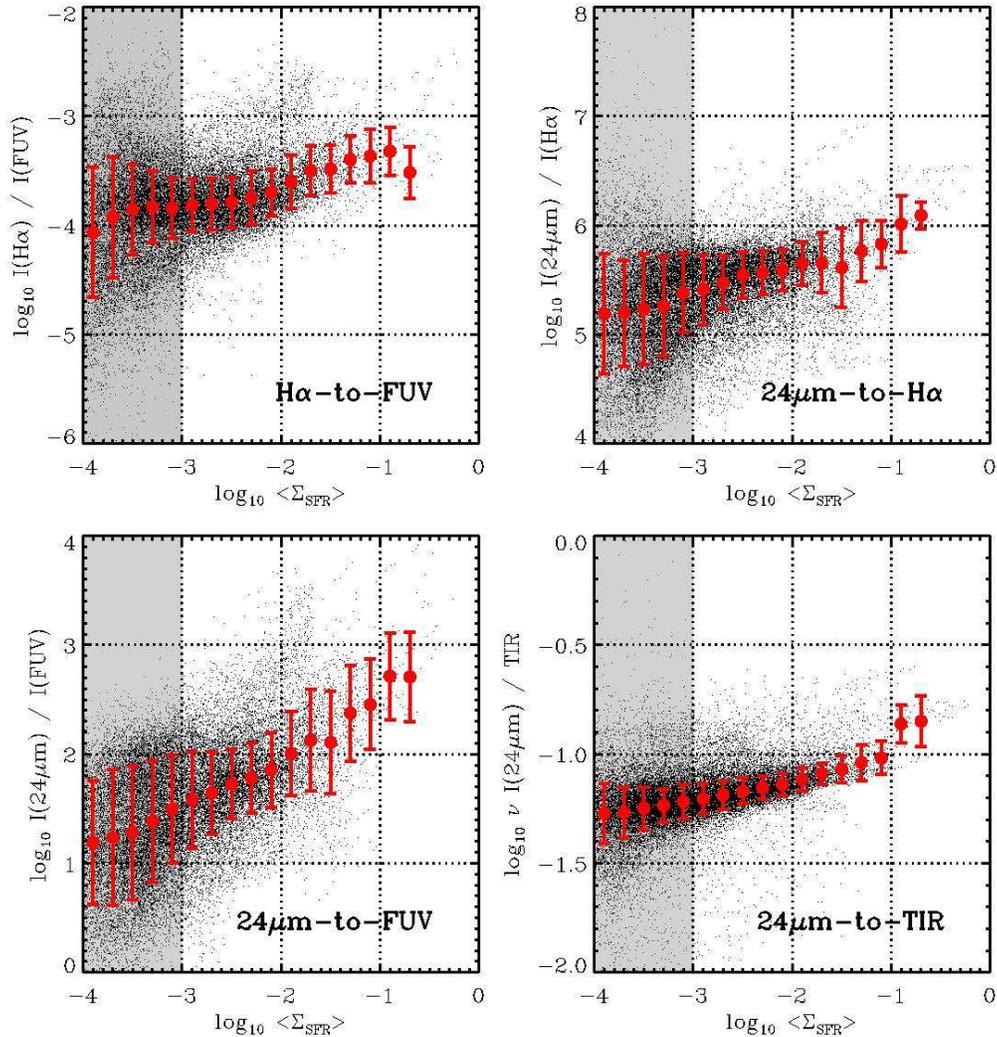}
\caption{{\em Ratios among terms in hybrid SFR tracers as a function
    of composite $\left< \Sigma_{\rm SFR} \right>$.} ({\em top left})
  Ratio of H$\alpha$ to FUV intensity along a line of sight. ({\em top
    right}) Ratio of 24$\mu$m to H$\alpha$ emission. ({\em bottom
    left}) Ratio of 24$\mu$m to FUV intensity. ({\em bottom right})
  Ratio of 24$\mu$m to TIR luminosity surface density ($\nu L_{\nu}$)
  calculated at $45\arcsec$ resolution ($\sim 2.2$~kpc at the median distance of our sample). The red points show median ratio in bins of $\left< \Sigma_{\rm SFR} \right>$ with error bars indicating $1\sigma$ scatter.}
\label{fig:sfr_colors}
\end{figure*}

\begin{figure*}
\plottwo{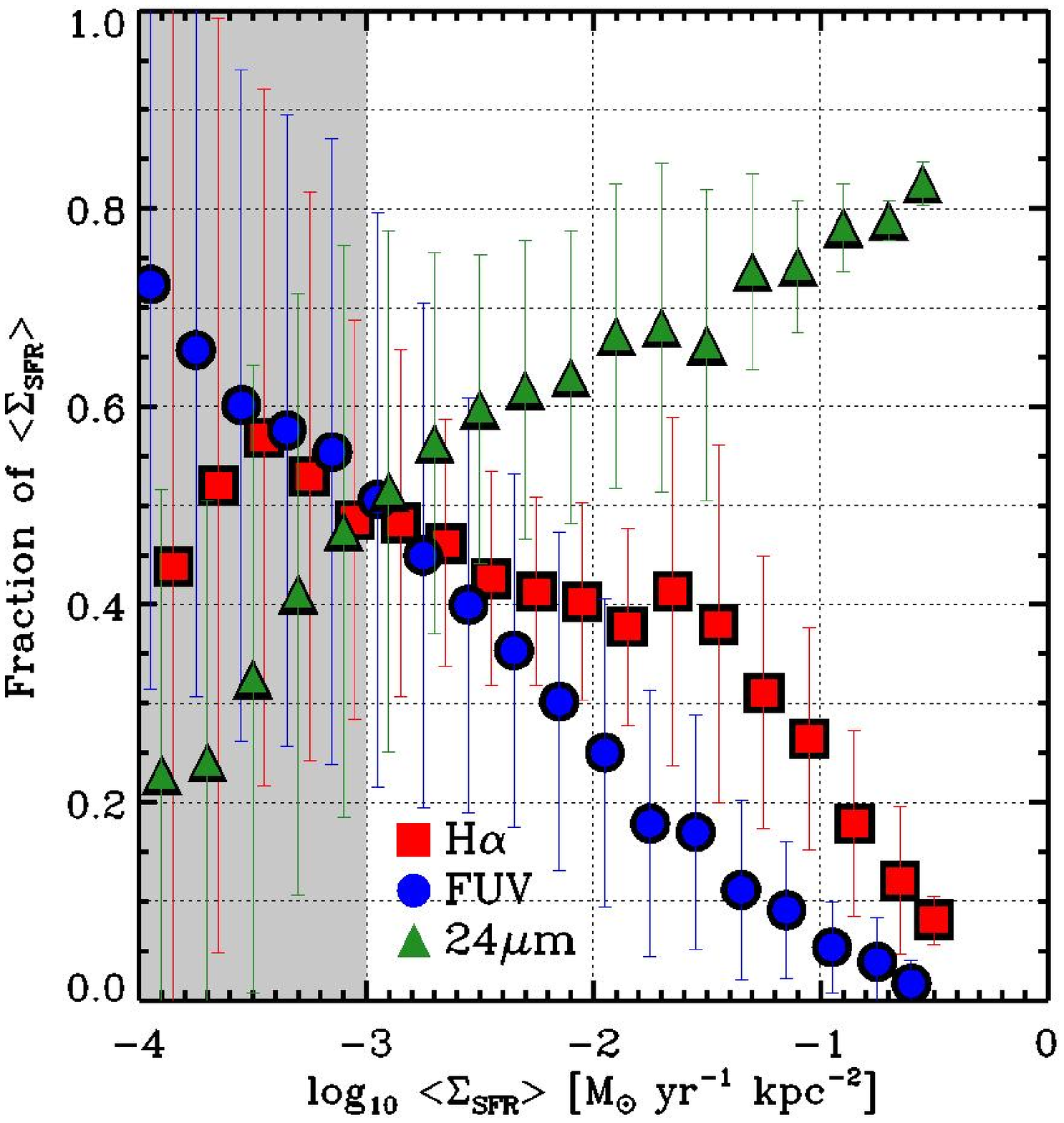}{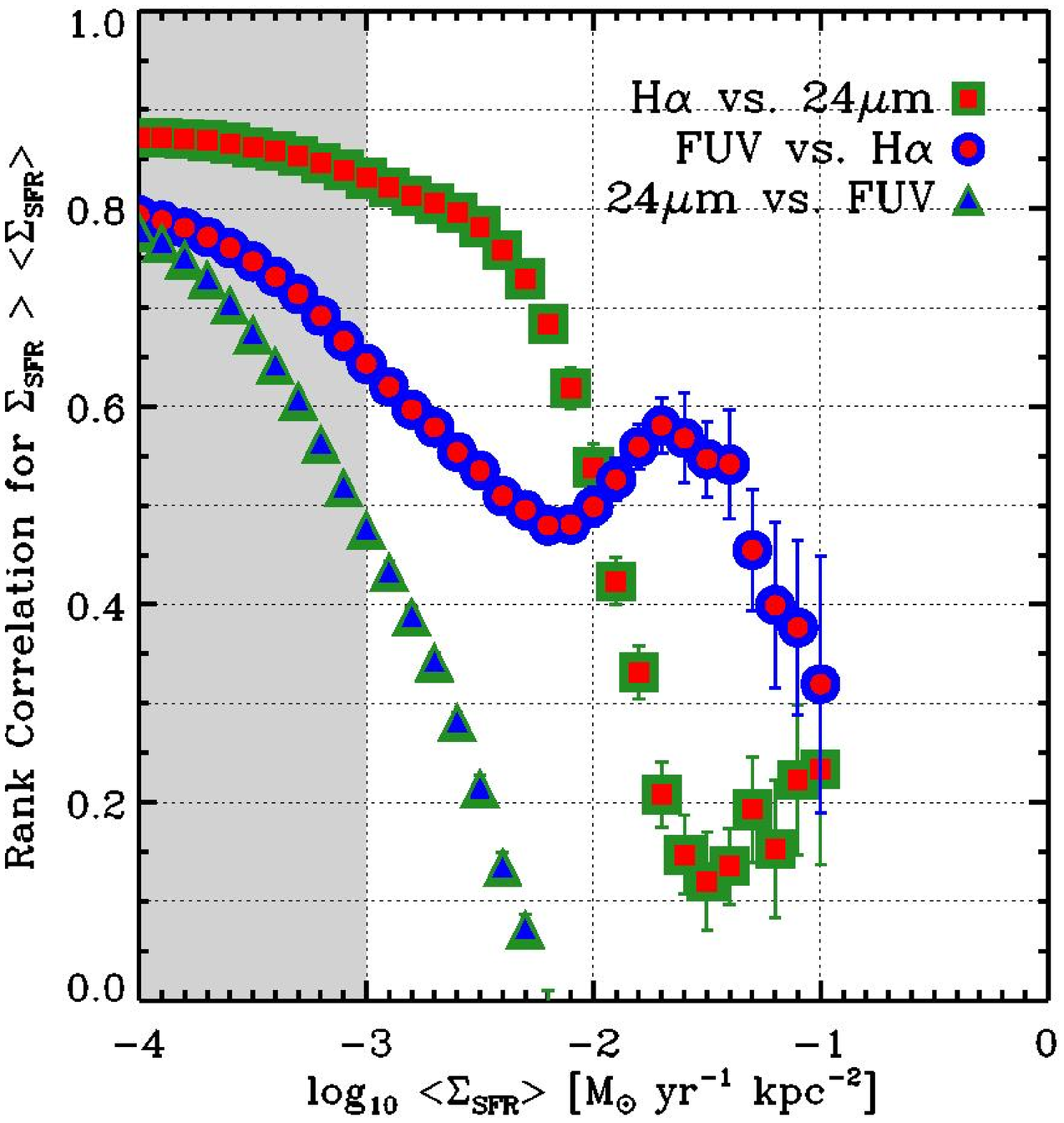}
\caption{{\em Fractional contribution and correlation among SFR
    terms:} ({\em left}) Median fractional contribution of different
  terms to $\Sigma_{\rm SFR}$ as a function of composite $\left<
  \Sigma_{\rm SFR} \right>$: ({\em red}) H$\alpha$, ({\em blue}) FUV,
  and ({\em green}) 24$\mu$m emission with $w = 1$ (so that the H$\alpha$ and 24$\mu$m components roughly sum to 1.0). ({\em right}) Rank
  correlation among these terms for all data above a threshold $\left<
  \Sigma_{\rm SFR} \right>$ as a function of that threshold.}
\label{fig:sfr_terms}
\end{figure*}

Dust absorbs H$\alpha$ and UV light, leaving only $\sim 20$--$40\%$ of
the original emission visible in a typical spiral galaxy
\citep[$\sim 1$--$2$~mag, e.g.,][]{KENNICUTT98B} and reprocessing the remainder into IR
emission. Correcting for this extinction represents a central
challenge to SFR estimation. 

Following extensive work in the literature, we will account for
extinction by combining H$\alpha$ and FUV with IR emission. Adding UV
and IR emission has an established pedigree
\citep[][]{BUAT92,MEURER95,MEURER99,CORTESE08}, while the combination
of H$\alpha$ and IR has been explored more recently
\citep[][]{CALZETTI07,KENNICUTT07,KENNICUTT09}. Both combinations
offer a powerful way to circumvent the problem of extinction.

Driven by the availability and quality of {\em Spitzer} data at
$\lambda = 24\mu$m, we work with linear combinations of H$\alpha$ and
FUV with IR emission at this wavelength. Several subtle effects
complicate the simple conversion of 24$\mu$m into SFR. First, the
24$\mu$m band does not capture a significant fraction of the total IR
luminosity for radiation fields $\lesssim 10^4$ times the Solar
Neighborhood value \citep{DRAINE07A}. Second, the exact fraction of
the total IR luminosity emitted near 24$\mu$m depends on the dust size
distribution, specifically the PAH fraction in the \citet{DRAINE07A}
models, and the mixture of radiation fields illuminating the
dust. Finally, producing 24$\mu$m does not require ionizing
photons. Indeed, stochastic heating by an older stellar population may
represent an important contribution in some regions \citep[e.g.,][]{MURPHY11}. This last point
is a general concern for the use of IR emission to estimate SFR, one
that may be somewhat alleviated at 24$\mu$m compared to longer
wavelengths. For this paper the critical point is that due to these
complicating factors, {\em the use of 24$\mu$m emission to infer SFRs
  has an empirical, not theoretical, foundation}. 

\subsection{Definitions and Approach}
\label{sec:w_defn}

We follow recent work in the field by defining a ``reference'' SFR,
which we assume to be correct, and bootstrapping the calibration of
more readily observed estimates to match this reference. We draw
reference SFRs from literature measurements at wavelengths where dust
is only a small concern, including radio continuum emission and Pa$\alpha$
emission \citep[][]{NIKLAS97,MURPHY11,CALZETTI07}. We also use UV and
H$\alpha$, corrected for extinction based on astrophysical
expectations, e.g., by the ``Balmer decrement'' or IRX-$\beta$
approach \citep[][]{STOREY95,OSTERBROCK06,KENNICUTT98B,CORTESE08}. In Section \ref{sec:w}, we give more details on these ``reference" measurements and use them to estimate the calibration of the 24$\mu$m term in hybrid H$\alpha$+24$\mu$m or UV+24$\mu$m tracers. Before doing so, we define our terms (this section), investigate the relationships among the components of these hybrid tracers (Section \ref{sec:comp}), and consider the effect of 24$\mu$m emission not associated with recent star formation (Section \ref{sec:cirrus}). 

Linear combinations of H$\alpha$+24$\mu$m or UV+24$\mu$m have the
desirable properties of working independent of
scale and breaking apart into ``obscured'' and ``unobscured''
terms in an easy-to-interpret way. This behavior comes at the cost of accuracy, as the conversion
of IR intensity to extinction may depend on $\Sigma_{\rm SFR}$ (i.e., luminosity), scale, or the
ratio of IR-to-unobscured tracer in a complex way. For example, see \citet[][]{CORTESE08} or the comparison among competing, often regime-dependent and non-linear, conversions of monochromatic 24$\mu$m intensity to SFR in \citet{CALZETTI10}.  We adopt the simple approach,
focusing on linear combinations in which the SFR is the sum of an
obscured and an unobscured term.

{\em A priori} we do not know the weight to apply to 24$\mu$m
intensity, $I_{\rm 24}$, in combination with an unobscured tracer,
$\Sigma_{\rm SFR,tracer}$, to recover $\Sigma_{\rm SFR}$. We solve for
this quantity, which we refer to as $w_{\rm tracer}$ and define as

\begin{equation}
\label{eq:w}
w_{\rm tracer} = \frac{(\Sigma_{\rm SFR,ref} - \Sigma_{\rm SFR,
    tracer})}{0.0025~I_{\rm 24}}~.
\end{equation}

\noindent Very simply, $w$ is the coefficient to convert 24$\mu$m into $\Sigma_{\rm SFR}$ in combination with a tracer of unobscured star formation.  It has units of M$_\odot$~yr$^{-1}$~kpc$^{-2}$~$\left( 400~{\rm MJy~sr}^{-1} \right)^{-1}$

Thus

\begin{equation}
\label{eq:w_sfr}
\Sigma_{\rm SFR} =  \Sigma_{\rm SFR, tracer} + w_{\rm tracer}~\frac{I_{\rm 24}}{400~{\rm MJy~sr}^{-1}}
\end{equation}

\noindent For integrated systems, one can derive $w$ by substituting
total SFR for $\Sigma_{\rm SFR}$ and $L_{\rm 24}$ for $I_{\rm
  24}$. For ease of comparison, we choose the normalization of $w$ so
that $w_{\rm H\alpha}=1$ recovers the \citet{CALZETTI07} result,
Equation \ref{eq:c07}.

It will be desirable to consider roughly how much star formation
occurs along a line of sight without too much emphasis on a particular
calibration. We define the ``composite'' $\Sigma_{\rm SFR}$, $\left<
\Sigma_{\rm SFR} \right>$, to be the median among $\Sigma_{\rm SFR}$
calculated in a variety of ways. The quantity depends on the adopted suite of
tracers, but represents a useful ordinate that we will plot throughout
the paper. Our $\left< \Sigma_{\rm SFR} \right>$ is the median of 1) from H$\alpha$ + $A_{\rm H\alpha}
= 1$~mag; 2-9)\footnote{$A_{\rm H\alpha}$ refers to the extinction (in magnitudes) of H$\alpha$ emission by dust. $A_{\rm FUV}$ is defined analogously for FUV emission in the GALEX FUV band.} from H$\alpha$+24$\mu$m and FUV+24$\mu$m with and
without 24$\mu$m cirrus subtraction, with only cirrus associated with
\hi\ subtracted, and with cirrus due to twice our adopted radiation
field (see Section \ref{sec:cirrus}). 10) FUV + $A_{FUV}$ estimated
from the UV spectral slope \citep[the ``IRX-$\beta$
  relation''][]{MUNOZMATEOS09B}; 11-12) FUV + $A_{\rm FUV}$ estimated
from the TIR/FUV ratio for young and middle-aged populations following
\citet{CORTESE08}. 

\subsubsection{Existing Work on 24$\mu$m-based ``Hybrid'' Tracers}

{\em H$\alpha$+24$\mu$m in ``\ion{H}{2} knots'':} \citet{CALZETTI07}
and \citet{KENNICUTT07} demonstrated that a linear combination of
H$\alpha$ and 24$\mu$m intensities recover ionizing fluxes inferred
from Pa$\alpha$ emission for ``HII knots,'' bright regions from $\sim
0.05$--$1.2$~kpc in size, and averages over the central parts of
galaxies. Assuming the Pa$\alpha$ to trace the true SFR,
\citet{CALZETTI07} found:Œ

\begin{eqnarray}
\label{eq:c07}
\Sigma_{\rm SFR} \left[ {\rm M}_\odot~{\rm yr}^{-1}~{\rm kpc}^{-2}
  \right] = 634~I_{\rm H\alpha}~\left[ {\rm erg~s}^{-1}~{\rm sr}^{-1} \right] + \\
\nonumber 0.0025~I_{\rm 24\mu m} \left[ {\rm MJy~sr}^{-1}\right]~.
\end{eqnarray}

\noindent The first term is Equation \ref{eq:sfr_ha}. The second term
attempts to account for H$\alpha$ emission obscured by dust. This
corresponds to $w_{\rm H\alpha} = 1$ (see our Equation \ref{eq:w_sfr}).

{\em H$\alpha$+24$\mu$m for Whole Galaxies:} \citet{KENNICUTT09}
compared combinations of 24$\mu$m and H$\alpha$ emission to H$\alpha$
fluxes corrected for extinction using the Balmer decrement. For whole
galaxies, they found $w_{\rm H\alpha} = 0.68$; for the SINGS galaxies
specifically, they found $w_{\rm H\alpha} = 0.52$. The Balmer
decrements used to calibrate these results appear uncertain, at least
for the SINGS galaxies \citep{MOUSTAKAS10}. We return to this result
below.

{\em FUV+24$\mu$m in Radial Profile:} \citet{LEROY08} proposed that
FUV+24$\mu$m emission could trace the recent SFR for large parts of
the disks of nearby galaxies \citep[see also][]{THILKER07}. They
suggested

\begin{eqnarray}
\label{eq:l08}
\Sigma_{\rm SFR} \left[ {\rm M}_\odot~{\rm yr}^{-1}~{\rm kpc}^{-2}
  \right] = 0.081~I_{\rm FUV}~\left[ {\rm MJy~sr}^{-1} \right] +
\\ \nonumber 0.0032~I_{\rm 24\mu m} \left[ {\rm MJy~sr}^{-1}\right]~.
\end{eqnarray}

\noindent They picked the 24$\mu$m coefficient, $w_{\rm FUV} = 1.3$,
to match other estimates in radial profile. In profile, this estimate
matched other tracers with $\approx 50\%$ scatter extending down to
$\Sigma_{\rm SFR}$ a few times
$10^{-4}$~M$_\odot$~yr$^{-1}$~kpc$^{-2}$.

\subsection{Relation Among Hybrid Tracer Components}
\label{sec:comp}

Figures \ref{fig:sfr_colors} and \ref{fig:sfr_terms} show how the
components of the hybrid tracers relate to one another. Figure
\ref{fig:sfr_colors} shows the ratios of H$\alpha$-to-FUV,
24$\mu$m-to-H$\alpha$, 24$\mu$m-to-FUV, and 24$\mu$m-to-TIR emission
as a function of $\left< \Sigma_{\rm SFR} \right>$. Black points show
individual kpc-resolution lines of sight. Red points plot the median
trend with error bars indicating $1\sigma$ scatter. Figure
\ref{fig:sfr_terms} shows the median fraction of the total $\left<
\Sigma_{\rm SFR} \right>$ contributed by each term as a function of
$\left< \Sigma_{\rm SFR} \right>$ (left) and the rank correlation
among the terms for data above a given $\left< \Sigma_{\rm SFR}
\right>$ as a function of that limiting value (right).

As $\left< \Sigma_{\rm SFR} \right>$ increases, FUV emission becomes fainter relative to both H$\alpha$ and
24$\mu$m emission. While FUV has about
the same magnitude as the 24$\mu$m term near $\left< \Sigma_{\rm SFR}
\right> \sim 10^{-3}$~M$_\odot$~yr$^{-1}$~kpc$^{-2}$, it contributes
negligibly, $\lesssim 10\%$, to the overall SFR by $\left< \Sigma_{\rm
  SFR} \right> \sim 10^{-1}$~M$_\odot$~yr$^{-1}$~kpc$^{-2}$. FUV
exhibits a significant correlation with H$\alpha$ even up to high
$\left< \Sigma_{\rm SFR} \right>$, while its correlation with 24$\mu$m
diminishes quickly, becoming consistent with no correlation or a weak
anti-correlation by $\left< \Sigma_{\rm SFR} \right> \sim
10^{-2}$~M$_\odot$~yr$^{-1}$~kpc$^{-2}$.

The ratio of 24$\mu$m to H$\alpha$ emission also increases with with increasing $\left< \Sigma_{\rm SFR} \right>$ though the trend is weaker than for the 24$\mu$m-to-FUV ratio. This is consistent with star formation becoming increasingly embedded in regions with high $\Sigma_{\rm SFR}$ and agrees with trends observed in individual SINGS galaxies by \citet{PRESCOTT07}. H$\alpha$ contributes $\sim 40$\% of the total SFR over the range, $10^{-3} < \left< \Sigma_{\rm SFR} \right> \sim 10^{-1}$~M$_\odot$~yr$^{-1}$~kpc$^{-2}$, equivalent to $\sim 1$~magnitude of extinction. We include few lines of sight with $\left< \Sigma_{\rm SFR} \right> > 10^{-1}$~M$_\odot$~yr$^{-1}$~kpc$^{-2}$, but in these regions H$\alpha$ contributes a small amount to the total SFR.

Over the whole data set, $\left< \Sigma_{\rm SFR} \right> > 10^{-3}$ M$_\odot$~yr$^{-1}$~kpc$^{-2}$, H$\alpha$ and 24$\mu$m exhibit the
strongest rank correlation among the terms. This correlation diminishes at high $\left< \Sigma_{\rm SFR} \right>$, with the correlation between H$\alpha$ and FUV becoming stronger than that between H$\alpha$ and 24$\mu$m at very high $\left< \Sigma_{\rm SFR} \right>$. However, we emphasize that again that we have little data at very high $\left< \Sigma_{\rm SFR} \right>$.

The bottom right panel of Figure \ref{fig:sfr_colors} shows fraction
of TIR emission emerging at 24$\mu$m increases from $\sim 5$ to $\sim
10\%$ across two decades in $\left< \Sigma_{\rm SFR} \right>$. This
magnitude roughly agrees with the calculations of \citet{DRAINE07A}
and the trend has the expected sense if dust heating increases with
$\left< \Sigma_{\rm SFR} \right>$. The small scatter in the
ratio suggests that 24$\mu$m can be used to estimate the the bolometric IR emission (though some of the narrowness in the scatter 
arises because we measure 24$\mu$m-to-TIR at the
coarse resolution of the 160$\mu$m data).

Thus Figures \ref{fig:sfr_colors} and \ref{fig:sfr_terms} illustrate the
utility of the 24$\mu$m band to help trace the SFR. It correlates
closely with H$\alpha$, traces the bolometric IR emission at higher
resolution and sensitivity than can be achieved near the peak of the IR SED, and offers a way to trace the dominant,
obscured component of $\Sigma_{\rm SFR}$. 

The figures also show how the FUV+24$\mu$m tracer operates. Unlike
H$\alpha$+24$\mu$m, where both terms contribute significantly across a
wide range of $\left< \Sigma_{\rm SFR} \right>$, the FUV+24$\mu$m
works as a true hybrid, blending tracers appropriate for different
regimes: FUV traces $\Sigma_{\rm SFR}$ at low levels where dust
extinction is weak and $24\mu$m traces $\Sigma_{\rm SFR}$ in dusty,
vigorously star-forming regions.

\section{A Physical Approach to 24$\mu$m ``Cirrus''}
\label{sec:cirrus}

\begin{figure}
\plotone{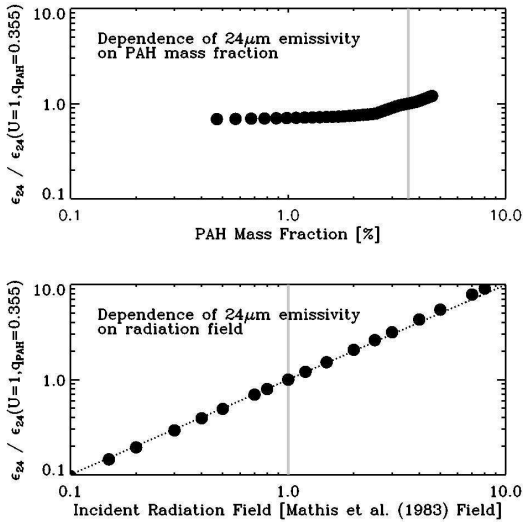}
\caption{{\em Dependence of 24$\mu$m emissivity on dust properties.}
  Dust emission per unit mass at $\lambda = 24\mu$m, $\epsilon_{\rm
    24}$, as a function of dust properties in the \citet{DRAINE07A}
  models. The upper panel shows the weak dependence of $\epsilon_{\rm
    24}$ on PAH mass fraction, a measure of the dust size
  distribution. The lower panel plots the dependence of $\epsilon_{\rm
    24}$ on the radiation field illuminating the dust. Gray vertical bars show the value at which we fix the quantity in the other panel. The dotted line in the lower panel
  shows a linear dependence for comparison.}
\label{fig:emis24}
\end{figure}

Weak radiation fields associated with older stellar populations may
still excite 24$\mu$m emission, leading to a potential 24$\mu$m
``cirrus.'' Here we define ``cirrus'' to mean infrared emission from dust heated by radiation fields not generated by recent star formation. Such radiation fields may illuminate either molecular or atomic gas, so that our ``cirrus" can come from dust anywhere in a galaxy (i.e., we do not use the term to refer to exclusively high-latitude or atomic gas). Figures \ref{fig:sfr_colors} and \ref{fig:sfr_terms}
suggest that this cirrus does not dominate the 24$\mu$m over our range
of interest. If the cirrus did dominate we would expect a breakdown in
the correlation between H$\alpha$ and 24$\mu$m emission and an
increase in the ratio of 24$\mu$m to other tracers at low $\left<
\Sigma_{\rm SFR} \right>$. Neither effect appears strong in the
data. However, even sub-dominant cirrus may represent an important
second order correction and a physical estimate of contamination is
important.

Proposals to identify and subtract cirrus emission tend to be
morphological, defining the cirrus as a smooth background and removing
it via image processing techniques such as iterative median filtering
\citep[e.g., see][]{GREENAWALT97,THILKER00,DALE07,RAHMAN10}. Such
approaches have very little utility at 1~kpc resolution, where the
contrast between star-forming regions 10--100~pc in size and the
embedding medium occurs almost entirely sub-resolution. We therefore
focus on an astrophysical cirrus estimate that leverages our
multiwavelength data. We estimate the amount of dust present in each
resolution element and the radiation field not associated with recent
star formation. Together, these give us an estimate of potential
cirrus contamination.

The 24$\mu$m cirrus will be the product of the amount of dust present,
$\Sigma_{\rm dust}$, and the emission per unit dust mass of dust
heated by sources other than recent star formation,
$\epsilon_{24}^{\rm cirrus}$. Then

\begin{equation}
\label{eq:cirrus_setup}
I_{\rm 24}^{\rm cirrus} = \epsilon_{24}^{\rm cirrus}~\Sigma_{\rm dust}~.
\end{equation}

\noindent The dust surface density depends on the gas surface density
and the dust-to-gas ratio, $DGR$. $DGR$, in turn, varies with
metallicity, at least to first order
\citep{DRAINE07B,MUNOZMATEOS09B,LEROY11}. The $24\mu$m emissivity
depends on the dust composition and incident radiation field. In
Figure \ref{fig:emis24} we plot these dependencies as they appear in
the \citet{DRAINE07A} dust models. The top panel shows the weak
dependence of $\epsilon_{24}^{\rm cirrus}$ on $q_{\rm PAH}$, the PAH
mass fraction. The lower panel shows how $\epsilon_{24}^{\rm cirrus}$
varies with the incident radiation field\footnote{Following
  \citet{DRAINE07A} we discuss $U$ in units of the local interstellar
  radiation field \citep[ISRF,][]{MATHIS83} and consider only changes
  to the intensity, not the color, of the field.}, $U$. To good
approximation over our regime of interest, $0.1 < U < 10$,
$\epsilon_{24}^{\rm cirrus}$ will vary linearly with the radiation
field not associated with recent star formation.

\subsection{Dust and $q_{\rm PAH}$ for Our Data}

To first order $I_{24}^{\rm cirrus}$ depends linearly on $\Sigma_{\rm
  dust}$ and $U$ and weakly on $q_{\rm PAH}$. We need to estimate each
of these terms for each line of sight.

From our fits using the \citet{DRAINE07A} models to the broadband IR
data, we know $\Sigma_{\rm dust}$ at the coarse ($\sim 40\arcsec$)
resolution of the 160$\mu$m data. To estimate $\Sigma_{\rm dust}$ at
our finer 1~kpc resolution we assume that the $DGR$ varies on large
spatial scales. With this assumption, we can take advantage of the
finer resolution of our ISM maps to estimate $\Sigma_{\rm dust}$ from
$DGR \times \Sigma_{\rm gas}$. Not much is known about the behavior of
the $DGR$ on kpc scales, but changes in metallicity do tend to be weak
over such scales \citep[e.g.,][]{ROSOLOWSKY08,MOUSTAKAS10}.

Using gas to trace dust on small scales also provides a breakdown
between dust associated with \hi\ and dust associated with
H$_2$. \hi\ is less directly associated with star formation than H$_2$
\citep[see references in][]{SCHRUBA11} and most studies of Milky Way
cirrus focus on \hi\ \citep[e.g.,][]{BOULANGER96}. We may be fairly
certain that dust associated with \hi\ and a weak radiation field is
not driven by star formation. We therefore experiment with cirrus
subtractions that use either only dust associated with \hi\ or all
dust.

Our fits using the \citet{DRAINE07A} models also yield $q_{\rm PAH}$,
which we assume to hold sub-resolution. This term only weakly
influences $\epsilon_{\rm 24}$ (Figure \ref{fig:emis24}, top panel) so
the approximation should have little impact.

\subsection{The Cirrus Radiation Field for Our Data}
\label{sec:u}

Estimating the radiation field not associated with recent star
formation represents the most challenging part of the
calculation. Following \citet{DRAINE07B}, we fit our data with a
two-component radiation field model. One component is heated by a
power law distribution of radiation fields extending to very high $U$,
presumably due to star formation. The other component is heated by a
single, weaker radiation field with magnitude represented by the free
parameter $U_{\rm min}$. In our fits, $U_{\rm min}$ exhibits a narrow
5--95\% range $U=0.7$--$4.0$, maximum $\approx 9$, and median $\approx
1.35$. It displays a positive correlation with $\left< \Sigma_{\rm
  SFR} \right>$ ($r_{\rm corr} \sim 0.4$) and the stellar mass surface
density ($r_{\rm corr} \sim 0.4$) and an anti-correlation with
galactocentric radius ($r_{\rm corr} \sim -0.3$). See
\citet{MUNOZMATEOS09B} for more details.

$U_{\rm min}$ can be interpreted as the ambient radiation field due
to old stars but this interpretation is not unique. In the fit this is
only the minimum radiation field illuminating the dust. A high $U_{\rm
  min}$ may still be driven by star formation so that discarding all
emission associated with $U_{\rm min}$ may represent an overestimate
of the cirrus. For instance, a square kiloparsec centered on the Sun would
include several OB associations \citep[Sco-Cen, Perseus, Orion;
  see][]{REIPURTH08A,REIPURTH08B}. These contribute substantially to
the average ISRF so that the dominant heating for our local $U=1$
cannot be said to come from a truly ``old'' population
\citep[see][]{MATHIS83}.

Because of this uncertainty, we undertake an independent estimate of the radiation field driving the cirrus. This investigation appears in Appendix \ref{sec:ufield}. We conclude that our data suggest a cirrus driven by a radiation field $U_{\rm cirrus} = 0.5 U_{\rm min}$ with typical magnitude $U_{\rm cirrus} \sim 0.6$. Such a field would drive an equilibrium dust temperature $\sim 1$~K lower than a $U=1$ field, consistent with the $\sim 16$--$19$~K observed in \hi\ and the outskirts of molecular clouds in the Solar Neighborhood \citep[e.g.,][]{BOULANGER96,SCHNEE08,LEE11}. When we refer to ``cirrus" or ``cirrus subtraction" throughout the rest paper, we mean that we have calculated the 24$\mu$m cirrus emission associated with cirrus powered by $U_{\rm cirrus} = 0.5 U_{\rm min}$. When we refer to ``double'' or ``twice" cirrus we calculated the 24$\mu$m cirrus emission associated with $U_{\rm cirrus} = U_{\rm min}$, i.e., twice our best estimate.

\subsection{Effect of Cirrus Subtraction}
\label{sec:cirruseffect}

We combine our dust SED fits (to radial profiles) and our gas maps to estimate the 24$\mu$m cirrus everywhere across our sample. We assume that $q_{\rm PAH}$ and the dust-to-gas ratio remain fixed within a radial profile and that $U_{\rm cirrus} = 0.5~U_{\rm min}$. Combined with our higher-resolution gas maps this gives us a local estimate of $I_{\rm 24}^{\rm cirrus}$ for each line of sight. When we refer to "cirrus-subtracted" or corrected data, this intensity has been subtracted from the observed 24$\mu$m intensity.

\begin{figure*}
 \plotone{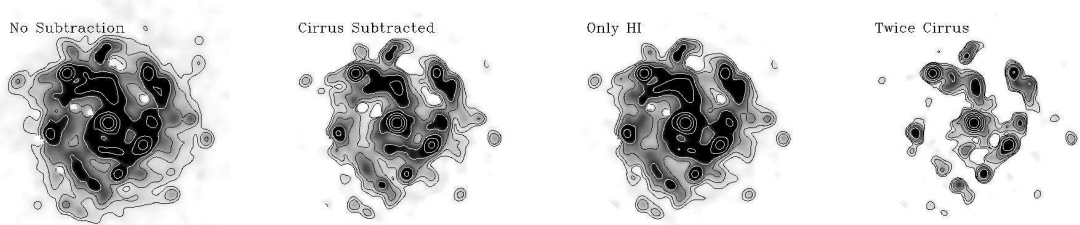}
 \caption{{\em Illustration of the cirrus subtraction for NGC~3184.}
   Infrared emission at $\lambda = 24\mu$m and 1~kpc
   resolution. Contours run from 0.1 to 6.4~MJy~sr$^{-1}$, stepping by
   factors of two. Contours $\geq 0.8$~MJy~sr$^{-1}$ are
   white. Individual panels show left to right: no subtraction,
   24$\mu$m emission after our cirrus subtraction, 24$\mu$m emission
   after our cirrus subtraction applied only for dust associated with
   \hi , and 24$\mu$m emission after cirrus subtraction using double
   the recommended radiation field.}
 \label{fig:showcirrus}
\end{figure*}

\begin{deluxetable}{lcccc}
\tablecaption{Effects of Cirrus Subtraction\tablenotemark{a}} 
\tablehead{ \colhead{Description} & \colhead{Median $f_{\rm cirr}$ + 67\% Range\tablenotemark{b}} &
\colhead{$\log_{10} \Sigma_{\rm SFR}^{0.95}$\tablenotemark{c}}}
\startdata
Best Cirrus Estimate & 0.19 (0.05--0.32) & \nodata \\
Only Dust with \hi\ & 0.10 (0.03--0.17) & \nodata \\ 
Twice Cirrus & 0.36 (0.10--0.58) & $-2.4$
\enddata
\tablenotetext{a}{See Section \ref{sec:cirruseffect}.}
\tablenotetext{b}{Median integrated cirrus fraction and 67\% range by galaxy.}
\tablenotetext{c}{$\log_{10} \left< \Sigma_{\rm SFR} \right>$ below which 95\% of IR is deleted.}
\label{tab:cirrus}
\end{deluxetable}

\begin{figure}
\plotone{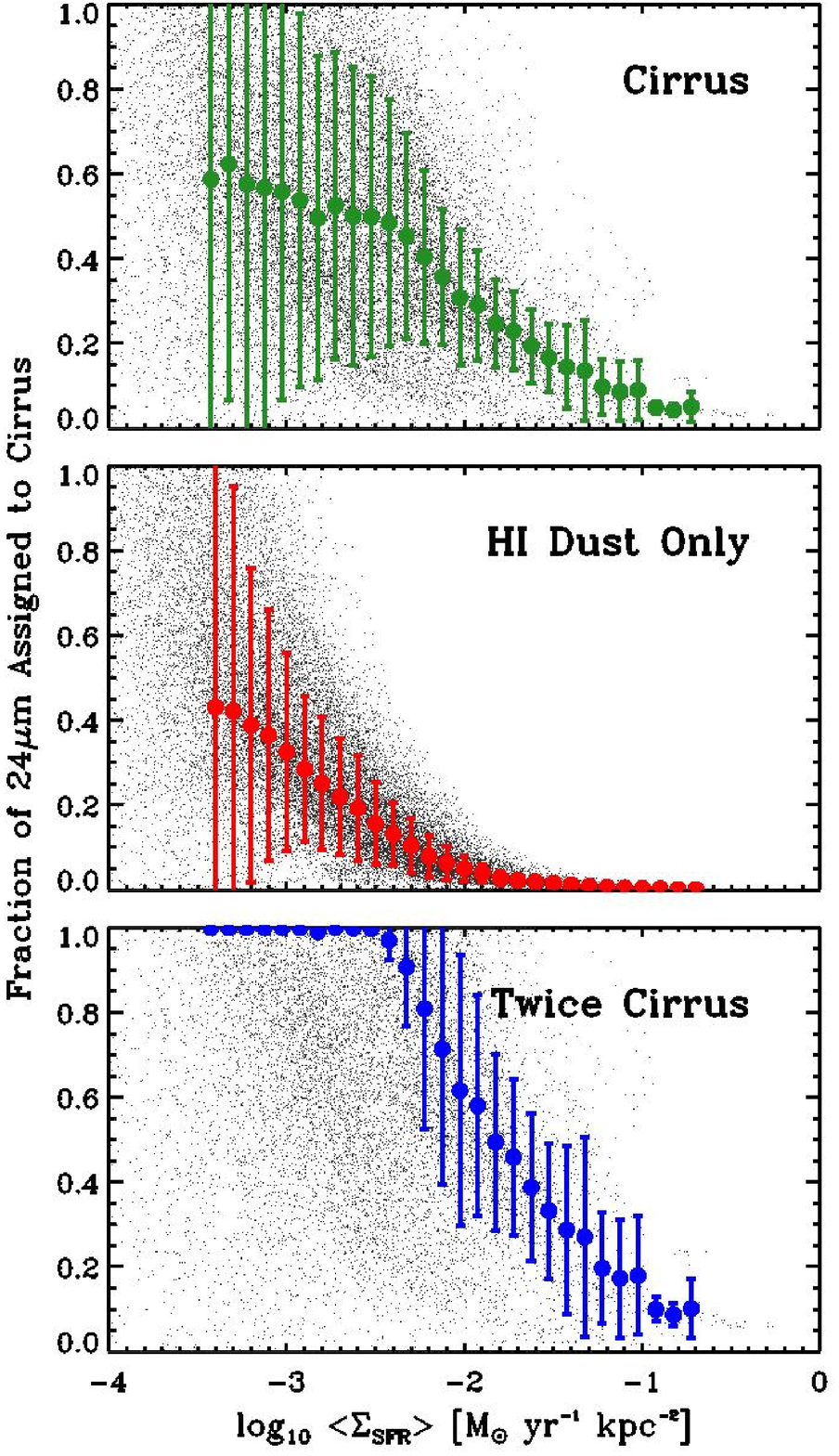}
\caption{{\em Fraction of 24$\mu$m emission identified as cirrus} as a
  function of $\left< \Sigma_{\rm SFR} \right>$ from top to bottom:
  ({\em green}) our best-estimate cirrus; ({\em red}) emission only from dust associated with \hi ; ({\em
    blue}) a cirrus powered by a radiation field twice our best-fit
  value. Individual
  points show results for 1~kpc resolution elements, colored points
  show the median and $1\sigma$ scatter after binning the data by
  $\left< \Sigma_{\rm SFR} \right>$.}
\label{fig:fcirrus}
\end{figure}

Figure \ref{fig:showcirrus} displays our cirrus subtraction
graphically for one face-on spiral. Figure \ref{fig:fcirrus} plots the
fraction of 24$\mu$m emission deemed cirrus by our calculation as a
function of composite $\left< \Sigma_{\rm SFR} \right>$. In both
plots, we illustrate the impact of our methodology by also plotting
results for two variant cirrus subtractions: one using only dust associated with
\hi\ and one using twice our adopted radiation field
(see Section \ref{sec:u}). Table \ref{tab:cirrus} synthesizes these
data into a few key numbers: the median and $1\sigma$ range of cirrus
emission subtracted by galaxy and the limiting $\left< \Sigma_{\rm
  SFR} \right>$ below which 95\% of lines of sight are blanked by each
cirrus approach.

The cirrus subtraction suppresses faint emission at large radius and
enhances the contrast between bright regions and their
surroundings. Subtracting only the component associated with
\hi\ removes the low-lying, extended component but has almost no
effect on the bright part of the galaxy where most of the star
formation occurs. Setting the radiation field to twice its nominal value
isolates only the brightest star formation, implying no extinction for
H$\alpha$ or UV emission over the rest of the galaxy.

The magnitude of the 24$\mu$m cirrus identified by our approach resembles but somewhat exceeds the fraction of emission identified as powered by old stars in an analysis of integrated SINGS and LVL SEDs by \citet{LAW11}. \citet{LAW11} found typically $\sim 7\%$ of 24$\mu$m luminosity in these galaxies, on average, to come from old stars. This resembles the fraction of \hi -associated cirrus in our data but is about half the total cirrus that we find. Qualitatively, both studies find that dust emission from old stars represents a second-order, but still potentially important, correction in nearby disk galaxies. 

\section{The $24\mu$m Term in Hybrid SFR Tracers}
\label{sec:w}

As described in Section \ref{sec:hybrid}, the calibration of the
24$\mu$m component in hybrid tracers is empirical and may vary with
scale, target, and ``unobscured'' tracer. Given this empirical
underpinning, any new approach to handling 24$\mu$m emission requires
that one re-derive, or at least verify, the adopted calibration $w$.  Recall that $w_{\rm tracer}$, defined in Equation \ref{eq:w}, is the factor to linearly scale $24\mu$m intensity in combination with an unobscured tracer (the ``tracer" indicated in the subscript) to account for dust-obscured star formation (Section \ref{sec:w_defn}). The lack of a ``gold-standard'' reference SFR
hampers this effort, but we are able to draw several independent
estimates of $\Sigma_{\rm SFR}$ from the literature to check
$w_{\rm H\alpha}$ and $w_{\rm FUV}$ for our approach and sample.

\subsection{Expectation in the IR-Dominated Case}
\label{sec:irdom}

From the typical ratio of 24$\mu$m-to-TIR emission we can estimate the
limiting $w$, regardless of unobscured tracer, by considering the case where the IR term dominates
$\Sigma_{\rm SFR}$. The median ratio of 24$\mu$m to TIR emission
across our sample is $\sim 0.07$ (Figure \ref{fig:sfr_colors}). If we
equate TIR emission with the reprocessed bolometric light from an
embedded, continuous starburst \citep[][corrected to our adopted
  IMF]{KENNICUTT98B} this implies $w \sim 2.4$. In fact, the
24$\mu$m-to-TIR ratio increases with increasing $\left< \Sigma_{\rm
  SFR} \right>$ so that for lines of sight where IR dominates it may
be $\sim 0.1$ or higher which implies a lower $w \sim 1.7$. Based on
this calculation, we expect $w \lesssim 2$. It would be possible, in principle, to have $w$ higher than this value for a specific regime but such a calibration could not successfully extend to the case where the overwhelming majority of star formation is embedded and visible only through the IR.

\subsection{H$\alpha$+24$\mu$m}

\subsubsection{Previous Work on the SINGS Sample}

\begin{figure}
\plotone{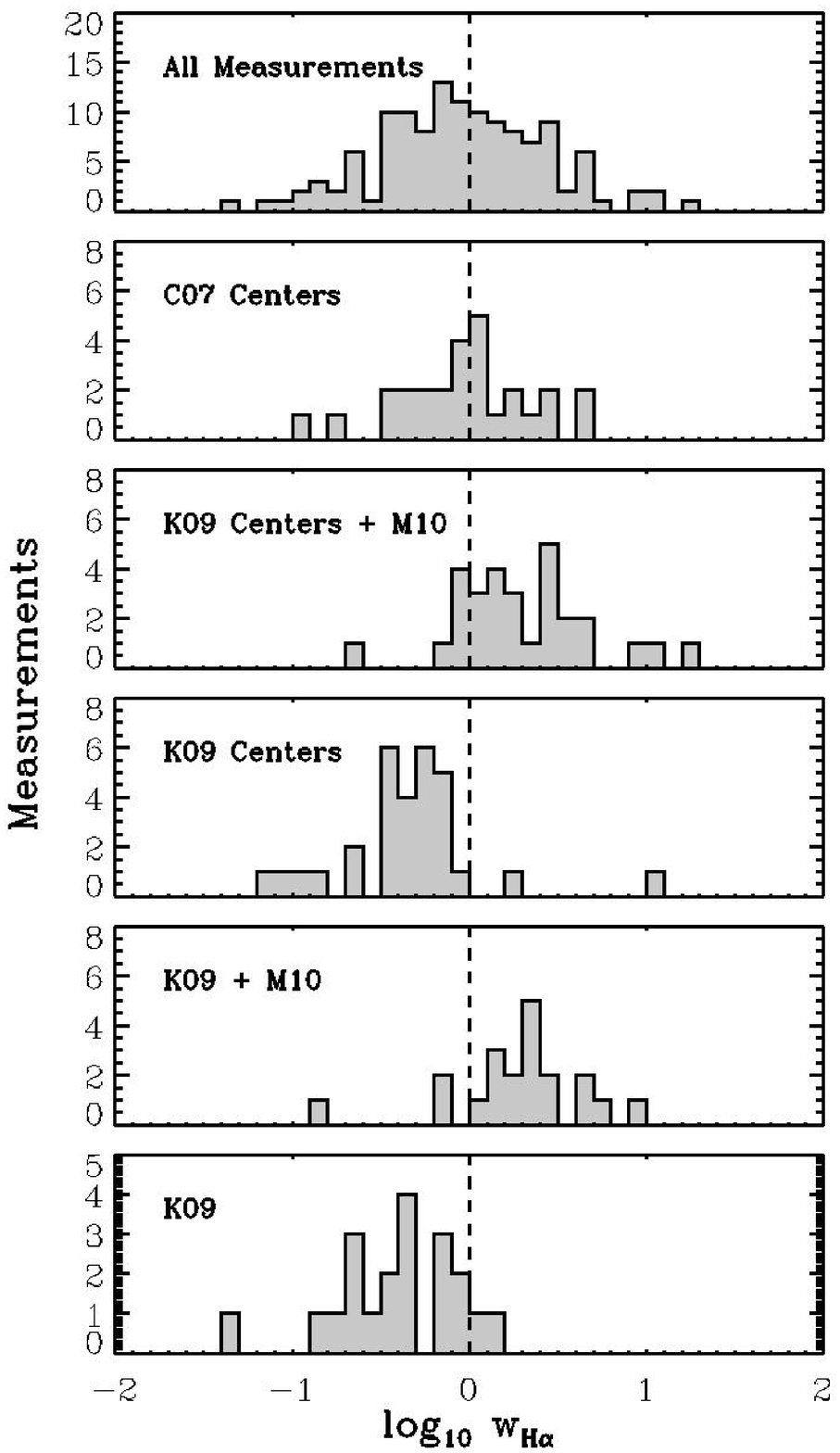}
\caption{{\em Literature calibrations of the H$\alpha+24\mu$m tracer
    in the SINGS sample.} The factor, $w_{\rm H\alpha}$, to be applied
  to 24$\mu$m emission to recover the recent star formation rate in
  linear combination with H$\alpha$ for integrals (see definition in Equation \ref{eq:w}) over large areas in
  SINGS galaxies. From top to bottom, the plots show histograms of $\log_{10} w_{\rm
    H\alpha}$ for (1) all measurements, allowing repeats among galaxies; (2) $50 \times 50\arcsec$
  central regions referenced to Pa$\alpha$ \citep[``C07
    Centers''][]{CALZETTI07}; (3)  $20 \times 20\arcsec$ central regions measured by \citet{KENNICUTT09} and referenced to
  ``circumnuclear'' Balmer decrements measured by \citet[][``K09
    Centers+M10'']{MOUSTAKAS10};  (4) the same central regions
  referenced to Balmer decrement-corrected H$\alpha$ from \citep[``K09
    Centers''][]{KENNICUTT09}; (5) whole galaxies measured by \citet[][]{KENNICUTT09} and corrected for extinction using the Balmer decrements measured by \citet[][``K09+M10'']{MOUSTAKAS10}. (6) whole galaxies applying the Balmer decrements measured by
  \citet[][``K09'']{KENNICUTT09}.}
\label{fig:sings_wha}
\end{figure}

\begin{deluxetable*}{lcccccc}
\tablecaption{$\log_{10} w_{\rm H\alpha}$ by Study for SINGS Galaxies\label{tab:sings_wha}} 
\tablehead{
\colhead{} & 
\colhead{C07} & 
\colhead{K07} & 
\colhead{K09} & 
\colhead{K09+M10\tablenotemark{a}} & 
\colhead{K09} & 
\colhead{K09+M10\tablenotemark{a}} \\
\colhead{} &
\colhead{Centers} & 
\colhead{Regions} &  
\colhead{Centers} & 
\colhead{Centers} & 
\colhead{Galaxies} & 
\colhead{Galaxies}
}
\startdata
Median $\log_{10} w_{\rm H\alpha}$\tablenotemark{b} & $0.0$ & $0.08$ & $-0.32$ & $0.25$ & $-0.32$ & $0.24$ \\
Scatter ($1\sigma$) $\log_{\rm 10} w_{\rm H\alpha}$ & 0.37 & 0.15 & 0.24 & 0.36 & 0.39 & 0.52 
\enddata
\tablenotetext{a}{Fluxes from \citet{KENNICUTT09} and \citet{DALE07} with \citet{MOUSTAKAS10} Balmer decrements.}
\tablenotetext{b}{Omitting data that yield negative or zero $w$.}
\end{deluxetable*}

Our targets heavily overlap the SINGS sample, which has acted as a
proving ground for combining H$\alpha$ and
24$\mu$m. \citet{CALZETTI07} and \citet{KENNICUTT07} determined
$w_{\rm H\alpha}$ for H$\alpha$ peaks using Pa$\alpha$ as a reference
SFR. \citet{KENNICUTT07} found $w_{\rm H\alpha} = 1.23$ and
\citet{CALZETTI07} derived $w_{\rm H\alpha} = 1.0$. This difference
presumably results from different geometry, escape fraction, and dust
properties between M51 and the larger \citet{CALZETTI07} sample, which
included M51. \citet{CALZETTI07} found that their 24$\mu$m coefficient
applies without modification to integrals over the central $50\arcsec$
of their targets. Subsequently, \citet{KENNICUTT09} considered
integrated SFRs for whole galaxies and found a 24$\mu$m term $w_{\rm
  H\alpha} = 0.68$, and $w_{\rm H\alpha} \approx 0.52$ for the SINGS
sample specifically. They also found significant scatter in $w_{\rm
  H\alpha}$ from galaxy-to-galaxy.

\citet{KENNICUTT09} interpreted their low $w_{\rm H\alpha}$ to imply
contamination of integrated measurements by a substantial cirrus
component. However, they based $w_{\rm H\alpha}$ on Balmer decrement
extinctions that were later revised to much higher values by
\citet{MOUSTAKAS10}. These revised estimates should represent an
improvement over those used in \citet{KENNICUTT09} (J. Moustakas,
priv. comm.). The revision affects both circumnuclear spectra and the
radial strips used for galaxy averages \footnote{\citet{KENNICUTT09}
  also work with a larger sample of IRAS 25$\mu$m fluxes and Balmer
  decrements measured by \citet{MOUSTAKAS06}. These measurements were
  not reconsidered with the improved approach of \citet{MOUSTAKAS10}
  but because of their higher S/N they may be less affected by the
  revision than the \citet{MOUSTAKAS10} radial strip measurements
  (R. C. Kennicutt, priv. comm.). ``Circumnuclear'' Balmer decrements
  also changed significantly from \citet{KENNICUTT09} to
  \citet{MOUSTAKAS10} and the quoted S/N for the \citet{MOUSTAKAS06}
  data is similar to that for the \citet{MOUSTAKAS10}
  ``Circumnuclear'' spectra.}.

Figure \ref{fig:sings_wha} and Table \ref{tab:sings_wha} show $w_{\rm
  H\alpha}$ for the central parts of galaxies measured by
\citet[][]{CALZETTI07} referencing to Pa$\alpha$ and for both
integrated measurements and the central parts of galaxies referenced
to Balmer decrement extinctions by \citet[][]{KENNICUTT09}. We also
plot $w_{\rm H\alpha}$ from applying the revised extinctions by
\citet{MOUSTAKAS10} to the \citet{KENNICUTT09} measurements. The top
panel of Figure \ref{fig:sings_wha} shows a histogram of all $w_{\rm
  H\alpha}$ measurements, allowing repeats among galaxies using
different reference SFRs. The bottom panel shows $w_{\rm H\alpha}$
broken down by study. Table \ref{tab:sings_wha} reports the median and
scatter in log$_{10}$ $w_{\rm H\alpha}$ by study. \citet{CALZETTI07}
did not publish their measurements of individual regions, but Table
\ref{tab:sings_wha} includes the M51 measurements by
\citet{KENNICUTT07}. However, both authors emphasize that because of
their heavy image processing these ``\ion{H}{2} knot'' calibrations
should not apply to large parts of galaxies.

The ensemble of measurements in Figure \ref{fig:sings_wha} yields
median $w_{\rm H\alpha} = 0.9$, just below the \citet{CALZETTI07}
value. The data exhibit significant scatter, $1\sigma$ in $\log_{\rm
  10} w_{\rm H\alpha}$ is $\approx0.45$~dex. This scatter includes
systematic shifts due to choice of reference SFR and so reflects both
an uncertainty and a true scatter. Even for a fixed reference SFR
$w_{\rm H\alpha}$ still scatters significantly ($1\sigma \approx
0.15$--$0.52$~dex) from galaxy to galaxy. The calibration of $w_{\rm
  H\alpha}$ for large parts of SINGS galaxies thus appears very
uncertain due to both uncertainty in the reference SFR and
galaxy-to-galaxy scatter.

\subsubsection{$w_{\rm H\alpha}$ in Our Data}

\begin{figure*}
\plotone{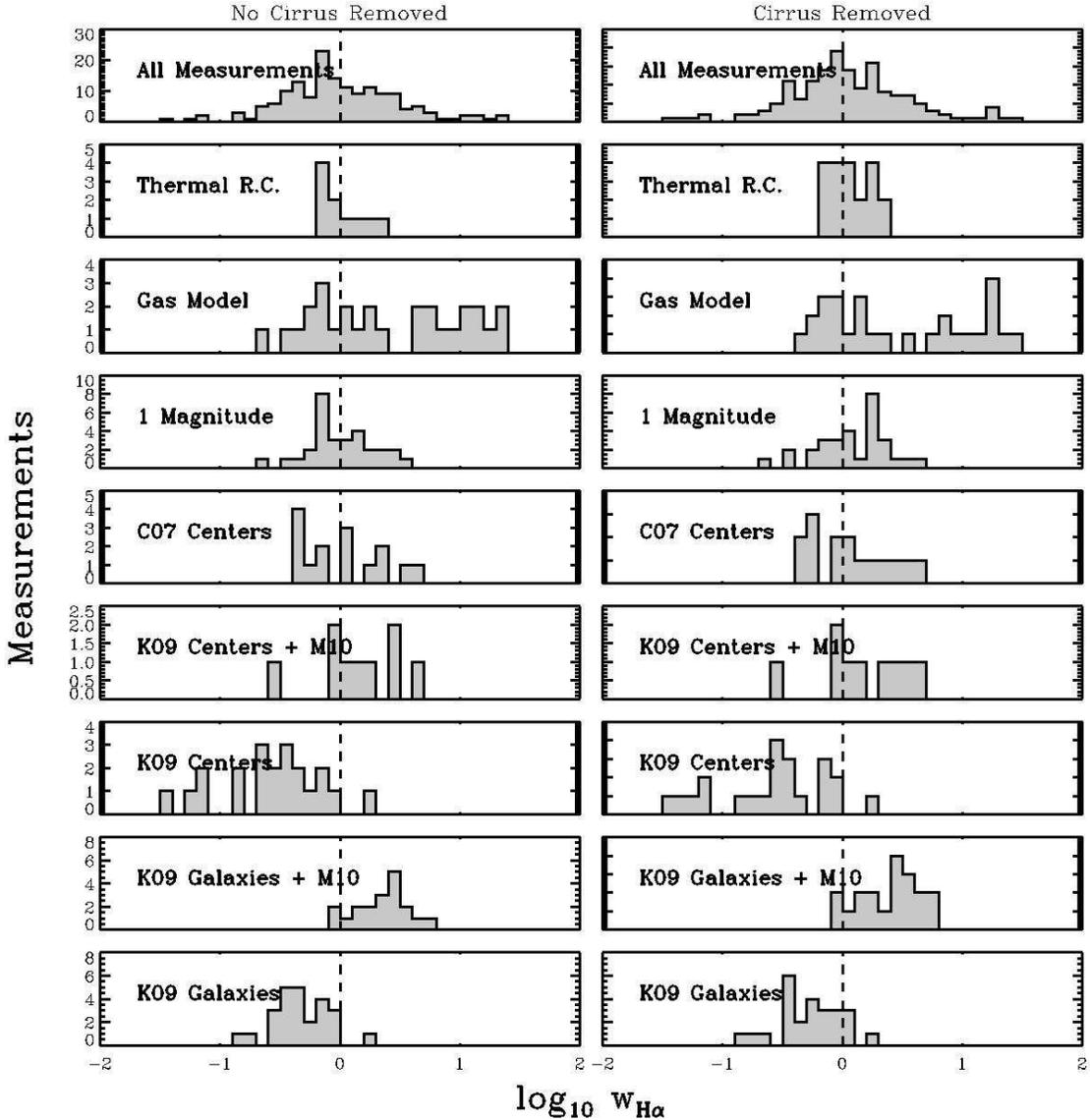}
\caption{{\em Calibration of the H$\alpha$+24$\mu$m tracer:} The
  factor, $w_{\rm H\alpha}$, to be applied to 24$\mu$m emission to
  recover the recent star formation rate in linear combination with
  H$\alpha$ (see definition in Equation \ref{eq:w}) for our targets with various estimates of the true
  extinction. The top row shows the ensemble of all $w_{\rm H\alpha}$ determinations for 24$\mu$m emission. The left column shows results for 24$\mu$m maps with no cirrus correction and the right panel shows $w_{\rm H\alpha}$ determined from 24$\mu$m maps that have been corrected for cirrus contamination. In addition to
  the estimates described in Figure \ref{fig:sings_wha} we use
  extinction estimates based on gas column \citep{WONG02}, thermal
  radio continuum \citep{NIKLAS97}, and a fixed $A_{\rm H\alpha} =
  1$~mag. Galaxies do not repeat in an individual determination but
  repeat among estimates and in the top row. A vertical line indicates $w_{\rm H\alpha} =
  1$.}
\label{fig:our_wha}
\end{figure*}

\begin{deluxetable}{lcc}
\tablecaption{$\log_{10} w_{\rm H\alpha}$ 24$\mu$m Term in the Hybrid H$\alpha$+24$\mu$m Tracer}
\tablehead{ \colhead{Reference SFR} & \colhead{No Cirrus} & \colhead{Cirrus}}
\startdata
Extinction-Free Estimates \\
... Pa-$\alpha$ Centers (C07) & $0.05 \pm 0.41$ & $0.08 \pm 0.44$\\
... Thermal R.C. (N95,N97) & $0.00 \pm 0.18$ & $0.08 \pm 0.19$ \\
\\
Balmer Decrements + H$\alpha$ \\
... Galaxy Centers (M10) & $0.13 \pm 0.34$ & $0.14 \pm 0.27$ \\
... Whole Galaxies (M10) & $0.40 \pm 0.16$ & $0.41 \pm 0.28$ \\
... Galaxy Centers (K09) & $-0.59 \pm 0.43$ & $-0.51 \pm 0.51$ \\
... Whole Galaxies (K09) & $-0.35 \pm 0.24$ & $-0.28 \pm 0.27$ \\
\\
Other Estimates \\
... Gas-Column\tablenotemark{a} + H$\alpha$ & $0.61 \pm 0.88$ & $0.58 \pm 0.97$ \\
... $A_{\rm H\alpha}=1$ Mag + H$\alpha$ & $-0.06 \pm 0.25$ & $0.14 \pm 0.28$ \\
\\
Average of All Estimates & $-0.07 \pm 0.47$ & $0.01 \pm 0.43$
\enddata
\tablenotetext{a}{Extinction based on column density of gas following "hybrid model" in \citet{WONG02}.}
\label{tab:wha}
\end{deluxetable}

\begin{figure}
\plotone{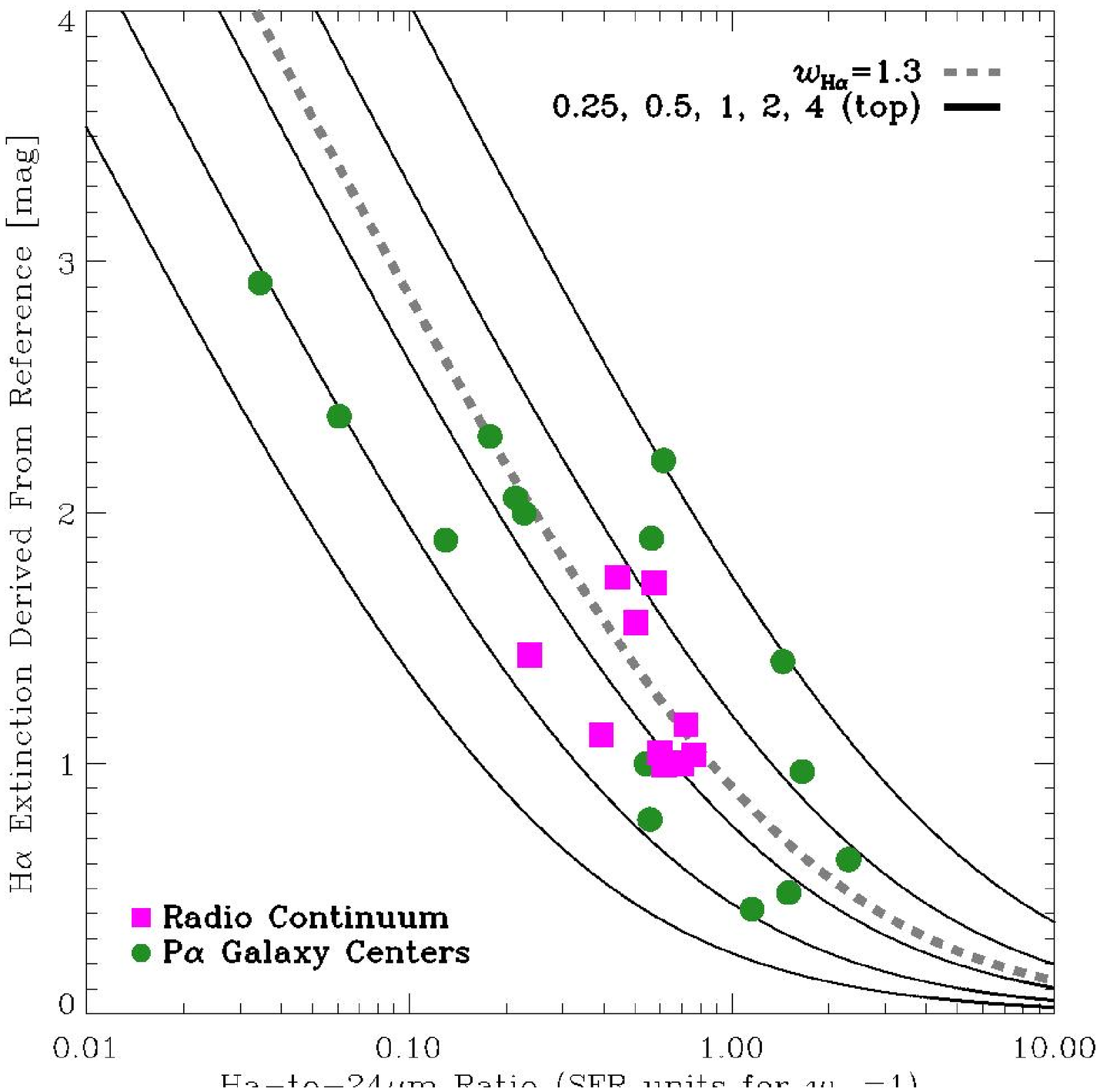} \plotone{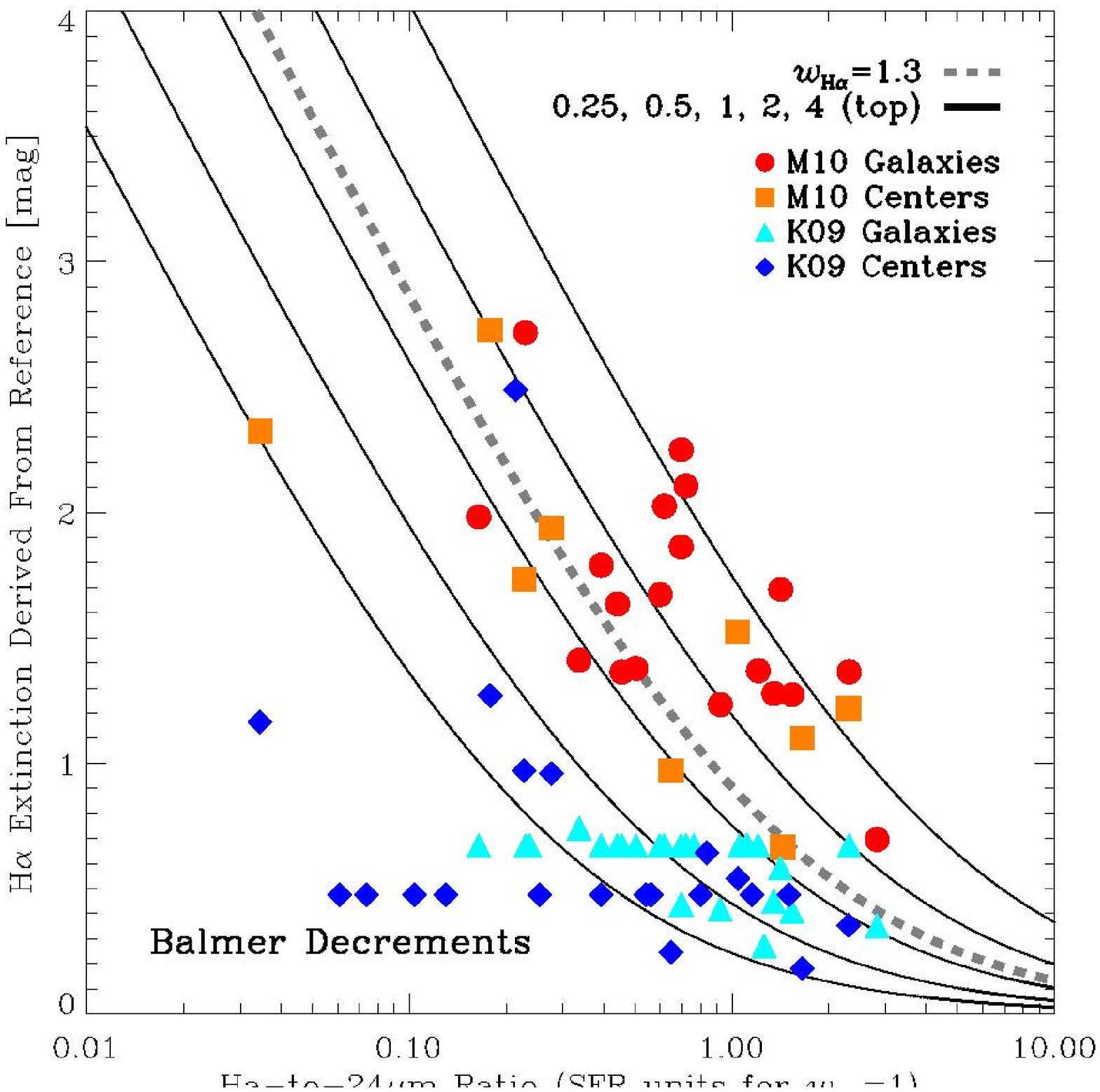}
\caption{{\em Extinction as a Function of H$\alpha$-to-24$\mu$m Ratio:} H$\alpha$ extinction drawn from the literature ($y$-axis) as a function of H$\alpha$-to-24$\mu$m ratio in our data ($x$-axis) measured in star formation surface density units (taking $w_{\rm H\alpha}=1$ for the ratio). Solid lines plot the expected relation for fixed values of $w_{\rm H\alpha}$ --- 0.25 (bottom), 0.5, 1.0, 2, and 4 (top); see the definition of $w$ in Equation \ref{eq:w}. The top panel shows measurements for extinction estimated from radio continuum and P$\alpha$ measurements. The bottom panel shows extinctions estimated from Balmer decrements, marked by study. All measured H$\alpha$-to-24$\mu$m ratios have been corrected for cirrus emission. The dashed gray line shows our adopted $w_{\rm H\alpha} = 1.3$. The different studies shown here all target the same set of galaxies, so that figure illustrates the uncertainty in the determination of $w_{\rm H\alpha}$ and the contradictory results achieved by recent studies.}
\label{fig:aha_vs_ha24}
\end{figure}

We calculate $w_{\rm H\alpha}$ for our approach and data set. To do so, we use a reference SFR calculated from H$\alpha$ plus an extinction correction. We draw the extinction correction from a variety of literature estimates based on , Balmer decrements, contrasting P$\alpha$- and H$\alpha$-emission, and contrasting radio continuum and H$\alpha$ emission. These are measured variously for either whole galaxies or central regions. For each extinction estimate, we calculate $w_{\rm H\alpha}$ following Equation \ref{eq:w}, reporting the results in
Table \ref{tab:wha} and Figure \ref{fig:our_wha}.

For the central parts of galaxies we use Pa$\alpha$-based extinctions
from \citet[][]{CALZETTI07} and Balmer decrements from \citet{KENNICUTT09} and \citet{MOUSTAKAS10}. For full galactic disks with use Balmer decrements from \citet{KENNICUTT09} and \citet{MOUSTAKAS10}, supplement these with extinctions estimated
from the gas column following \citet[][their ``hybrid'' model]{WONG02}
and extinctions inferred from comparing thermal radio continuum flux estimates
from \citet{NIKLAS95,NIKLAS97} to our H$\alpha$ maps
\citep[following][]{WONG02}. For full galactic disks, we also benchmark against $A_{\rm
  H\alpha} = 1$~mag, though this is a typical extinction rather than a
true estimate \citep{KENNICUTT98B}.

In the ``centers'' of galaxies, we work with average surface
brightnesses in the central 1~kpc. For whole galaxies we work within $r_{\rm 25}$ and we assume that the \citet{NIKLAS95} fluxes come
mostly from inside this area. Except in the case of the radio
continuum, we adopt extinctions rather than fluxes from the
literature, which somewhat obviates the need to precisely match apertures. Nonetheless we expect our reliance on published measurements rather than the original data to introduce scatter into our results. We do not consider central measurements from NGC~4736 or NGC~4569 because of their Seyfert nuclei. We also omit Balmer decrement measures for NGC~2841 and radio continuum estimates for NGC3351, NGC 5457, and NGC 6946 because they do not return sensible extinction estimates (they yield negative extinction).

As in the SINGS literature, the choice of reference SFR and 24$\mu$m
treatment affects $w_{\rm H\alpha}$, leading to $\sim 0.2$~dex
systematic variation. Even for a fixed reference SFR and cirrus
approach, we find significant scatter among galaxies, typically $\sim
0.3$~dex ($1\sigma$). Although the 24$\mu$m offers a powerful,
sensitive SFR tracer, its exact calibration remains uncertain. At
present it appears that any adopted $w_{\rm H\alpha}$ should be
associated with a factor of $\approx2$ uncertainty when applied to a
specific galaxy while the average $w_{\rm H\alpha}$ remains uncertain
by $\approx 50\%$.

Figure \ref{fig:aha_vs_ha24} presents an alternative, more direct, visualizations of our constraints on $w_{\rm H\alpha}$. We plot the H$\alpha$ extinction drawn from the literature, $A_{\rm H\alpha}$, as a function of the ratio of H$\alpha$-to-24$\mu$m intensity. These two quantities directly track one another for the case of a fixed $w_{\rm H\alpha}$ and the solid curves show the expected relation for $w_{\rm H\alpha} = 0.25,$ (bottom) $0.5,1,2,$ and $4$ (top). The two panels show extinction estimates based on radio continuum and P$\alpha$ emission (top panel) and Balmer decrements (bottom panel) with the individual studies labeled. If a fixed $w_{\rm H\alpha}$ perfectly described our data, we would expect to see them strung out along one of the solid lines in Figure \ref{fig:aha_vs_ha24}. This view highlights the contradictions among recent studies and the large scatter in $w_{\rm H\alpha}$ measurements for individual studies and galaxies.

Treating all determinations equally, we find an average $w_{\rm
  H\alpha} \approx 0.85$ with $0.47$~dex scatter among individual
determinations. Taking $A_{\rm H\alpha} = 1$~mag and no cirrus
subtraction also suggests $w_{\rm H\alpha}$ about this magnitude, confirming it
as a reasonable starting point. This ensemble approach agrees with the
average $w_{\rm H\alpha}$ from the SINGS literature. We recommend $w_{\rm H\alpha} \sim 0.9$ as a starting point for most analyses if
$w_{\rm H\alpha}$ is otherwise unconstrained. However, given the enormous scatter in results, we can not significantly distinguish this $w_{\rm H\alpha}$ from the \citet{CALZETTI07} value, $w_{\rm Halpha} = 1$, and that also represents a reasonable assumption.

When correcting for 24$\mu$m cirrus contamination, we adopt a higher
$w_{\rm H\alpha} = 1.3$, driven by the Pa$\alpha$ and radio continuum
estimates.  These robust, extinction-free independent tracers of
ionizing photons suggest $w_{\rm H\alpha} \approx 1.3$,
similar to the $w_{\rm H\alpha} = 1.23$ found for \hii\ knots in M51 by \citet{KENNICUTT07}.
Again this roughly matches the $A_{\rm H\alpha} \sim 1$~mag case. The upper panel in Figure \ref{fig:aha_vs_ha24} shows the overall consistency of the  radio continuum and P$\alpha$ measurements with this adopted $w_{\rm H\alpha}=1.3$, but also highlights the large scatter among even these "good" data. With better reference measurements, it should be possible to distinguish whether the scatter reflects the inadequacy of a linear hybrid to predict extinction or simply inconsistency among challenging measurements.

As in the SINGS literature, the results using Balmer decrement
extinctions are confusing and contradictory. The \citet{KENNICUTT09}
values yield the low-outlying $w_{\rm H\alpha}$ while the
\citet{MOUSTAKAS10} values yield among the highest $w_{\rm
  H\alpha}$. This difference holds for whole galaxies and galaxy
centers, an so does not appear exclusively driven by weighting or S/N
issues. The issues with these measurements can be seen directly in the lower panel of Figure \ref{fig:aha_vs_ha24}, where matched H$\alpha$-to-24$\mu$m ratios correspond to dramatically different extinction estimates. We incline towards interpreting this wide spread to indicate
that these data do not offer a robust reference SFR. However, if one
accepts that the \citet{MOUSTAKAS10} values accurately reflect the
true extinctions, the Balmer decrements argue for a high $w_{\rm
  H\alpha} \approx 1.4$--$2.4$, near the limiting case expected for
heavily embedded star formation (Section \ref{sec:irdom}).

\subsection{FUV+24$\mu$m}

\begin{figure}
\plotone{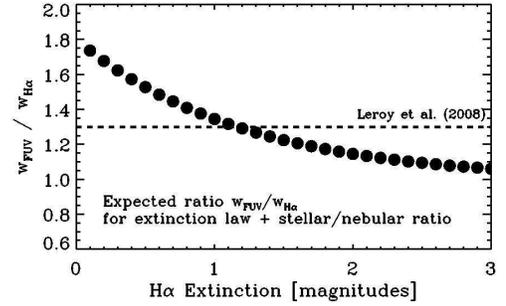}
\caption{{\em Expectation for $w_{\rm FUV}/w_{\rm H\alpha}$ as a
    function of H$\alpha$ extinction.} We plot the ratio of the calibration of
  24$\mu$m term in combination with FUV, $w_{\rm FUV}$,  to the calibration of the 24$\mu$m term in combination with H$\alpha$, $w_{\rm H\alpha}$, 
  as a function of the H$\alpha$ extinction along the line of sight. See the definition of $w$ in Equation \ref{eq:w}. The calculation assumes an
  extinction law with $A_{\rm FUV} / A_{\rm R} = 8.24 / 2.33$ and a
  fixed stellar-to-nebular extinction ratio of $A_{\rm H\alpha} / A_R
  \approx 2$.}
\label{fig:wfuvwha}
\end{figure}

\subsubsection{Expectation Relative to H$\alpha$}

Given an extinction curve and a typical stellar-to-nebular extinction
ratio, one can relate the FUV and H$\alpha$ extinctions, $A_{\rm FUV}$
and $A_{\rm H\alpha}$. \citet{LEROY08} adopted an $R$-band stellar-to-nebular
extinction ratio of $A_{\rm H\alpha} / A_R \approx 2$
\citep[]{CALZETTI94,ROUSSEL05} and an extinction law where $A_{\rm
  FUV} / A_{\rm R} = 8.24 / 2.33$ \citep[][using their filter definitions for R and the GALEX FUV band]{CARDELLI89,WYDER07}. They
then solved for $w_{\rm FUV} / w_{\rm H\alpha}$ as a function of
extinction (their Equation D9), which we plot in Figure
\ref{fig:wfuvwha}. At high extinctions, virtually all emission is
embedded and for FUV+24$\mu$m to match H$\alpha$+24$\mu$m then $w_{\rm
  FUV} \sim w_{\rm H\alpha}$. At low extinction, the ratio $w_{\rm
  FUV} / w_{\rm H\alpha} \sim 1.8$, which is the ratio of FUV
extinction to H$\alpha$ for the adopted extinction law and
stellar-to-nebular ratio. In the intermediate regime, near the $A_{\rm
  H\alpha} \sim 1$~magnitude typical of our sample, we expect $w_{\rm
  FUV} / w_{\rm H\alpha} \sim 1.3$.

\subsubsection{$w_{\rm FUV}$ in Our Data}

\begin{figure}
\plotone{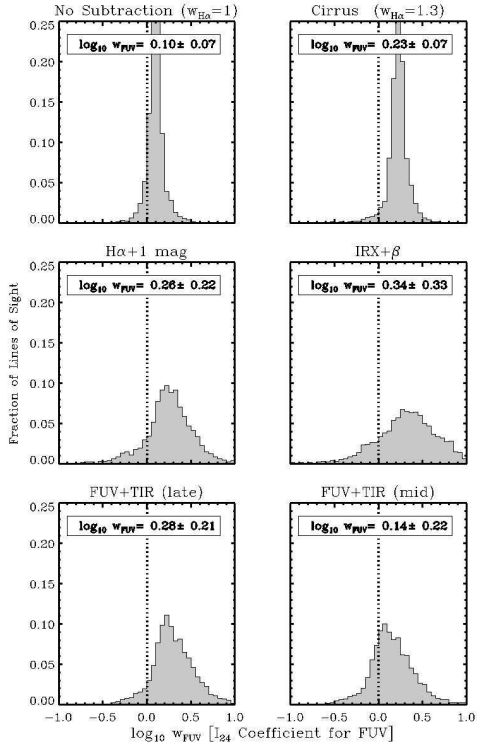}
\caption{{\em Calibration of the FUV+24$\mu$m tracer}: The factor
  $w_{\rm FUV}$ needed to combine FUV and 24$\mu$m, calculated for
  individual lines of sight (see the definition of $w$ in Equation \ref{eq:w}). The top row shows $w_{\rm FUV}$ derived
  referencing to H$\alpha$+24$\mu$m with and without cirrus. The
  middle row shows $w_{\rm FUV}$ derived from H$\alpha$ plus a typical
  1 magnitude of extinction and FUV plus extinction estimated via the
  IRX-$\beta$ relation of \citet{MUNOZMATEOS09B}. The bottom row shows
  $w_{\rm FUV}$ derived from FUV+TIR following \citet{CORTESE08} for a
  young (``late'') and intermediate age (``mid'') population,
  bootstrapping TIR from the 24$\mu$m. We list the median $\log_{\rm
    10} w_{\rm FUV}$ and $1\sigma$ scatter among lines of sight for
  each approach. A vertical line indicates $w_{\rm FUV} = 1$.}
\label{fig:our_wfuv}
\end{figure}

We also calculate $w_{\rm FUV}$. For this exercise, we expand the
suite of reference SFRs and carry out the calculation for each line of
sight with $\left< \Sigma_{\rm SFR} \right> > 3 \times
10^{-3}$~M$_\odot$~yr$^{-1}$~kpc$^{-2}$. Figure \ref{fig:our_wfuv}
shows histograms of $w_{\rm FUV}$ for different reference SFRs and
reports the median and scatter for each approach. The first two panels
show $w_{\rm FUV}$ from referencing to H$\alpha$+24$\mu$m using the
indicated 24$\mu$m processing and $w_{\rm H\alpha}$; we use the same
24$\mu$m approach for the H$\alpha$+24$\mu$m reference and the
solution for $w_{\rm FUV}$. The last four panels reference to
H$\alpha$ with $A_{\rm H\alpha} = 1$~mag, FUV corrected using $A_{\rm
  FUV}$ inferred from the UV spectral slope following \citet[][the
  so-called ``IRX-$\beta$'' relation]{MUNOZMATEOS09B}, and FUV
corrected using $A_{\rm FUV}$ inferred from the TIR/FUV ratio from
\citet{CORTESE08}. \citet{CORTESE08} consider stellar populations with
different ages. We plot results for their youngest population, which
we call ``late'' (their $\tau \geq 8$~Gyr case), and an intermediate
age case, which we call ``mid'' ($\tau = 5$~Gyr).

Referencing to H$\alpha$+24$\mu$m after cirrus subtraction, we find
$w_{\rm FUV} \approx 1.7$. The average of the two \citet{CORTESE08}
treatments give a similar value, $w_{\rm FUV} \approx 1.6$. This is higher than the $w_{\rm FUV}$
recommended by \citet{LEROY08} but note that we have also increased
$w_{\rm H\alpha}$ to $1.3$ from the $w_{\rm H\alpha} = 1.0$ used in
\citet{LEROY08}. The revised $w_{\rm FUV}$ is still $\sim 1.3~w_{\rm
  H\alpha}$, in accordance with our expectations (Figure
\ref{fig:wfuvwha}). Because this value can be derived from the
\citet{CORTESE08} FUV+TIR approach alone, this calculation establishes
the FUV+24$\mu$m as somewhat independent of H$\alpha$+24$\mu$m, and
not a purely bootstrapped tracer.

Different approaches yield $w_{\rm FUV}$ from $1.3$--$2.1$. The
scatter in $w_{\rm FUV}$ appears lower than in $w_{\rm H\alpha}$ but
this is mostly due to the fact that our references often already
include 24$\mu$m. To the degree that we bootstrap the calibration from
either our $w_{\rm H\alpha}$ or the \citet{CORTESE08} approach,
$w_{\rm FUV}$ carries the same $\sim 0.2$--$0.3$~dex uncertainty
associated with those approaches.

\section{Comparison Among Tracers and Uncertainty}
\label{sec:compare}

\begin{figure*}
\epsscale{1.0}
\plotone{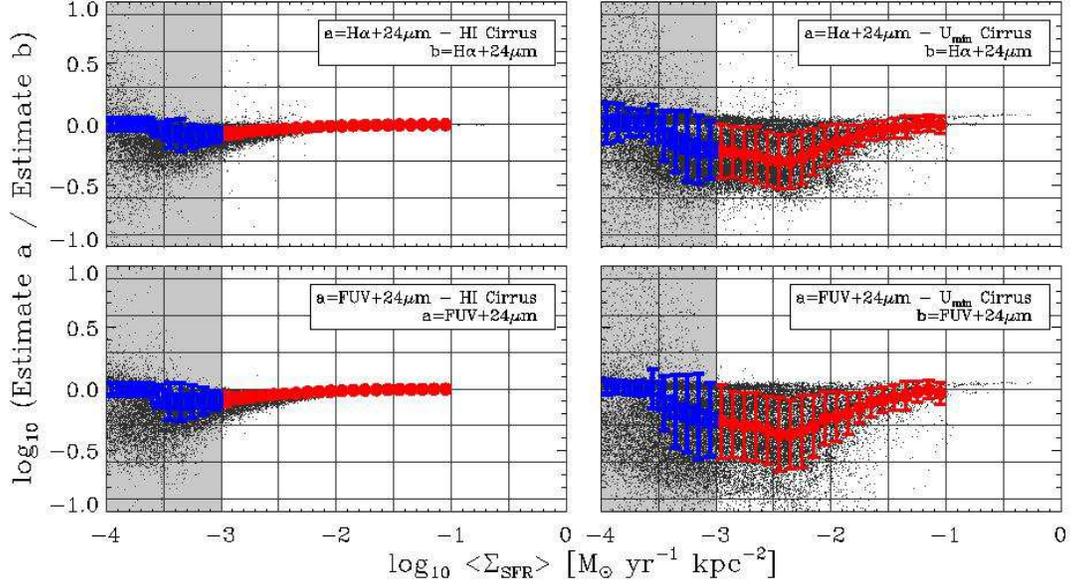}
\caption{{\em Hybrid tracers with and without cirrus:} Ratio
  ($\log_{\rm 10}$) of hybrid star formation rate tracers as a
  function of composite $\left< \Sigma_{\rm SFR} \right>$. The gray
  area shows where $\left< \Sigma_{\rm SFR} \right> <
  10^{-3}$~M$_\odot$~yr$^{-1}$~kpc$^{-2}$, our lower limit for a
  reliable $\Sigma_{\rm SFR}$. Red circles show the median ratio in
  bins of $\left< \Sigma_{\rm SFR} \right>$ with error bars indicating
  $1\sigma$ scatter in that bin; blue circles show the same where
   $\left< \Sigma_{\rm SFR} \right> < 10^{-3}$~M$_\odot$~yr$^{-1}$~kpc$^{-2}$ and we consider $\Sigma_{\rm SFR}$ unreliable. Black points show individual lines of sight. The top shows the effect of subtracting the IR cirrus on the
  integrated estimate. The bottom row compares H$\alpha$ and FUV-based
  tracers for matched cirrus approaches}
\label{fig:sfrcomp_cirrus}
\end{figure*}

\begin{figure*}
\epsscale{1.0}
\plotone{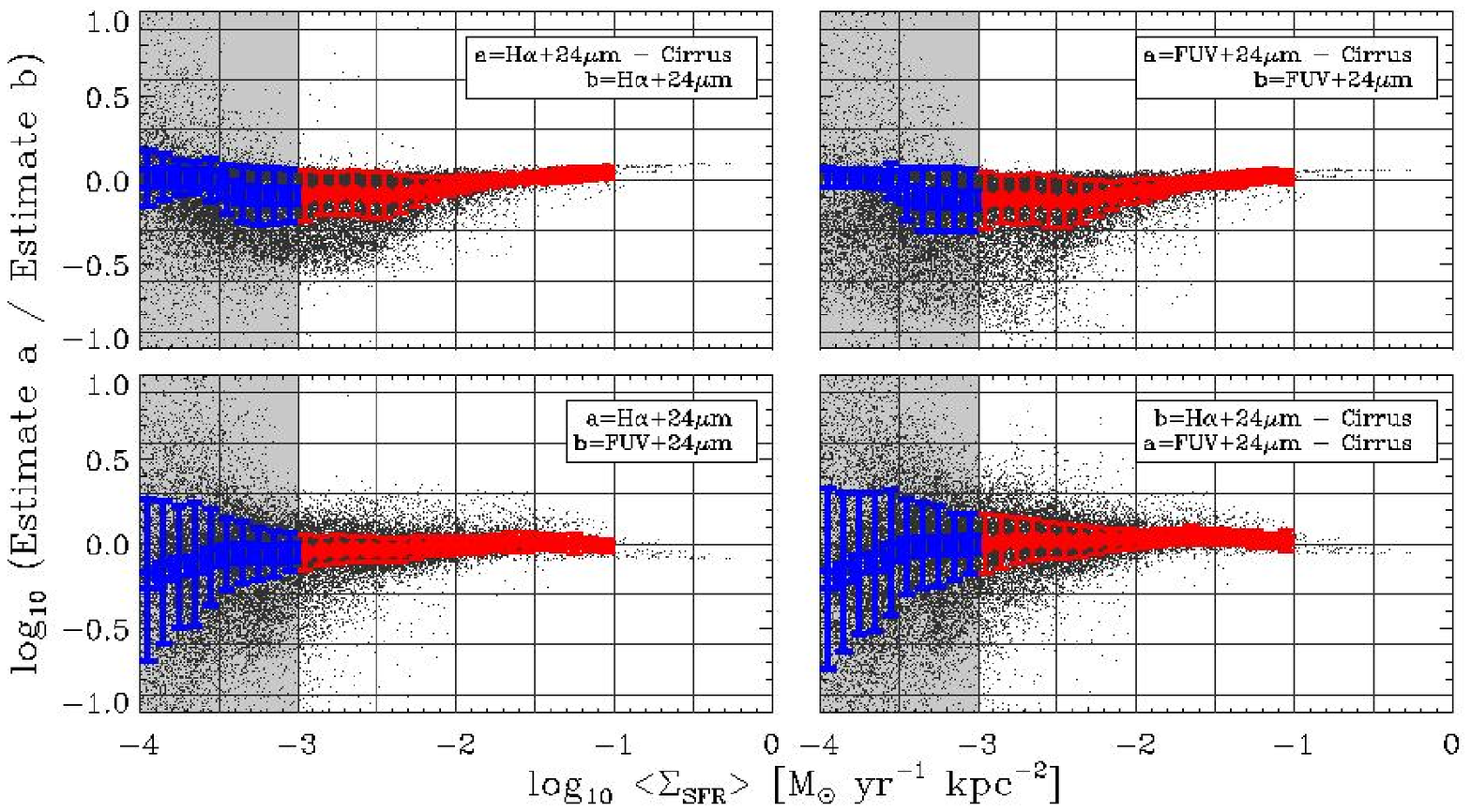}
\caption{{\em Effect of varying the cirrus approach:} Ratio
  ($\log_{\rm 10}$) of star formation rate tracers as a function of
  composite $\left< \Sigma_{\rm SFR} \right>$. The gray area shows
  where $\left< \Sigma_{\rm SFR} \right> <
  10^{-3}$~M$_\odot$~yr$^{-1}$~kpc$^{-2}$, our lower limit for a
  reliable $\Sigma_{\rm SFR}$. Red circles show the median ratio in
  bins of $\left< \Sigma_{\rm SFR} \right>$ with error bars indicating
  $1\sigma$ scatter in that bin; blue circles show the same where
   $\left< \Sigma_{\rm SFR} \right> < 10^{-3}$~M$_\odot$~yr$^{-1}$~kpc$^{-2}$ and we consider $\Sigma_{\rm SFR}$ unreliable. Black points show individual lines of sight. The left panels show the effect of subtracting only cirrus
  emission from dust associated with \hi . The right panel shows the
  effect of setting the radiation field to double our adopted value.}
\label{fig:sfrcomp_hybrid}
\end{figure*}

\begin{figure*}
\epsscale{1.0}
\plotone{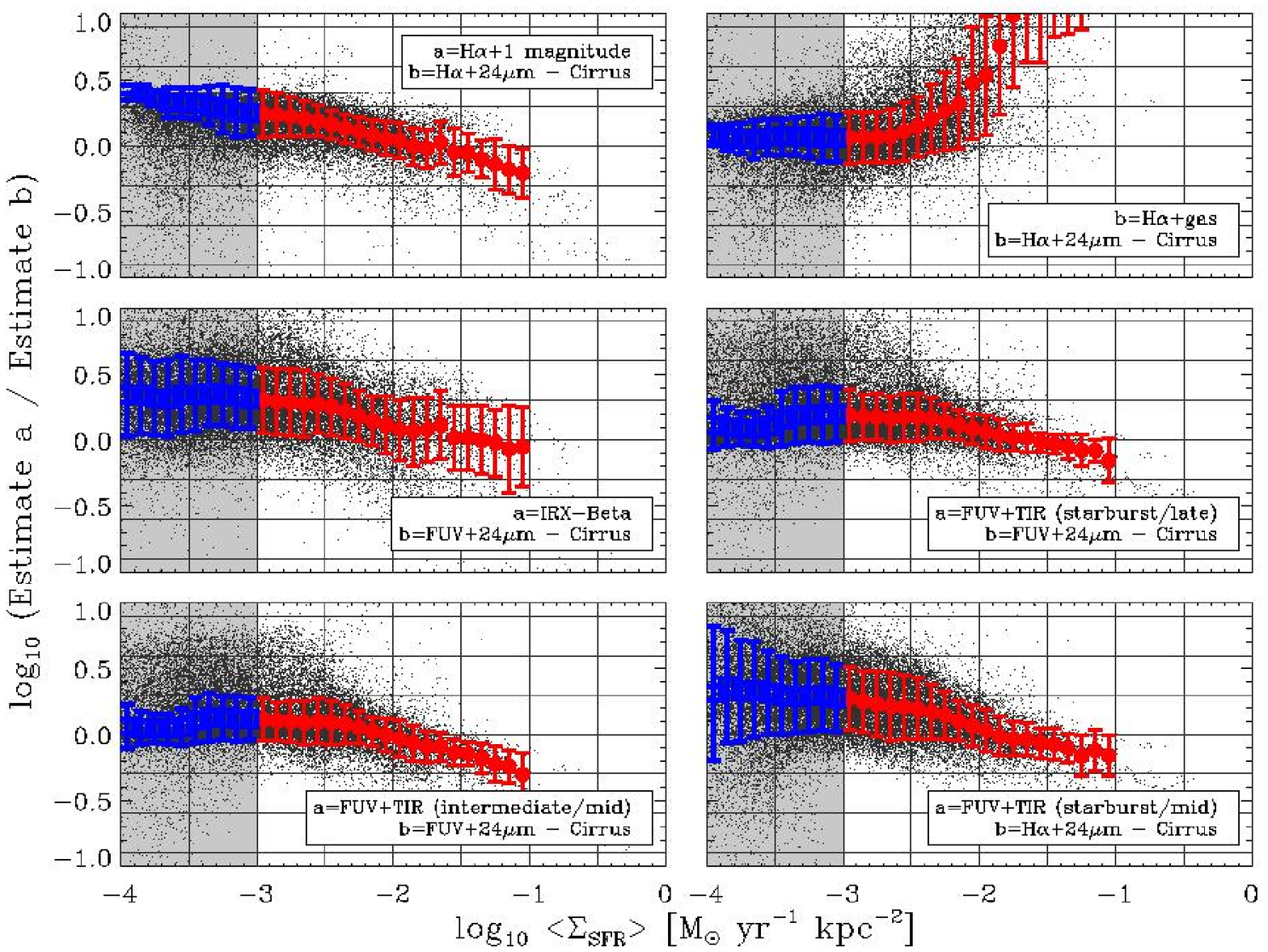}
\caption{{\em Comparison with other tracers:} Ratio ($\log_{\rm 10}$)
  of star formation rate tracers as a function of composite $\left<
  \Sigma_{\rm SFR} \right>$. The gray area shows where $\left<
  \Sigma_{\rm SFR} \right> < 10^{-3}$~M$_\odot$~yr$^{-1}$~kpc$^{-2}$,
  our lower limit for a reliable $\Sigma_{\rm SFR}$. Red circles show
  the median ratio in bins of $\left< \Sigma_{\rm SFR} \right>$ with
  error bars indicating $1\sigma$ scatter in that bin; blue circles show the same where
   $\left< \Sigma_{\rm SFR} \right> < 10^{-3}$~M$_\odot$~yr$^{-1}$~kpc$^{-2}$ and we consider $\Sigma_{\rm SFR}$ unreliable. Black points
  show individual lines of sight. Each panel compares one of our
  hybrid tracers to another approach.}
\label{fig:sfrcomp_other}
\end{figure*}

Having examined the constituent terms, contamination, and calibration
of hybrid tracers, we now gauge the practical impact of our choices on
$\Sigma_{\rm SFR}$. Figures \ref{fig:sfrcomp_cirrus} --
\ref{fig:sfrcomp_other} compare pairs of tracers as a function of
$\left< \Sigma_{\rm SFR} \right>$.

Figure \ref{fig:sfrcomp_cirrus} shows the effect of our cirrus
subtraction and compares FUV and H$\alpha$-based tracers. The bottom
row shows that H$\alpha$ and FUV based approaches agree well for
matched treatments \citep[see also][]{LEROY08}. The top row shows that
the cirrus subtraction has a net effect of 10s of percent, but less
than 100\% across the range $10^{-3} < \left< \Sigma_{\rm
  SFR} \right> < 10^{-1}$~M$_\odot$~yr$^{-1}$~kpc$^{-2}$. Below this
range, the offsets can become much more severe as the H$\alpha$ maps
become unreliable and the cirrus correction has removed virtually all
24$\mu$m emission.

Figure \ref{fig:sfrcomp_hybrid} shows the effect of pushing the cirrus
subtraction to its limits. The most conservative approach, removing
only emission from dust associated with \hi , has a very small impact
on the net $\Sigma_{\rm SFR}$. If we adopt double our nominal cirrus estimate then the 24$\mu$m component of the tracer is largely
removed below $10^{-2}$~M$_\odot$~yr$^{-1}$~kpc$^{-2}$, suppressing
$\Sigma_{\rm SFR}$ by a factor of $\sim 2$ (recall the typical $A_{\rm
  H\alpha} \sim 1$~mag).

Figure \ref{fig:sfrcomp_other} presents comparisons with
``independent'' tracers: a fixed $A_{\rm H\alpha} = 1$~mag; FUV +
$A_{\rm FUV}$ inferred from the UV spectral slope
\citep[][``IRX-$\beta$'']{MUNOZMATEOS09B}; $A_{\rm H\alpha}$ inferred
using the ``hybrid'' gas model of \citet{WONG02}; and FUV combined
with $A_{\rm FUV}$ estimated from the TIR/FUV ratio for a young and
middle-aged population \citep{CORTESE08}.

Some ``tilt" can be seen in the comparisons, particularly the comparison with "independent" tracers. This usually has the sense that our IR-based $\Sigma_{\rm SFR}$ estimate recovers higher $\Sigma_{\rm SFR}$ than the comparison indicated in high $\Sigma_{\rm SFR}$, presumably heavily embedded regions.
In any case magnitude of systematic differences rarely exceeds a factor
of 2 (0.3 dex) over 2 orders of magnitude in dynamic range.  Only
using gas to infer the extinction diverges dramatically at high
$\left< \Sigma_{\rm SFR}\right>$, an issue also noted by \citet{LEROY08} that
presumably results from complex geometry.

\begin{figure}
\plotone{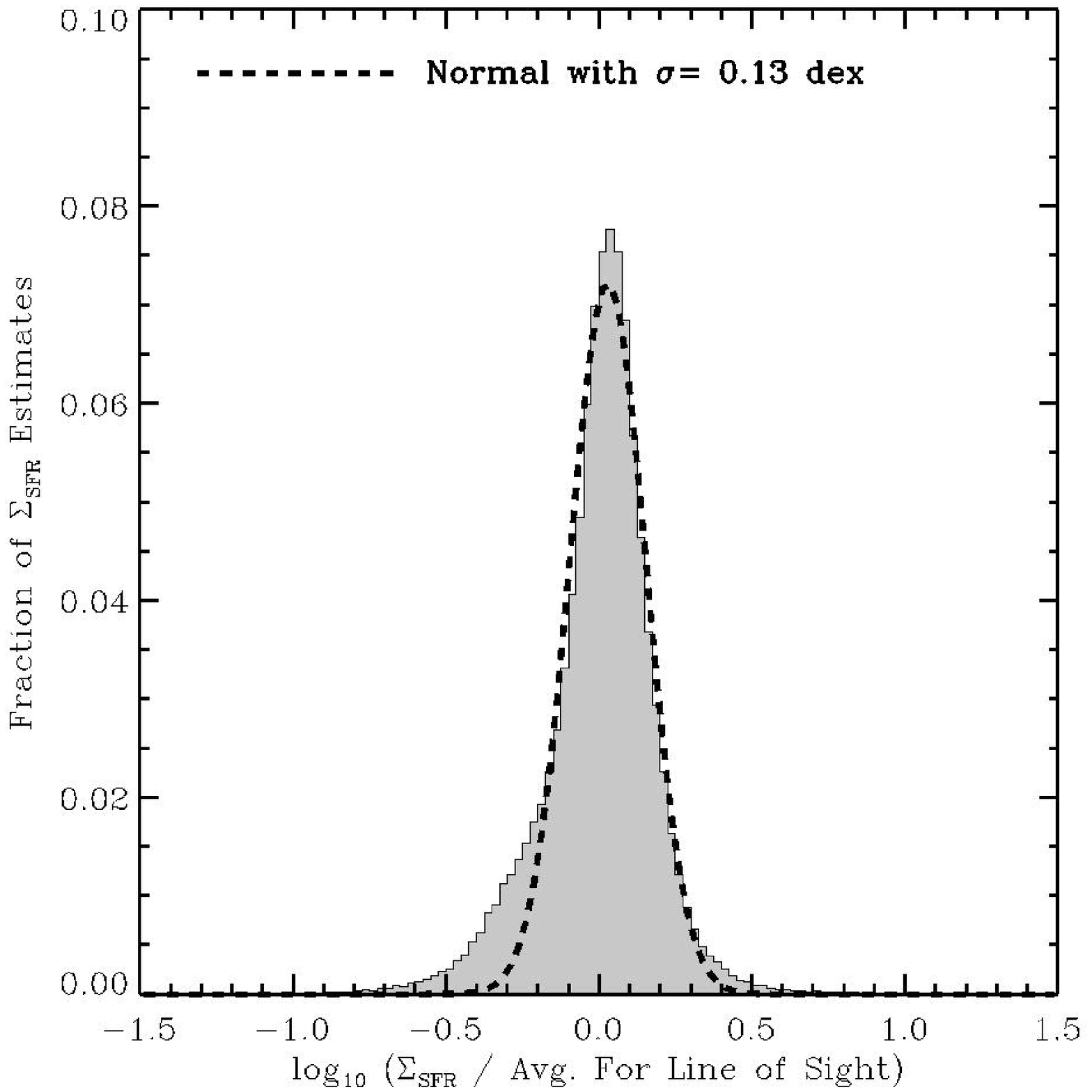}
\caption{{\em Uncertainty estimated From scatter among tracers:} Distribution of $\Sigma_{\rm SFR} - \left< \Sigma_{\rm SFR}\right>$, the difference between each individual SFR estimate in our data set and the mean $\Sigma_{\rm SFR}$ estimate for that line of sight. The histogram includes all lines of sight with $\Sigma_{\rm SFR} > 10^{-3}$~M$_\odot$~yr$^{-1}$~kpc$^{-2}$. A typical scatter across the sample is $0.13$~dex, which gives a rough empirical estimate of the uncertainty in individual $\Sigma_{\rm SFR}$ estimates.}
\label{fig:unc}
\end{figure}

The scatter among different tracers in Figures
\ref{fig:sfrcomp_cirrus}--\ref{fig:sfrcomp_other} allows us to
estimate the uncertainty in any given estimate of $\Sigma_{\rm
  SFR}$. For each line of sight in our data set, we take the median absolute deviation-based
scatter of our ensemble of $\Sigma_{\rm SFR}$ estimates about the mean
value for that point. That is, we measure the scatter of $\Sigma_{\rm SFR}$ estimates about the mean value for each point. This folds some systematic uncertainty into the estimate because some of our approaches contradict one
another. Inasmuch as each approach represents a reasonable method to
estimate $\Sigma_{\rm SFR}$, however, this simple calculation should
yield a good idea of how uncertain a particular $\Sigma_{\rm SFR}$
estimate actually is. Overall for $\left< \Sigma_{\rm SFR} \right> >
10^{-3}$~M$_{\odot}$~yr$^{-1}$~kpc$^{-2}$ we find a scatter of $\sim
0.13$~dex. We illustrate this in Figure \ref{fig:unc}, plotting a histogram of the scatter of individual estimates about the median $\Sigma_{\rm SFR}$ for that line of sight.

We have also argued that the calibration of the 24$\mu$m term in the
hybrid tracers appears systematically uncertain by a factor of $\sim
2$. The 24$\mu$m contributes $\sim 60\%$ of the SFR on average, so
this suggests a systematic uncertainty of order of $0.15$--$0.2$~dex
in the magnitude of the total SFR. Inasmuch as we have aimed for a
self-consistent stable of tracers, this overall calibration
uncertainty exists in addition to the point-by-point uncertainty
derived from Figure \ref{fig:unc}.

\subsection{Limiting $\left< \Sigma_{\rm SFR} \right>$}

Our H$\alpha$ maps are unreliable below $\left< \Sigma_{\rm SFR} \right> \lesssim
10^{-3}$~M$_\odot$~yr$^{-1}$~kpc$^{-2}$ and the details of the cirrus methodology dictates the
24$\mu$m distribution below this level. We therefore adopt $\Sigma_{\rm SFR} =
10^{-3}$~M$_\odot$~yr$^{-1}$~kpc$^{-2}$ as a limiting surface density
for a robust estimate. The 24$\mu$m and FUV maps recover signal well below this value, but their translation into $\Sigma_{\rm SFR}$ becomes more uncertain. The low-$\Sigma_{\rm SFR}$ regime can help constrain many interesting
astrophysical processes. Our general recommendation in this regime is
to carefully focus on the constituent parts of the hybrid tracers
(i.e., FUV, H$\alpha$, and 24$\mu$m separately) both as intensities
and $\Sigma_{\rm SFR}$ estimates \citep[e.g.,
  see][]{BIGIEL10,SCHRUBA11}. FUV and H$\alpha$ alone should provide
robust lower limits to the overall $\Sigma_{\rm SFR}$.

\subsection{Comparison to Integrated Values}

\begin{figure}
\plotone{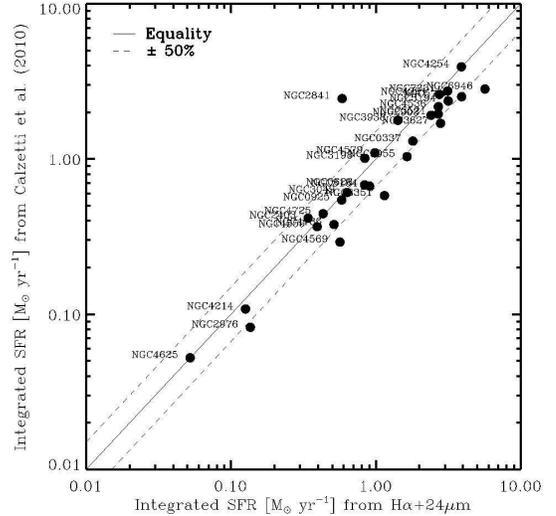}
\caption{{\em Comparison to integrated SFRs:} Comparison of SFRs
  derived integrating our $\Sigma_{\rm SFR}$ ($x$-axis) to integrated
  SFRs from \citet{CALZETTI10} ($y$-axis). The solid line shows
  equality and dashed lines indicate deviations of $\pm 50\%$.}
\label{fig:sfrcomp_int}
\end{figure}

A key property of well-formed $\Sigma_{\rm SFR}$ estimates that span a
large part of a galaxy is that they integrate to a sensible total
SFR. Figure \ref{fig:sfrcomp_int} shows the results of integrating our
maps, here H$\alpha$ plus cirrus-corrected 24$\mu$m, and comparing
them to the galaxy-average $\Sigma_{\rm SFR}$ of
\citet{CALZETTI10}. Note that the \citet{CALZETTI10} results are partially anchored to the \citet{KENNICUTT09}
results, which we consider somewhat uncertain for these targets. The
overall match appears very good given this uncertainty, with our estimates tending to be slightly higher in high SFR galaxy because of our higher adopted $w_{\rm H\alpha}$.

\subsection{Other Uncertainties}

We have attempted a thorough investigation of $\Sigma_{\rm SFR}$
estimation on kpc scales but several important uncertainties
remain. We discuss these briefly here.

{\em Diffuse Ionized Gas and Escape of Photons from \hii\ Regions:}
Photons may escape from \ion{H}{2} regions, producing H$\alpha$
emission away from the emitting stellar population. Estimates of the
fraction of such ``diffuse'' H$\alpha$ emission in galaxies are often
$\sim 30\%$ \citep[see][]{RAHMAN10}. Most such emission still reflects
ionizing photons produced by massive stars and so {\em must} be
accounted in the SFR if kpc-sized estimates are to match integrated
estimates. In theory this emission should thus be added back to the
likely parent region \citep[e.g.,][]{BLANC09} and should not be entirely
removed from the map.

Given the cross section of \hi\ to ionizing photons, the path length
of ionizing photons near the Lyman limit will be extraordinarily short
in the presence of any neutral gas. Despite this, diffusion over
scales of a few hundred pc clearly occurs. We observe a warm ionized
medium in our own Galaxy and diffuse ionized gas in other
galaxies. However given that this flux must be conserved and our kpc
resolution this systematic should represent a decidedly second-order
concern.

{\em Dust Absorption of Ionizing Photons:} The complementary concern
may also confuse our measurements. Dust inside \hii\ regions may
absorb ionizing photons before they reach a hydrogen atom. Such
photons will be missed by recombination line observations, including
Pa$\alpha$, though any bolometric tracer will account for this light
as it is reradiated by the dust. Robust estimates of the
magnitude of this effect remain scarce, but \citet{INOUE01A} and
\citet{INOUE01B} studied star-forming regions in the Milky Way and
Local Group galaxies and suggested the dust may absorb as many as half
of the ionizing photons. Calculations by \citet{DOPITA03} also suggest
plausible magnitudes of several times 10\%\, with the effect strongest
for ultracompact \hii\ regions \citep[see also][]{GROVES08}.

We find good overall consistency between H$\alpha$-based approaches,
which will be sensitive to this effect, and FUV-based approaches,
which should not be affected. However, to some degree we also
calibrate the FUV-based hybrids to match the H$\alpha$-based
hybrids. The bottom right panel in Figure \ref{fig:sfrcomp_other}
compares H$\alpha$+24$\mu$m to FUV+TIR and does reveal a bias towards
higher $\Sigma_{\rm SFR}$ for the bolometric FUV+TIR tracer. The bias
becomes worse at lower $\left< \Sigma_{\rm SFR} \right>$ while we
expect the opposite for dust absorption of ionizing photons. Therefore
we tend to attribute the disagreement mostly to the cirrus issue, but
the data could certainly be consistent with non-negligible absorption
of photons. The ideal test will be to compare a robust ionizing photon
production map, e.g., thermal radio continuum emission at 1--10~mm or
IR recombination lines, to high resolution bolometric luminosity maps
and stellar populations.

{\em UV Emission From Older Populations:} Even a relatively evolved
population may still produce some FUV emission, contaminating our
$\Sigma_{\rm SFR}$ estimates. We estimate potential contamination by
measuring the ratio of FUV to median-filtered $3.6\mu$m emission, an
approximate tracer of stellar mass \citep[see appendices
  in][]{LEROY08}, in regions that appear dominated by old stellar
populations. We measure ratios for the bulges of M~81, M~94 (``the
Sombrero''), NGC~4725, and NGC~3351\footnote{M~81 and M~94 are
  otherwise not included in this paper.}, all of which show bright
stellar structures with little evidence of current star formation. We
also measure the ratio across our sample for all lines of sight with
very low star formation, $\left< \Sigma_{\rm SFR} \right> < 5 \times
10^{-4}$~M$_\odot$~yr$^{-1}$~kpc$^{-2}$, but significant stellar mass,
$\Sigma_* \gtrsim 100$~M$_\odot$~pc$^{-2}$. These all suggest a ratio
$I_{\rm FUV} / \Sigma_* \sim 0.7$--$1.9 \times
10^{-5}$~MJy~sr$^{-1}$~(M$_\odot$~pc$^{-2}$)$^{-1}$.

We test the effect of correcting the FUV or $\Sigma_{\rm SFR}$ maps
using a ratio at the upper end of this range, $2 \times
10^{-5}$~MJy~sr$^{-1}$~(M$_\odot$~pc$^{-2}$)$^{-1}$. Our adopted
mass-to-light ratio is

\begin{equation}
\Sigma_* \left[{\rm M_\odot~kpc}^{-2}\right] = 200~I_{3.6}~{\rm
  MJy}~{\rm sr}^{-1}~,
\end{equation}

\noindent similar to \citet[][]{LEROY08}, though revised slightly to
match the maps of \citet{ZIBETTI09}. Therefore this ratio corresponds
to $I_{\rm FUV} \left[ {\rm MJy~sr}^{-1} \right] \sim 0.004$~I$_{\rm
  3.6} \left[ {\rm MJy~sr}^{-1}\right]$. From Equation
\ref{eq:sfr_fuv} the implied contamination in units of $\Sigma_{\rm
  SFR}$ will be

\begin{equation}
\label{eq:olduv}
\Sigma_{\rm SFR}^{old~UV} \left[ {\rm M}_\odot~{\rm yr}^{-1}~{\rm
    kpc}^{-2} \right] \approx 1.6 \times 10^{-6}~\Sigma_* \left[ {\rm
    M~pc}^{-2} \right]~.
\end{equation}

\noindent Equation \ref{eq:olduv} implies that observations of stellar
bulges ($\Sigma_* \gtrsim 10^2$~M$_\odot$~pc$^{-2}$) that measure
$\Sigma_{\rm SFR}$ only a few times
10$^{-4}$~M$_\odot$~yr$^{-1}$~kpc$^{-2}$ should be viewed with
suspicion. The contamination will be lower in the outer disks of
galaxies ($\Sigma_* \sim 10$~M$_\odot$~pc$^{-2}$) in proportion to the
lower stellar surface density. However, for our sample and working
$\Sigma_{\rm SFR}$ limit the contribution of this truly old UV
component can be safely neglected. The SFR associated with FUV
emission from old stars appears to be only $\approx 3$\% of the total
SFR where $\Sigma_{\rm SFR} > 10^{-3}$~M$_\odot$~yr$^{-1}$~kpc$^{-2}$
and contributes only $\approx 10\%$ of the FUV light.

Contributions from intermediate age populations represent a more
complex problem requiring detailed stellar population analysis (a more
complex extension of Section \ref{sec:uvha}) that is beyond the scope
of this paper.

{\em Stochasticity and IMF Variations:} At 1~kpc resolution we expect
to average several star-forming regions in each element. For
$\Sigma_{\rm SFR} > 10^{-3}$~M$_\odot$~yr$^{-1}$~kpc$^{-2}$, we expect
$M_* \gtrsim 5 \times 10^{3}$~M$_\odot$ formed over the last $\sim
5$~Myr in each element. This minimizes but does not completely
eliminate concerns about sampling the stellar initial mass function
(IMF). Changes in the IMF will directly impact our estimates but
concrete evidence for such variations remains amibiguous at best
\citep{BASTIAN10}. We do not consider either effect.

\section{Discussion}
\label{sec:disc}

\subsection{Hybrid SFR Tracers}

Hybrid tracers that combine an unobscured and an obscured (i.e., IR) component have many
advantages. The two tracers complement one another, yielding an
approach that can apply, in principle, to a wide range of
environments. Maps of UV, H$\alpha$, and IR emission are widely
available. IR emission directly probes reprocessed photons, avoiding
the need for less reliable extinction tracers. Even given the
uncertainty in the IR term calibration, using a direct measure of
reprocessed starlight makes for much more robust approaches than
optical, UV, or gas-based approaches. Tracers constructed from linear
combinations of these bands will be reasonably scale-independent and
can be easily interpreted. Despite the significant uncertainties
discussed in this paper, we argue that hybrid IR+UV or IR+H$\alpha$
tracers do offer the best current combination of widely available
data and robustness.

{\em 24$\mu$m Cirrus:} The issue of infrared ``cirrus'', emission due
to heating by an older stellar population, remains a challenge for
IR-based SFR estimates. Here we have suggested an approach based on
physical dust models, with the cirrus emission mainly dependent on the
dust abundance and illuminating radiation field. Several points emerge
from our calculation. First, cirrus emission associated with molecular
gas can make an important contribution especially in the bright parts
of star-forming galaxies. This is emission from dust mixed with molecular gas but heated by weak radiation fields. Because the dust emitting in such a component is mixed with the clumpy molecular ISM this contribution may easily be missed by
morphological approaches (e.g., median filtering) designed to reject a smooth component. 

Second, estimating the radiation field $U_{\rm cirrus}$ that is not associated with
recent star formation is challenging. Doing so one must be careful to
both avoid oversubtraction and catch all of the cirrus. We present a
detailed attempt to estimate the appropriate $U$. Important checks on
this quantity will come from high resolution fits to the dust SED in
quiescent regions. Though not totally trivial to interpret,
these will help inform our understanding of the ``non-star-forming''
$U$ in galaxies. Our sample has been observed by the {\em Herschel}
Key Program KINGFISH at higher resolution and we expect follow-up
analyses to further illuminate this issue. Finally, though the
IR-cirrus represents an important issue one must bear in mind that
other systematics with similar magnitude persist, including the
absolute calibration of the IR term in hybrid tracers, losses of
ionizing photons due to dust or escape, and departures from the
``continuous star formation'' approximation that underlies the whole
concept of star formation rates.

{\em Calibration and the Need For A Sample of Wide-Field,
  Extinction-Robust $\Sigma_{\rm SFR}$ Maps:} We have shown that
substantial calibration uncertainties still plague the IR portion of
hybrid SFR tracers. These lingering uncertainties and the other
systematics we discuss highlight the pressing need for wide-field,
extinction-robust maps of the distribution of recent star
formation. Though observationally expensive, such maps are
indispensible to calibrate more widely used tracers like those we
discuss here. Wide-field maps are a necessity because
understanding galaxy evolution requires $\Sigma_{\rm SFR}$ estimates
that span whole galaxies, which in turn require understanding faint
regions and characterizing diffuse emission, not only studying
\hii\ peaks. Perhaps the best prospect for a large sample of such maps
is using ALMA, the EVLA, or the GBT to map thermal radio continuum
emission at 1--10~mm \citep{NIKLAS97,MURPHY11}. Such maps avoid
extinction concerns completely and can reveal the distribution of
ionizing photons in galaxies in exquisite detail.

{\em Other Uncertainties and Limiting Returns:} We have noted several
other important uncertainties: escape of ionizing photons from
\hii\ regions, absorption of ionizing photons by dust, contribution of
an intermediate age stellar population, stochasticity or IMF
variations, and the closely related age of discrete stellar
populations. Several of these have a likely magnitude $\approx 10\%$
and remain difficult to constrain. As an ensemble these suggest that
one should bear an uncertainty of at least a few times 10\% for any
star formation rate estimates. Indeed at small scales the age-based
scatter we discuss implies that the very concept of a star formation
rate will not be precise beyond this level.

\subsection{Recommended Approach}

We recommend estimating $\Sigma_{\rm SFR}$ at $\sim$kpc resolution
over the range $10^{-3}$~M$_\odot$~yr$^{-1}$~kpc$^{-2}$~$< \Sigma_{\rm SFR} <
10^{-1}$~M$_\odot$~yr$^{-1}$~kpc$^{-2}$ in the following way. Use the
broadband dust SED along with the models of \citet{DRAINE07A} to
estimate $\Sigma_{\rm dust}$, $U_{\rm min}$, and $q_{\rm PAH}$. The
empirical fits of \citet{MUNOZMATEOS09B} may offer a helpful
simplification or check. Derive the 24$\mu$m emissivity per unit dust
mass for $U_{\rm cirrus} = 0.5~U_{\rm min}$\footnote{At very different resolution than this study, $U_{\rm cirrus} \approx 0.6$ may be a safer choice.} from the \citet{DRAINE07A} models and subtract this emissivity times $\Sigma_{\rm dust}$ from the observed
24$\mu$m map (setting negative regions to have $I_{\rm 24} = 0$). If
using only {\em Spitzer} data, gas maps and the assumption of a weakly
varying $DGR$ can provide a useful way to work at resolution higher
than the {\em Spitzer} 160$\mu$m PSF. Combine the resulting 24$\mu$m
intensity with an unobscured tracer, either H$\alpha$ (Equation
\ref{eq:sfr_ha}) or FUV (Equation \ref{eq:sfr_fuv}) emission, using
$w_{\rm H\alpha} = 1.3$ or $w_{\rm FUV} = 1.7$, respectively (Equation
\ref{eq:w_sfr}).

One may also attempt to correct the FUV for contamination by emission
from old stars. Based on the colors of bulges and low-$\Sigma_{\rm
  SFR}$, high $\Sigma_*$ regions we suggest correcting the FUV-related
term in proportion to the old stellar population by $\Sigma_{\rm SFR}
= 1.6 \times 10^{-6}~\Sigma_{*}$ (Equation \ref{eq:olduv}). In our
sample we find this correction to have a small impact, but caution
that it only accounts for old populations.

This approach removes a physically plausible, locally estimated cirrus
component. It is quantitatively and qualitatively consistent with
other approaches over the range $10^{-3}$~M$_\odot$~yr$^{-1}$~kpc$^{-2}$~$\leq \Sigma_{\rm SFR} \leq
10^{-1}$~M$_\odot$~yr$^{-1}$~kpc$^{-2}$. It yields maps that integrate
into sensible integrated SFR maps. We suggest an uncertainty of $\sim
0.15$~dex in individual measurements, $\sim 0.3$~dex in the
calibration of the 24$\mu$m term for any given system, and $\sim 0.15$
dex for the average calibration of the 24$\mu$m term.

{\em In the Absence of Long-Wavelength IR Data:} If one has only gas,
24$\mu$m, and H$\alpha$ or FUV data one can still attempt a
first-order cirrus estimate. Across our sample, 24$\mu$m emissivity
per unit gas mass normalized to $U=0.6$ and $DGR=0.01$ is

\begin{equation}
I_{24}~\left( U\approx0.6, DGR\approx0.01 \right) \left[ {\rm
    MJy}~{\rm sr}^{-1} \right] \approx 1.2 \times
10^{-2}~\Sigma_{\rm gas}~\left[ {\rm M~pc}^{-2} \right]~.
\end{equation}

\noindent Where the gas surface density, $\Sigma_{\rm gas}$, combines
\hi\ and H$_2$ and includes a contribution from helium. Taking a typical
cirrus $U$ field and $DGR$ one can then attempt a somewhat less
rigorous cirrus correction and proceed as above.

If only \hi\ data are present these can be used for a partial
correction following the methodology above. In this case, we suggest a
lower $w_{\rm H\alpha} = 1.0$ or $w_{\rm FUV} = 1.3$. This will
somewhat offset the lack of correction for cirrus associated with
H$_2$.

Note that these suggestions assume the typical dust-to-gas ratio for
our sample. For low-luminosity, low-metallicity systems the expected
cirrus emission will be lower by a factor that is to first-order
proportional to the metallicity.

{\em With No Multiwavelength Data:} In the absence of multiwavelength
data one can leverage the fact that in star-forming galaxies the
deprojected \hi\ surface density, $\Sigma_{\rm HI}$, appears
relatively flat. Across our data the median $\Sigma_{\rm HI} \approx
6$~M$_\odot$~pc$^{-2}$ with a factor of $2$ scatter. In the absence of
any knowledge of the gas distribution and provided that one is
studying a ``normal'' $z=0$ star-forming galaxy, adopting $\Sigma_{\rm
  HI} \sim 6$~M$_\odot$~pc$^{-2}$ and then proceeding as above will
allow a rough cirrus correction.

\subsection{Implication for SFR-Gas Comparisons}

Our calculations have several implications for measurements of the
relationship between gas and star formation in galaxies. The most
basic implication is that we find broad consistency among different
estimates of $\Sigma_{\rm SFR}$ at kpc resolution across a sample of
$30$ nearby galaxies. We demonstrate sensible scaling relations among
hybrid tracer components, motivate a physical approach to separate
24$\mu$m emission not associated with star formation, and verify the
calibration of the IR component of hybrid tracers for our approach.

{\em Calibration Uncertainty and Gas Depletion Time:} The uncertainty
in the absolute calibration of the dominant 24$\mu$m term implies a
corresponding uncertainty in the ratio of gas to star formation. This
quantity can be phrased as the gas depletion time, $\tau_{\rm Dep} =
M_{\rm gas}/{\rm SFR}$, or its inverse, the star formation efficiency
${\rm SFE} = SFR/M_{\rm gas}$. In either case, our factor of $\approx
1.5$ uncertainty on the 24$\mu$m term in a hybrid SFR estimate implies
an uncertainty of $\approx 25\%$ on the absolute determination of the
SFR and thus the ratio. Comparison to integrated SFRs from
\citet{CALZETTI10} bears out this estimate; these differ from our
estimates only in treatment and calibration of the IR emission and
display a median $\approx 11\%$ offset with $\approx 30\%$
scatter. This uncertainty will be compounded by other uncertainties in
physical parameter estimation: the CO-to-H$_2$ conversion factor,
\hi\ opacity, stellar initial mass function, adopted star formation
history, etc. However, as this is a calibration uncertainty, one
should still be able to carry out internally consistent comparisons
below this level.

{\em Uncertainty in Scaling Relations:} From intercomparison of
tracers we find that the systematic ``tilt'' among $\Sigma_{\rm SFR}$
estimates does not exceed $0.3$~dex across the range $10^{-3} < \left<
\Sigma_{\rm SFR} \right> < 10^{-1}$~M$_\odot$~yr$^{-1}$~kpc$^{-2}$ and
is usually substantially lower. Here ``tilt'' means systematic change
in the ratio of a pair of $\Sigma_{\rm SFR}$ estimates across this
range. The systematic uncertainty in the index of any power law
scaling relation involving $\Sigma_{\rm SFR}$ will be $\sim {\rm
  tilt}/\Delta x$ where $\Delta x$ is the dynamic range of the other,
non-$\Sigma_{\rm SFR}$ variable over the range $10^{-3} < \left<
\Sigma_{\rm SFR} \right> < 10^{-1}$~M$_\odot$~yr$^{-1}$~kpc$^{-2}$. If
$\Sigma_{\rm SFR} \sim x^n$ then the uncertainty in the the index $n$
due to choice of $\Sigma_{\rm SFR}$ esimate will be $\sim 0.15 \times
n$. Thus for any nearly linear relationship \citep[e.g., $\Sigma_{\rm
    SFR}$ vs. $\Sigma_{\rm H2}$,][]{BIGIEL08,BIGIEL11}, the implied uncertainty
in the power law index due to uncertainty in $\Sigma_{\rm SFR}$ is
$\sim 0.15$. For a steeper relationship like the $\Sigma_{\rm SFR}
\sim \Sigma_{\rm HI}^2$ found in the outer parts of galaxies
\citep[][]{BIGIEL10} the implied uncertainty is $\sim 0.3$. The
normalization of the scaling relation will have the overall
calibration uncertainty already discussed.

{\em Small-Scale Scatter in Scaling Relations:} High resolution
observations and simulations have begun to consider the scatter in the
ratios of SFR tracers to gas as a function of scale
\citep{SCHRUBA10,ONODERA10,FELDMANN11}. This scatter encodes key
information on the evolution of star forming regions, the small scale
correlation of star formation, and intrinsic variations in the star
formation efficiency. On the scale of individual clouds or clusters,
one cannot escape the fact that star formation occurs in discrete
events. We considered the intrinsic scatter in H$\alpha$ or FUV
emission from the same population at different, but still young, ages.
These suggest that as one isolates individual populations one will
find intrinsic scatter $\sim 0.3$~dex in H$\alpha$ intensities and
$\sim 0.45$~dex in FUV intensity for otherwise identical regions
with different ages and $\sim 0.5$~dex scatter. This scatter
represents a minimum and will be compounded by any uncertainties in
the estimation of gas mass and extinction or evolution of the
region. The limited measurements available so far appear consistent
with these rough calculations \citep[see also][]{BLANC09}.

\section{Summary}
\label{sec:conc}

We estimate the surface density of recent star formation, $\Sigma_{\rm
  SFR}$, at 1~kpc resolution in the HERACLES sample and examine the
underpinnings of empirical hybrid tracers (H$\alpha$+24$\mu$m and
FUV+24$\mu$m). We look at the magnitudes and correlations among
constituent parts of the hybrid tracers and consider the effects of
discrete events on the derivation of SFRs. Using dust models and
multiwavelength data, we make a physically motivated estimate of the
contamination of the 24$\mu$m band by emission not associated with
star formation. Benchmarking to reference SFRs drawn from the
literature we derive the appropriate calibration to use 24$\mu$m with
H$\alpha$ or FUV given our cirrus approach and physical resolution. We
then compare a wide range of $\Sigma_{\rm SFR}$ estimates and check
the integrated SFRs for our targets. We present our recommended
approach and discuss uncertainties and implications for SFR-gas
comparisons.

We highlight the following points:
\begin{enumerate}

\item Starburst99 simulations of an evolving single stellar population imply an intrinsic scatter
  of $\sim 0.3$~dex in H$\alpha$-based SFR estimates and $\sim
  0.5$~dex in FUV based SFR estimates when isolating a single stellar
  population.

\item We use the dust models of \citet{DRAINE07A} to make physical
  estimates of the 24$\mu$m unassociated with star formation. These models suggest that
  emission along a line of sight depends weakly on the PAH mass
  fraction and linearly on both the dust-to-gas ratio, gas column density,
  and the radiation field not associated with recent star formation.

\item By maximizing the impact of the cirrus subtraction on low-lying
  emission without oversubtracting we find $U_{\rm cirrus} \approx 0.5~U_{\rm min}$
  to be an appropriate radiation field not associated with recent star
  formation. Here $U_{\rm min}$ is the fixed, pervasive radiation
  field illuminating all dust derived from fitting the
  \citet{DRAINE07A} models. Our best-fit field has typical magnitude
  $U=0.6$ times the solar neighborhood interstellar radiation
  field. The resulting cirrus subtraction removes $\approx 20$\% of
  the 24$\mu$m, on average, across our sample.

\item Subtracting only cirrus emitted by dust associated with \hi\ does not offer an appreciable correction in the parts of galaxies where most star formation occurs. IR cirrus from molecular gas must represent an important term, especially given that most dust is mixed with molecular gas in the inner, H$_2$ dominated parts of spiral galaxies. This emission will likely exhibit the same clumpy morphology as CO emission, making it difficult to estimate using image processing techniques.

\item Lack of a robust, wide-field reference SFR renders the absolute
  calibration and universality of the 24$\mu$m term in hybrid tracers
  like H$\alpha$+24$\mu$m and FUV+24$\mu$m uncertain, even for the
  well-studied SINGS sample. We demonstrate this uncertainty in both the SINGS literature and our own measurements.

\item Although there is substantial scatter in the available reference
  SFRs, we verify the appropriate calibration for the 24$\mu$m portion
  of H$\alpha$- and UV-based hybrid tracers for our cirrus approach
  and 1~kpc resolution. Based on references to literature extinction
  and SFR estimates, we recommend $w_{\rm H\alpha} = 1.3$ and $w_{\rm
    FUV} = 1.7$ for our resolution and cirrus approach. We arrive at
  the FUV calibration referencing to both H$\alpha$+24$\mu$m and the
  FUV+TIR approach of \citet{CORTESE08}. The latter means that the
  FUV+24$\mu$m approach is not only derived from the
  H$\alpha$+24$\mu$m calibration.

\item A wide variety of UV, H$\alpha$, and IR-based approaches to
  estimate $\Sigma_{\rm SFR}$ at 1~kpc resolution yield results with
  systematic disagreement of a factor of 2 or less above $\Sigma_{\rm
    SFR} = 10^{-3}$~M$_\odot$~yr$^{-1}$~kpc$^{-2}$.

\item Below $\Sigma_{\rm SFR} =
  10^{-3}$~M$_\odot$~yr$^{-1}$~kpc$^{-2}$ the H$\alpha$ data become unreliable and our results depend strongly on the adopted approach to the IR cirrus. We recommend this value as a useful lower limit for robust $\Sigma_{\rm SFR}$
  estimation using hybrid tracers. Studies targeting lower levels
  should carefully consider the constituent terms.
\end{enumerate}

Our work also has implications for SFR-gas comparisons, which have
heavily leveraged these tracers. We emphasize the requirement that a
well-constructed $\Sigma_{\rm SFR}$ map integrate to a sensible SFR
and note:

\begin{enumerate} 
\item The uncertainty in the calibration of the 24$\mu$m term implies
  a corresponding uncertainty in the gas depletion time, $\tau_{\rm
    dep} = M_{\rm gas}/{\rm SFR}$, of $\approx 25\%$.

\item Systematic variations among tracers are less than 0.3 dex over
  the range $10^{-3} < \left< \Sigma_{\rm SFR} \right> <
  10^{-1}$~M$_\odot$~yr$^{-1}$~kpc$^{-2}$. Over this range, the
  implied uncertainty in the power law slope of any scaling relation
  $\Sigma_{\rm SFR} \sim x^n$ will be $\leq 0.15 \times n$.

\item We expect an intrinsic SFR scatter of $\sim 0.3$~dex (H$\alpha$)
  -- $0.5$~dex (FUV) to emerge at small scales due to age
  effects. Observations that isolate single stellar populations should
  observe such scatter.
\end{enumerate}

\acknowledgments We thank the referee for a detailed, constructive report that improved the quality of this paper. We thank the {\em GALEX} NGS, SINGS, and LVL teams
for making their outstanding datasets available. We acknowledge
helpful correspondence and discussion with K.-F. Schuster, 
R. C. Kennicutt, J. Moustakas, S. Schnee, and K. Sheth. We thank staff
of the IRAM 30m for their assistance carrying out the HERACLES
survey. We thank Deidre Hunter for sharing her H$\alpha$ image of
NGC~4214. F.B., A.K.L., and F.W. gratefully acknowledge the Aspen
Center for Physics, where part of this work was carried out.  Support
for A.K.L. for part of this project was provided by NASA through
Hubble Fellowship grant HST-HF-51258.01-A awarded by the Space
Telescope Science Institute, which is operated by the Association of
Universities for Research in Astronomy, Inc., for NASA, under contract
NAS 5-26555. J.C.M.M. acknowledges financial support from NASA
JPL/{\em Spitzer} grant RSA 1374189 provided for the S4G project.  A.B. wishes to acknowledge support from NSF AST-0955836 and a Cottrell Scholar
award from the Research Corporation for Science Advancement. We
have made use of the Extragalactic Database (NED), which is operated
by the Jet Propulsion Laboratory, California Institute of Technology,
under contract with the National Aeronautics and Space
Administration. We also acknowledge use of the Lyon Extragalactic
Database (LEDA) and NASA's Astrophysics Data System (ADS). The
National Radio Astronomy Observatory is a facility of the National
Science Foundation operated under cooperative agreement by Associated
Universities, Inc.
% \bibliography{/Users/aleroy/bib/akl}

\begin{thebibliography}{93}
\expandafter\ifx\csname natexlab\endcsname\relax\def\natexlab#1{#1}\fi

\bibitem[{{Abdo} {et~al.}(2010){Abdo}, {Ackermann}, {Ajello}, {Baldini},
  {Ballet}, {Barbiellini}, {Bastieri}, {Baughman}, {Bechtol}, {Bellazzini},
  {Berenji}, {Bloom}, {Bonamente}, {Borgland}, {Bregeon}, {Brez}, {Brigida},
  {Bruel}, {Burnett}, {Buson}, {Caliandro}, {Cameron}, {Caraveo}, {Casandjian},
  {Cecchi}, {{\c C}elik}, {Chekhtman}, {Cheung}, {Chiang}, {Ciprini}, {Claus},
  {Cohen-Tanugi}, {Cominsky}, {Conrad}, {Dermer}, {de Palma}, {Digel}, {Silva},
  {Drell}, {Dubois}, {Dumora}, {Farnier}, {Favuzzi}, {Fegan}, {Focke},
  {Fortin}, {Frailis}, {Fukazawa}, {Funk}, {Fusco}, {Gargano}, {Gehrels},
  {Germani}, {Giavitto}, {Giebels}, {Giglietto}, {Giordano}, {Glanzman},
  {Godfrey}, {Grenier}, {Grondin}, {Grove}, {Guillemot}, {Guiriec}, {Harding},
  {Hayashida}, {Horan}, {Hughes}, {Jackson}, {J{\'o}hannesson}, {Johnson},
  {Johnson}, {Kamae}, {Katagiri}, {Kataoka}, {Kawai}, {Kerr}, {Kn{\"o}dlseder},
  {Kuss}, {Lande}, {Latronico}, {Lemoine-Goumard}, {Longo}, {Loparco}, {Lott},
  {Lovellette}, {Lubrano}, {Makeev}, {Mazziotta}, {McEnery}, {Meurer},
  {Michelson}, {Mitthumsiri}, {Mizuno}, {Monte}, {Monzani}, {Morselli},
  {Moskalenko}, {Murgia}, {Nolan}, {Norris}, {Nuss}, {Ohsugi}, {Okumura},
  {Omodei}, {Orlando}, {Ormes}, {Paneque}, {Pelassa}, {Pepe}, {Pesce-Rollins},
  {Piron}, {Porter}, {Rain{\`o}}, {Rando}, {Razzano}, {Reimer}, {Reimer},
  {Reposeur}, {Rodriguez}, {Ryde}, {Sadrozinski}, {Sanchez}, {Sander}, {Saz
  Parkinson}, {Sgr{\`o}}, {Siskind}, {Smith}, {Spandre}, {Spinelli}, {Starck},
  {Strickman}, {Strong}, {Suson}, {Takahashi}, {Tanaka}, {Thayer}, {Thayer},
  {Thompson}, {Tibaldo}, {Torres}, {Tosti}, {Tramacere}, {Uchiyama}, {Usher},
  {Vasileiou}, {Vilchez}, {Vitale}, {Waite}, {Wang}, {Winer}, {Wood}, {Ylinen},
  \& {Ziegler}}]{ABDOXCO}
{Abdo}, A.~A., {Ackermann}, M., {Ajello}, M., {Baldini}, L., {Ballet}, J.,
  {Barbiellini}, G., {Bastieri}, D., {Baughman}, B.~M., {Bechtol}, K.,
  {Bellazzini}, R., {Berenji}, B., {Bloom}, E.~D., {Bonamente}, E., {Borgland},
  A.~W., {Bregeon}, J., {Brez}, A., {Brigida}, M., {Bruel}, P., {Burnett},
  T.~H., {Buson}, S., {Caliandro}, G.~A., {Cameron}, R.~A., {Caraveo}, P.~A.,
  {Casandjian}, J.~M., {Cecchi}, C., {{\c C}elik}, {\"O}., {Chekhtman}, A.,
  {Cheung}, C.~C., {Chiang}, J., {Ciprini}, S., {Claus}, R., {Cohen-Tanugi},
  J., {Cominsky}, L.~R., {Conrad}, J., {Dermer}, C.~D., {de Palma}, F.,
  {Digel}, S.~W., {Silva}, E.~d.~C.~e., {Drell}, P.~S., {Dubois}, R., {Dumora},
  D., {Farnier}, C., {Favuzzi}, C., {Fegan}, S.~J., {Focke}, W.~B., {Fortin},
  P., {Frailis}, M., {Fukazawa}, Y., {Funk}, S., {Fusco}, P., {Gargano}, F.,
  {Gehrels}, N., {Germani}, S., {Giavitto}, G., {Giebels}, B., {Giglietto}, N.,
  {Giordano}, F., {Glanzman}, T., {Godfrey}, G., {Grenier}, I.~A., {Grondin},
  M., {Grove}, J.~E., {Guillemot}, L., {Guiriec}, S., {Harding}, A.~K.,
  {Hayashida}, M., {Horan}, D., {Hughes}, R.~E., {Jackson}, M.~S.,
  {J{\'o}hannesson}, G., {Johnson}, A.~S., {Johnson}, W.~N., {Kamae}, T.,
  {Katagiri}, H., {Kataoka}, J., {Kawai}, N., {Kerr}, M., {Kn{\"o}dlseder}, J.,
  {Kuss}, M., {Lande}, J., {Latronico}, L., {Lemoine-Goumard}, M., {Longo}, F.,
  {Loparco}, F., {Lott}, B., {Lovellette}, M.~N., {Lubrano}, P., {Makeev}, A.,
  {Mazziotta}, M.~N., {McEnery}, J.~E., {Meurer}, C., {Michelson}, P.~F.,
  {Mitthumsiri}, W., {Mizuno}, T., {Monte}, C., {Monzani}, M.~E., {Morselli},
  A., {Moskalenko}, I.~V., {Murgia}, S., {Nolan}, P.~L., {Norris}, J.~P.,
  {Nuss}, E., {Ohsugi}, T., {Okumura}, A., {Omodei}, N., {Orlando}, E.,
  {Ormes}, J.~F., {Paneque}, D., {Pelassa}, V., {Pepe}, M., {Pesce-Rollins},
  M., {Piron}, F., {Porter}, T.~A., {Rain{\`o}}, S., {Rando}, R., {Razzano},
  M., {Reimer}, A., {Reimer}, O., {Reposeur}, T., {Rodriguez}, A.~Y., {Ryde},
  F., {Sadrozinski}, H., {Sanchez}, D., {Sander}, A., {Saz Parkinson}, P.~M.,
  {Sgr{\`o}}, C., {Siskind}, E.~J., {Smith}, P.~D., {Spandre}, G., {Spinelli},
  P., {Starck}, J., {Strickman}, M.~S., {Strong}, A.~W., {Suson}, D.~J.,
  {Takahashi}, H., {Tanaka}, T., {Thayer}, J.~B., {Thayer}, J.~G., {Thompson},
  D.~J., {Tibaldo}, L., {Torres}, D.~F., {Tosti}, G., {Tramacere}, A.,
  {Uchiyama}, Y., {Usher}, T.~L., {Vasileiou}, V., {Vilchez}, N., {Vitale}, V.,
  {Waite}, A.~P., {Wang}, P., {Winer}, B.~L., {Wood}, K.~S., {Ylinen}, T., \&
  {Ziegler}, M. 2010, \apj, 710, 133

\bibitem[{{Alonso-Herrero} {et~al.}(2006){Alonso-Herrero}, {Rieke}, {Rieke},
  {Colina}, {P{\'e}rez-Gonz{\'a}lez}, \& {Ryder}}]{ALONSOHERRERO06}
{Alonso-Herrero}, A., {Rieke}, G.~H., {Rieke}, M.~J., {Colina}, L.,
  {P{\'e}rez-Gonz{\'a}lez}, P.~G., \& {Ryder}, S.~D. 2006, \apj, 650, 835

\bibitem[{{Bastian} {et~al.}(2010){Bastian}, {Covey}, \& {Meyer}}]{BASTIAN10}
{Bastian}, N., {Covey}, K.~R., \& {Meyer}, M.~R. 2010, \araa, 48, 339

\bibitem[{{Bigiel} {et~al.}(2010){Bigiel}, {Bolatto}, {Leroy}, {Blitz},
  {Walter}, {Rosolowsky}, {Lopez}, \& {Plambeck}}]{BIGIEL10}
{Bigiel}, F., {Bolatto}, A., {Leroy}, A., {Blitz}, L., {Walter}, F.,
  {Rosolowsky}, E., {Lopez}, L., \& {Plambeck}, R. 2010, ArXiv e-prints

\bibitem[{{Bigiel} {et~al.}(2008){Bigiel}, {Leroy}, {Walter}, {Brinks}, {de
  Blok}, {Madore}, \& {Thornley}}]{BIGIEL08}
{Bigiel}, F., {Leroy}, A., {Walter}, F., {Brinks}, E., {de Blok}, W.~J.~G.,
  {Madore}, B., \& {Thornley}, M.~D. 2008, \aj, 136, 2846

\bibitem[{{Bigiel} {et~al.}(2011){Bigiel}, {Leroy}, {Walter}, {Brinks}, {de
  Blok}, {Kramer}, {Rix}, {Schruba}, {Schuster}, {Usero}, \&
  {Wiesemeyer}}]{BIGIEL11}
{Bigiel}, F., {Leroy}, A.~K., {Walter}, F., {Brinks}, E., {de Blok}, W.~J.~G.,
  {Kramer}, C., {Rix}, H.~W., {Schruba}, A., {Schuster}, K.-F., {Usero}, A., \&
  {Wiesemeyer}, H.~W. 2011, \apjl, 730, L13+

\bibitem[{{Blanc} {et~al.}(2009){Blanc}, {Heiderman}, {Gebhardt}, {Evans}, \&
  {Adams}}]{BLANC09}
{Blanc}, G.~A., {Heiderman}, A., {Gebhardt}, K., {Evans}, N.~J., \& {Adams}, J.
  2009, \apj, 704, 842

\bibitem[{{Boselli} \& {Gavazzi}(2002)}]{BOSELLI02B}
{Boselli}, A., \& {Gavazzi}, G. 2002, \aap, 386, 124

\bibitem[{{Boulanger} {et~al.}(1996){Boulanger}, {Abergel}, {Bernard},
  {Burton}, {Desert}, {Hartmann}, {Lagache}, \& {Puget}}]{BOULANGER96}
{Boulanger}, F., {Abergel}, A., {Bernard}, J.-P., {Burton}, W.~B., {Desert},
  F.-X., {Hartmann}, D., {Lagache}, G., \& {Puget}, J.-L. 1996, \aap, 312, 256

\bibitem[{{Buat}(1992)}]{BUAT92}
{Buat}, V. 1992, \aap, 264, 444

\bibitem[{{Calzetti} {et~al.}(2007){Calzetti}, {Kennicutt}, {Engelbracht},
  {Leitherer}, {Draine}, {Kewley}, {Moustakas}, {Sosey}, {Dale}, {Gordon},
  {Helou}, {Hollenbach}, {Armus}, {Bendo}, {Bot}, {Buckalew}, {Jarrett}, {Li},
  {Meyer}, {Murphy}, {Prescott}, {Regan}, {Rieke}, {Roussel}, {Sheth}, {Smith},
  {Thornley}, \& {Walter}}]{CALZETTI07}
{Calzetti}, D., {Kennicutt}, R.~C., {Engelbracht}, C.~W., {Leitherer}, C.,
  {Draine}, B.~T., {Kewley}, L., {Moustakas}, J., {Sosey}, M., {Dale}, D.~A.,
  {Gordon}, K.~D., {Helou}, G.~X., {Hollenbach}, D.~J., {Armus}, L., {Bendo},
  G., {Bot}, C., {Buckalew}, B., {Jarrett}, T., {Li}, A., {Meyer}, M.,
  {Murphy}, E.~J., {Prescott}, M., {Regan}, M.~W., {Rieke}, G.~H., {Roussel},
  H., {Sheth}, K., {Smith}, J.~D.~T., {Thornley}, M.~D., \& {Walter}, F. 2007,
  \apj, 666, 870

\bibitem[{{Calzetti} {et~al.}(2005){Calzetti}, {Kennicutt}, {Bianchi},
  {Thilker}, {Dale}, {Engelbracht}, {Leitherer}, {Meyer}, {Sosey}, {Mutchler},
  {Regan}, {Thornley}, {Armus}, {Bendo}, {Boissier}, {Boselli}, {Draine},
  {Gordon}, {Helou}, {Hollenbach}, {Kewley}, {Madore}, {Martin}, {Murphy},
  {Rieke}, {Rieke}, {Roussel}, {Sheth}, {Smith}, {Walter}, {White}, {Yi},
  {Scoville}, {Polletta}, \& {Lindler}}]{CALZETTI05}
{Calzetti}, D., {Kennicutt}, Jr., R.~C., {Bianchi}, L., {Thilker}, D.~A.,
  {Dale}, D.~A., {Engelbracht}, C.~W., {Leitherer}, C., {Meyer}, M.~J.,
  {Sosey}, M.~L., {Mutchler}, M., {Regan}, M.~W., {Thornley}, M.~D., {Armus},
  L., {Bendo}, G.~J., {Boissier}, S., {Boselli}, A., {Draine}, B.~T., {Gordon},
  K.~D., {Helou}, G., {Hollenbach}, D.~J., {Kewley}, L., {Madore}, B.~F.,
  {Martin}, D.~C., {Murphy}, E.~J., {Rieke}, G.~H., {Rieke}, M.~J., {Roussel},
  H., {Sheth}, K., {Smith}, J.~D., {Walter}, F., {White}, B.~A., {Yi}, S.,
  {Scoville}, N.~Z., {Polletta}, M., \& {Lindler}, D. 2005, \apj, 633, 871

\bibitem[{{Calzetti} {et~al.}(1994){Calzetti}, {Kinney}, \&
  {Storchi-Bergmann}}]{CALZETTI94}
{Calzetti}, D., {Kinney}, A.~L., \& {Storchi-Bergmann}, T. 1994, \apj, 429, 582

\bibitem[{{Calzetti} {et~al.}(2010){Calzetti}, {Wu}, {Hong}, {Kennicutt},
  {Lee}, {Dale}, {Engelbracht}, {van Zee}, {Draine}, {Hao}, {Gordon},
  {Moustakas}, {Murphy}, {Regan}, {Begum}, {Block}, {Dalcanton}, {Funes}, {Gil
  de Paz}, {Johnson}, {Sakai}, {Skillman}, {Walter}, {Weisz}, {Williams}, \&
  {Wu}}]{CALZETTI10}
{Calzetti}, D., {Wu}, S., {Hong}, S., {Kennicutt}, R.~C., {Lee}, J.~C., {Dale},
  D.~A., {Engelbracht}, C.~W., {van Zee}, L., {Draine}, B.~T., {Hao}, C.,
  {Gordon}, K.~D., {Moustakas}, J., {Murphy}, E.~J., {Regan}, M., {Begum}, A.,
  {Block}, M., {Dalcanton}, J., {Funes}, J., {Gil de Paz}, A., {Johnson}, B.,
  {Sakai}, S., {Skillman}, E., {Walter}, F., {Weisz}, D., {Williams}, B., \&
  {Wu}, Y. 2010, \apj, 714, 1256

\bibitem[{{Cardelli} {et~al.}(1989){Cardelli}, {Clayton}, \&
  {Mathis}}]{CARDELLI89}
{Cardelli}, J.~A., {Clayton}, G.~C., \& {Mathis}, J.~S. 1989, \apj, 345, 245

\bibitem[{{Chabrier}(2003)}]{CHABRIER03}
{Chabrier}, G. 2003, \pasp, 115, 763

\bibitem[{{Cortese} {et~al.}(2008){Cortese}, {Boselli}, {Franzetti}, {Decarli},
  {Gavazzi}, {Boissier}, \& {Buat}}]{CORTESE08}
{Cortese}, L., {Boselli}, A., {Franzetti}, P., {Decarli}, R., {Gavazzi}, G.,
  {Boissier}, S., \& {Buat}, V. 2008, \mnras, 386, 1157

\bibitem[{{Dale} {et~al.}(2009){Dale}, {Cohen}, {Johnson}, {Schuster},
  {Calzetti}, {Engelbracht}, {Gil de Paz}, {Kennicutt}, {Lee}, {Begum},
  {Block}, {Dalcanton}, {Funes}, {Gordon}, {Johnson}, {Marble}, {Sakai},
  {Skillman}, {van Zee}, {Walter}, {Weisz}, {Williams}, {Wu}, \& {Wu}}]{DALE09}
{Dale}, D.~A., {Cohen}, S.~A., {Johnson}, L.~C., {Schuster}, M.~D., {Calzetti},
  D., {Engelbracht}, C.~W., {Gil de Paz}, A., {Kennicutt}, R.~C., {Lee}, J.~C.,
  {Begum}, A., {Block}, M., {Dalcanton}, J.~J., {Funes}, J.~G., {Gordon},
  K.~D., {Johnson}, B.~D., {Marble}, A.~R., {Sakai}, S., {Skillman}, E.~D.,
  {van Zee}, L., {Walter}, F., {Weisz}, D.~R., {Williams}, B., {Wu}, S., \&
  {Wu}, Y. 2009, \apj, 703, 517

\bibitem[{{Dale} {et~al.}(2007){Dale}, {Gil de Paz}, {Gordon}, {Hanson},
  {Armus}, {Bendo}, {Bianchi}, {Block}, {Boissier}, {Boselli}, {Buckalew},
  {Buat}, {Burgarella}, {Calzetti}, {Cannon}, {Engelbracht}, {Helou},
  {Hollenbach}, {Jarrett}, {Kennicutt}, {Leitherer}, {Li}, {Madore}, {Martin},
  {Meyer}, {Murphy}, {Regan}, {Roussel}, {Smith}, {Sosey}, {Thilker}, \&
  {Walter}}]{DALE07}
{Dale}, D.~A., {Gil de Paz}, A., {Gordon}, K.~D., {Hanson}, H.~M., {Armus}, L.,
  {Bendo}, G.~J., {Bianchi}, L., {Block}, M., {Boissier}, S., {Boselli}, A.,
  {Buckalew}, B.~A., {Buat}, V., {Burgarella}, D., {Calzetti}, D., {Cannon},
  J.~M., {Engelbracht}, C.~W., {Helou}, G., {Hollenbach}, D.~J., {Jarrett},
  T.~H., {Kennicutt}, R.~C., {Leitherer}, C., {Li}, A., {Madore}, B.~F.,
  {Martin}, D.~C., {Meyer}, M.~J., {Murphy}, E.~J., {Regan}, M.~W., {Roussel},
  H., {Smith}, J.~D.~T., {Sosey}, M.~L., {Thilker}, D.~A., \& {Walter}, F.
  2007, \apj, 655, 863

\bibitem[{{Dame} {et~al.}(2001){Dame}, {Hartmann}, \& {Thaddeus}}]{DAME01}
{Dame}, T.~M., {Hartmann}, D., \& {Thaddeus}, P. 2001, \apj, 547, 792

\bibitem[{{Dopita} {et~al.}(2003){Dopita}, {Groves}, {Sutherland}, \&
  {Kewley}}]{DOPITA03}
{Dopita}, M.~A., {Groves}, B.~A., {Sutherland}, R.~S., \& {Kewley}, L.~J. 2003,
  \apj, 583, 727

\bibitem[{{Draine} {et~al.}(2007){Draine}, {Dale}, {Bendo}, {Gordon}, {Smith},
  {Armus}, {Engelbracht}, {Helou}, {Kennicutt}, {Li}, {Roussel}, {Walter},
  {Calzetti}, {Moustakas}, {Murphy}, {Rieke}, {Bot}, {Hollenbach}, {Sheth}, \&
  {Teplitz}}]{DRAINE07B}
{Draine}, B.~T., {Dale}, D.~A., {Bendo}, G., {Gordon}, K.~D., {Smith},
  J.~D.~T., {Armus}, L., {Engelbracht}, C.~W., {Helou}, G., {Kennicutt}, Jr.,
  R.~C., {Li}, A., {Roussel}, H., {Walter}, F., {Calzetti}, D., {Moustakas},
  J., {Murphy}, E.~J., {Rieke}, G.~H., {Bot}, C., {Hollenbach}, D.~J., {Sheth},
  K., \& {Teplitz}, H.~I. 2007, \apj, 663, 866

\bibitem[{{Draine} \& {Li}(2007)}]{DRAINE07A}
{Draine}, B.~T., \& {Li}, A. 2007, \apj, 657, 810

\bibitem[{{Feldmann} {et~al.}(2011){Feldmann}, {Gnedin}, \&
  {Kravtsov}}]{FELDMANN11}
{Feldmann}, R., {Gnedin}, N.~Y., \& {Kravtsov}, A.~V. 2011, \apj, 732, 115

\bibitem[{{Galliano} {et~al.}(2011){Galliano}, {Hony}, {Bernard}, {Bot},
  {Madden}, {Roman-Duval}, {Galametz}, {Li}, {Meixner}, {Engelbracht},
  {Lebouteiller}, {Misselt}, {Montiel}, {Panuzzo}, {Reach}, \&
  {Skibba}}]{GALLIANO11}
{Galliano}, F., {Hony}, S., {Bernard}, J.~., {Bot}, C., {Madden}, S.~C.,
  {Roman-Duval}, J., {Galametz}, M., {Li}, A., {Meixner}, M., {Engelbracht},
  C.~W., {Lebouteiller}, V., {Misselt}, K., {Montiel}, E., {Panuzzo}, P.,
  {Reach}, W.~T., \& {Skibba}, R. 2011, ArXiv e-prints

\bibitem[{{Gavazzi} {et~al.}(2003){Gavazzi}, {Boselli}, {Donati}, {Franzetti},
  \& {Scodeggio}}]{GAVAZZI03}
{Gavazzi}, G., {Boselli}, A., {Donati}, A., {Franzetti}, P., \& {Scodeggio}, M.
  2003, \aap, 400, 451

\bibitem[{{Genzel} {et~al.}(2010){Genzel}, {Tacconi}, {Gracia-Carpio},
  {Sternberg}, {Cooper}, {Shapiro}, {Bolatto}, {Bouch{\'e}}, {Bournaud},
  {Burkert}, {Combes}, {Comerford}, {Cox}, {Davis}, {Schreiber},
  {Garcia-Burillo}, {Lutz}, {Naab}, {Neri}, {Omont}, {Shapley}, \&
  {Weiner}}]{GENZEL10}
{Genzel}, R., {Tacconi}, L.~J., {Gracia-Carpio}, J., {Sternberg}, A., {Cooper},
  M.~C., {Shapiro}, K., {Bolatto}, A., {Bouch{\'e}}, N., {Bournaud}, F.,
  {Burkert}, A., {Combes}, F., {Comerford}, J., {Cox}, P., {Davis}, M.,
  {Schreiber}, N.~M.~F., {Garcia-Burillo}, S., {Lutz}, D., {Naab}, T., {Neri},
  R., {Omont}, A., {Shapley}, A., \& {Weiner}, B. 2010, \mnras, 407, 2091

\bibitem[{{Gil de Paz} {et~al.}(2007){Gil de Paz}, {Boissier}, {Madore},
  {Seibert}, {Joe}, {Boselli}, {Wyder}, {Thilker}, {Bianchi}, {Rey}, {Rich},
  {Barlow}, {Conrow}, {Forster}, {Friedman}, {Martin}, {Morrissey}, {Neff},
  {Schiminovich}, {Small}, {Donas}, {Heckman}, {Lee}, {Milliard}, {Szalay}, \&
  {Yi}}]{GILDEPAZ07}
{Gil de Paz}, A., {Boissier}, S., {Madore}, B.~F., {Seibert}, M., {Joe}, Y.~H.,
  {Boselli}, A., {Wyder}, T.~K., {Thilker}, D., {Bianchi}, L., {Rey}, S.-C.,
  {Rich}, R.~M., {Barlow}, T.~A., {Conrow}, T., {Forster}, K., {Friedman},
  P.~G., {Martin}, D.~C., {Morrissey}, P., {Neff}, S.~G., {Schiminovich}, D.,
  {Small}, T., {Donas}, J., {Heckman}, T.~M., {Lee}, Y.-W., {Milliard}, B.,
  {Szalay}, A.~S., \& {Yi}, S. 2007, \apjs, 173, 185

\bibitem[{{Gini}(1912)}]{GINI1912}
{Gini}, C. 1912, {Memorie di Metodologia Statistica}, ed. {E. Pizetti \& Ti.
  Selvemini}

\bibitem[{{Greenawalt} {et~al.}(1998){Greenawalt}, {Walterbos}, {Thilker}, \&
  {Hoopes}}]{GREENAWALT97}
{Greenawalt}, B., {Walterbos}, R.~A.~M., {Thilker}, D., \& {Hoopes}, C.~G.
  1998, \apj, 506, 135

\bibitem[{{Groves} {et~al.}(2008){Groves}, {Dopita}, {Sutherland}, {Kewley},
  {Fischera}, {Leitherer}, {Brandl}, \& {van Breugel}}]{GROVES08}
{Groves}, B., {Dopita}, M.~A., {Sutherland}, R.~S., {Kewley}, L.~J.,
  {Fischera}, J., {Leitherer}, C., {Brandl}, B., \& {van Breugel}, W. 2008,
  \apjs, 176, 438

\bibitem[{{Helfer} {et~al.}(2003){Helfer}, {Thornley}, {Regan}, {Wong},
  {Sheth}, {Vogel}, {Blitz}, \& {Bock}}]{HELFER03}
{Helfer}, T.~T., {Thornley}, M.~D., {Regan}, M.~W., {Wong}, T., {Sheth}, K.,
  {Vogel}, S.~N., {Blitz}, L., \& {Bock}, D.~C.-J. 2003, \apjs, 145, 259

\bibitem[{{Helou} {et~al.}(2004){Helou}, {Roussel}, {Appleton}, {Frayer},
  {Stolovy}, {Storrie-Lombardi}, {Hurt}, {Lowrance}, {Makovoz}, {Masci},
  {Surace}, {Gordon}, {Alonso-Herrero}, {Engelbracht}, {Misselt}, {Rieke},
  {Rieke}, {Willner}, {Pahre}, {Ashby}, {Fazio}, \& {Smith}}]{HELOU04}
{Helou}, G., {Roussel}, H., {Appleton}, P., {Frayer}, D., {Stolovy}, S.,
  {Storrie-Lombardi}, L., {Hurt}, R., {Lowrance}, P., {Makovoz}, D., {Masci},
  F., {Surace}, J., {Gordon}, K.~D., {Alonso-Herrero}, A., {Engelbracht},
  C.~W., {Misselt}, K., {Rieke}, G., {Rieke}, M., {Willner}, S.~P., {Pahre},
  M., {Ashby}, M.~L.~N., {Fazio}, G.~G., \& {Smith}, H.~A. 2004, \apjs, 154,
  253

\bibitem[{{Heyer} {et~al.}(2009){Heyer}, {Krawczyk}, {Duval}, \&
  {Jackson}}]{HEYER09}
{Heyer}, M., {Krawczyk}, C., {Duval}, J., \& {Jackson}, J.~M. 2009, \apj, 699,
  1092

\bibitem[{{Hoopes} {et~al.}(2001){Hoopes}, {Walterbos}, \& {Bothun}}]{HOOPES01}
{Hoopes}, C.~G., {Walterbos}, R.~A.~M., \& {Bothun}, G.~D. 2001, \apj, 559, 878

\bibitem[{{Hunter} \& {Elmegreen}(2004)}]{HUNTER04}
{Hunter}, D.~A., \& {Elmegreen}, B.~G. 2004, \aj, 128, 2170

\bibitem[{{Inoue}(2001)}]{INOUE01A}
{Inoue}, A.~K. 2001, \aj, 122, 1788

\bibitem[{{Inoue} {et~al.}(2001){Inoue}, {Hirashita}, \& {Kamaya}}]{INOUE01B}
{Inoue}, A.~K., {Hirashita}, H., \& {Kamaya}, H. 2001, \apj, 555, 613

\bibitem[{{Kennicutt} {et~al.}(2011){Kennicutt}, {Calzetti}, {Aniano},
  {Appleton}, {Armus}, {Beir{\~a}o}, {Bolatto}, {Brandl}, {Crocker}, {Croxall},
  {Dale}, {Meyer}, {Draine}, {Engelbracht}, {Galametz}, {Gordon}, {Groves},
  {Hao}, {Helou}, {Hinz}, {Hunt}, {Johnson}, {Koda}, {Krause}, {Leroy}, {Li},
  {Meidt}, {Montiel}, {Murphy}, {Rahman}, {Rix}, {Roussel}, {Sandstrom},
  {Sauvage}, {Schinnerer}, {Skibba}, {Smith}, {Srinivasan}, {Vigroux},
  {Walter}, {Wilson}, {Wolfire}, \& {Zibetti}}]{KENNICUTT11}
{Kennicutt}, R.~C., {Calzetti}, D., {Aniano}, G., {Appleton}, P., {Armus}, L.,
  {Beir{\~a}o}, P., {Bolatto}, A.~D., {Brandl}, B., {Crocker}, A., {Croxall},
  K., {Dale}, D.~A., {Meyer}, J.~D., {Draine}, B.~T., {Engelbracht}, C.~W.,
  {Galametz}, M., {Gordon}, K.~D., {Groves}, B., {Hao}, C.-N., {Helou}, G.,
  {Hinz}, J., {Hunt}, L.~K., {Johnson}, B., {Koda}, J., {Krause}, O., {Leroy},
  A.~K., {Li}, Y., {Meidt}, S., {Montiel}, E., {Murphy}, E.~J., {Rahman}, N.,
  {Rix}, H.-W., {Roussel}, H., {Sandstrom}, K., {Sauvage}, M., {Schinnerer},
  E., {Skibba}, R., {Smith}, J.~D.~T., {Srinivasan}, S., {Vigroux}, L.,
  {Walter}, F., {Wilson}, C.~D., {Wolfire}, M., \& {Zibetti}, S. 2011, \pasp,
  123, 1347

\bibitem[{{Kennicutt} {et~al.}(2009){Kennicutt}, {Hao}, {Calzetti},
  {Moustakas}, {Dale}, {Bendo}, {Engelbracht}, {Johnson}, \&
  {Lee}}]{KENNICUTT09}
{Kennicutt}, R.~C., {Hao}, C., {Calzetti}, D., {Moustakas}, J., {Dale}, D.~A.,
  {Bendo}, G., {Engelbracht}, C.~W., {Johnson}, B.~D., \& {Lee}, J.~C. 2009,
  \apj, 703, 1672

\bibitem[{{Kennicutt}(1998)}]{KENNICUTT98B}
{Kennicutt}, Jr., R.~C. 1998, \araa, 36, 189

\bibitem[{{Kennicutt} {et~al.}(2003){Kennicutt}, {Armus}, {Bendo}, {Calzetti},
  {Dale}, {Draine}, {Engelbracht}, {Gordon}, {Grauer}, {Helou}, {Hollenbach},
  {Jarrett}, {Kewley}, {Leitherer}, {Li}, {Malhotra}, {Regan}, {Rieke},
  {Rieke}, {Roussel}, {Smith}, {Thornley}, \& {Walter}}]{KENNICUTT03}
{Kennicutt}, Jr., R.~C., {Armus}, L., {Bendo}, G., {Calzetti}, D., {Dale},
  D.~A., {Draine}, B.~T., {Engelbracht}, C.~W., {Gordon}, K.~D., {Grauer},
  A.~D., {Helou}, G., {Hollenbach}, D.~J., {Jarrett}, T.~H., {Kewley}, L.~J.,
  {Leitherer}, C., {Li}, A., {Malhotra}, S., {Regan}, M.~W., {Rieke}, G.~H.,
  {Rieke}, M.~J., {Roussel}, H., {Smith}, J.-D.~T., {Thornley}, M.~D., \&
  {Walter}, F. 2003, \pasp, 115, 928

\bibitem[{{Kennicutt} {et~al.}(2007){Kennicutt}, {Calzetti}, {Walter}, {Helou},
  {Hollenbach}, {Armus}, {Bendo}, {Dale}, {Draine}, {Engelbracht}, {Gordon},
  {Prescott}, {Regan}, {Thornley}, {Bot}, {Brinks}, {de Blok}, {de Mello},
  {Meyer}, {Moustakas}, {Murphy}, {Sheth}, \& {Smith}}]{KENNICUTT07}
{Kennicutt}, Jr., R.~C., {Calzetti}, D., {Walter}, F., {Helou}, G.,
  {Hollenbach}, D.~J., {Armus}, L., {Bendo}, G., {Dale}, D.~A., {Draine},
  B.~T., {Engelbracht}, C.~W., {Gordon}, K.~D., {Prescott}, M.~K.~M., {Regan},
  M.~W., {Thornley}, M.~D., {Bot}, C., {Brinks}, E., {de Blok}, E., {de Mello},
  D., {Meyer}, M., {Moustakas}, J., {Murphy}, E.~J., {Sheth}, K., \& {Smith},
  J.~D.~T. 2007, \apj, 671, 333

\bibitem[{{Kennicutt} {et~al.}(2008){Kennicutt}, {Lee}, {Funes}, {Sakai}, \&
  {Akiyama}}]{KENNICUTT08}
{Kennicutt}, Jr., R.~C., {Lee}, J.~C., {Funes}, Jos{\'e}~G., S.~J., {Sakai},
  S., \& {Akiyama}, S. 2008, \apjs, 178, 247

\bibitem[{{Knapen} {et~al.}(2004){Knapen}, {Stedman}, {Bramich}, {Folkes}, \&
  {Bradley}}]{KNAPEN04}
{Knapen}, J.~H., {Stedman}, S., {Bramich}, D.~M., {Folkes}, S.~L., \&
  {Bradley}, T.~R. 2004, \aap, 426, 1135

\bibitem[{{Kroupa}(2001)}]{KROUPA01}
{Kroupa}, P. 2001, \mnras, 322, 231

\bibitem[{{Kuno} {et~al.}(2007){Kuno}, {Sato}, {Nakanishi}, {Hirota}, {Tosaki},
  {Shioya}, {Sorai}, {Nakai}, {Nishiyama}, \& {Vila-Vilar{\'o}}}]{KUNO07}
{Kuno}, N., {Sato}, N., {Nakanishi}, H., {Hirota}, A., {Tosaki}, T., {Shioya},
  Y., {Sorai}, K., {Nakai}, N., {Nishiyama}, K., \& {Vila-Vilar{\'o}}, B. 2007,
  \pasj, 59, 117

\bibitem[{{Law} {et~al.}(2011){Law}, {Gordon}, \& {Misselt}}]{LAW11}
{Law}, K.-H., {Gordon}, K.~D., \& {Misselt}, K.~A. 2011, \apj, 738, 124

\bibitem[{{Lawton} {et~al.}(2010){Lawton}, {Gordon}, {Babler}, {Block},
  {Bolatto}, {Bracker}, {Carlson}, {Engelbracht}, {Hora}, {Indebetouw},
  {Madden}, {Meade}, {Meixner}, {Misselt}, {Oey}, {Oliveira}, {Robitaille},
  {Sewilo}, {Shiao}, {Vijh}, \& {Whitney}}]{LAWTON10}
{Lawton}, B., {Gordon}, K.~D., {Babler}, B., {Block}, M., {Bolatto}, A.~D.,
  {Bracker}, S., {Carlson}, L.~R., {Engelbracht}, C.~W., {Hora}, J.~L.,
  {Indebetouw}, R., {Madden}, S.~C., {Meade}, M., {Meixner}, M., {Misselt}, K.,
  {Oey}, M.~S., {Oliveira}, J.~M., {Robitaille}, T., {Sewilo}, M., {Shiao}, B.,
  {Vijh}, U.~P., \& {Whitney}, B. 2010, \apj, 716, 453

\bibitem[{{Lee} {et~al.}(2011){Lee}, {Stanimirovic}, {Douglas}, {Knee}, {Di
  Francesco}, {Gibson}, {Begum}, {Grcevich}, {Heiles}, {Korpela}, {Leroy},
  {Peek}, {Pingel}, {Putman}, \& {Saul}}]{LEE11}
{Lee}, M.-Y., {Stanimirovic}, S., {Douglas}, K.~A., {Knee}, L.~B.~G., {Di
  Francesco}, J., {Gibson}, S.~J., {Begum}, A., {Grcevich}, J., {Heiles}, C.,
  {Korpela}, E.~J., {Leroy}, A.~K., {Peek}, J.~E.~G., {Pingel}, N., {Putman},
  M.~E., \& {Saul}, D. 2011, ArXiv e-prints

\bibitem[{{Leitherer} {et~al.}(1999){Leitherer}, {Schaerer}, {Goldader},
  {Gonz{\'a}lez Delgado}, {Robert}, {Kune}, {de Mello}, {Devost}, \&
  {Heckman}}]{LEITHERER99}
{Leitherer}, C., {Schaerer}, D., {Goldader}, J.~D., {Gonz{\'a}lez Delgado},
  R.~M., {Robert}, C., {Kune}, D.~F., {de Mello}, D.~F., {Devost}, D., \&
  {Heckman}, T.~M. 1999, \apjs, 123, 3

\bibitem[{{Leroy} {et~al.}(2009){Leroy}, {Bolatto}, {Bot}, {Engelbracht},
  {Gordon}, {Israel}, {Rubio}, {Sandstrom}, \& {Stanimirovi{\'c}}}]{LEROY09}
{Leroy}, A.~K., {Bolatto}, A., {Bot}, C., {Engelbracht}, C.~W., {Gordon}, K.,
  {Israel}, F.~P., {Rubio}, M., {Sandstrom}, K., \& {Stanimirovi{\'c}}, S.
  2009, \apj, 702, 352

\bibitem[{{Leroy} {et~al.}(2011){Leroy}, {Bolatto}, {Gordon}, {Sandstrom},
  {Gratier}, {Rosolowsky}, {Engelbracht}, {Mizuno}, {Corbelli}, {Fukui}, \&
  {Kawamura}}]{LEROY11}
{Leroy}, A.~K., {Bolatto}, A., {Gordon}, K., {Sandstrom}, K., {Gratier}, P.,
  {Rosolowsky}, E., {Engelbracht}, C.~W., {Mizuno}, N., {Corbelli}, E.,
  {Fukui}, Y., \& {Kawamura}, A. 2011, ArXiv e-prints

\bibitem[{{Leroy} {et~al.}(2008){Leroy}, {Walter}, {Brinks}, {Bigiel}, {de
  Blok}, {Madore}, \& {Thornley}}]{LEROY08}
{Leroy}, A.~K., {Walter}, F., {Brinks}, E., {Bigiel}, F., {de Blok}, W.~J.~G.,
  {Madore}, B., \& {Thornley}, M.~D. 2008, \aj, 136, 2782

\bibitem[{{Martin} {et~al.}(2005){Martin}, {Fanson}, {Schiminovich},
  {Morrissey}, {Friedman}, {Barlow}, {Conrow}, {Grange}, {Jelinsky},
  {Milliard}, {Siegmund}, {Bianchi}, {Byun}, {Donas}, {Forster}, {Heckman},
  {Lee}, {Madore}, {Malina}, {Neff}, {Rich}, {Small}, {Surber}, {Szalay},
  {Welsh}, \& {Wyder}}]{MARTIN05}
{Martin}, D.~C., {Fanson}, J., {Schiminovich}, D., {Morrissey}, P., {Friedman},
  P.~G., {Barlow}, T.~A., {Conrow}, T., {Grange}, R., {Jelinsky}, P.~N.,
  {Milliard}, B., {Siegmund}, O.~H.~W., {Bianchi}, L., {Byun}, Y.-I., {Donas},
  J., {Forster}, K., {Heckman}, T.~M., {Lee}, Y.-W., {Madore}, B.~F., {Malina},
  R.~F., {Neff}, S.~G., {Rich}, R.~M., {Small}, T., {Surber}, F., {Szalay},
  A.~S., {Welsh}, B., \& {Wyder}, T.~K. 2005, \apjl, 619, L1

\bibitem[{{Mathis} {et~al.}(1983){Mathis}, {Mezger}, \& {Panagia}}]{MATHIS83}
{Mathis}, J.~S., {Mezger}, P.~G., \& {Panagia}, N. 1983, \aap, 128, 212

\bibitem[{{McKee} \& {Williams}(1997)}]{MCKEE97}
{McKee}, C.~F., \& {Williams}, J.~P. 1997, \apj, 476, 144

\bibitem[{{Meurer} {et~al.}(1999){Meurer}, {Heckman}, \& {Calzetti}}]{MEURER99}
{Meurer}, G.~R., {Heckman}, T.~M., \& {Calzetti}, D. 1999, \apj, 521, 64

\bibitem[{{Meurer} {et~al.}(1995){Meurer}, {Heckman}, {Leitherer}, {Kinney},
  {Robert}, \& {Garnett}}]{MEURER95}
{Meurer}, G.~R., {Heckman}, T.~M., {Leitherer}, C., {Kinney}, A., {Robert}, C.,
  \& {Garnett}, D.~R. 1995, \aj, 110, 2665

\bibitem[{{Moustakas} \& {Kennicutt}(2006)}]{MOUSTAKAS06}
{Moustakas}, J., \& {Kennicutt}, Jr., R.~C. 2006, \apjs, 164, 81

\bibitem[{{Moustakas} {et~al.}(2010){Moustakas}, {Kennicutt}, {Tremonti},
  {Dale}, {Smith}, \& {Calzetti}}]{MOUSTAKAS10}
{Moustakas}, J., {Kennicutt}, Jr., R.~C., {Tremonti}, C.~A., {Dale}, D.~A.,
  {Smith}, J., \& {Calzetti}, D. 2010, \apjs, 190, 233

\bibitem[{{Mu{\~n}oz-Mateos} {et~al.}(2009{\natexlab{a}}){Mu{\~n}oz-Mateos},
  {Gil de Paz}, {Boissier}, {Zamorano}, {Dale}, {P{\'e}rez-Gonz{\'a}lez},
  {Gallego}, {Madore}, {Bendo}, {Thornley}, {Draine}, {Boselli}, {Buat},
  {Calzetti}, {Moustakas}, \& {Kennicutt}}]{MUNOZMATEOS09B}
{Mu{\~n}oz-Mateos}, J.~C., {Gil de Paz}, A., {Boissier}, S., {Zamorano}, J.,
  {Dale}, D.~A., {P{\'e}rez-Gonz{\'a}lez}, P.~G., {Gallego}, J., {Madore},
  B.~F., {Bendo}, G., {Thornley}, M.~D., {Draine}, B.~T., {Boselli}, A.,
  {Buat}, V., {Calzetti}, D., {Moustakas}, J., \& {Kennicutt}, R.~C.
  2009{\natexlab{a}}, \apj, 701, 1965

\bibitem[{{Mu{\~n}oz-Mateos} {et~al.}(2009{\natexlab{b}}){Mu{\~n}oz-Mateos},
  {Gil de Paz}, {Zamorano}, {Boissier}, {Dale}, {P{\'e}rez-Gonz{\'a}lez},
  {Gallego}, {Madore}, {Bendo}, {Boselli}, {Buat}, {Calzetti}, {Moustakas}, \&
  {Kennicutt}}]{MUNOZMATEOS09A}
{Mu{\~n}oz-Mateos}, J.~C., {Gil de Paz}, A., {Zamorano}, J., {Boissier}, S.,
  {Dale}, D.~A., {P{\'e}rez-Gonz{\'a}lez}, P.~G., {Gallego}, J., {Madore},
  B.~F., {Bendo}, G., {Boselli}, A., {Buat}, V., {Calzetti}, D., {Moustakas},
  J., \& {Kennicutt}, Jr., R.~C. 2009{\natexlab{b}}, \apj, 703, 1569

\bibitem[{{Murphy} {et~al.}(2011){Murphy}, {Condon}, {Schinnerer}, {Kennicutt},
  {Calzetti}, {Armus}, {Helou}, {Turner}, {Aniano}, {Beir{\~a}o}, {Bolatto},
  {Brandl}, {Croxall}, {Dale}, {Donovan Meyer}, {Draine}, {Engelbracht},
  {Hunt}, {Hao}, {Koda}, {Roussel}, {Skibba}, \& {Smith}}]{MURPHY11}
{Murphy}, E.~J., {Condon}, J.~J., {Schinnerer}, E., {Kennicutt}, Jr., R.~C.,
  {Calzetti}, D., {Armus}, L., {Helou}, G., {Turner}, J.~L., {Aniano}, G.,
  {Beir{\~a}o}, P., {Bolatto}, A.~D., {Brandl}, B.~R., {Croxall}, K.~V.,
  {Dale}, D.~A., {Donovan Meyer}, J.~L., {Draine}, B.~T., {Engelbracht}, C.,
  {Hunt}, L.~K., {Hao}, C.~., {Koda}, J., {Roussel}, H., {Skibba}, R., \&
  {Smith}, J.~.~T. 2011, ArXiv e-prints

\bibitem[{{Niklas} {et~al.}(1995){Niklas}, {Klein}, {Braine}, \&
  {Wielebinski}}]{NIKLAS95}
{Niklas}, S., {Klein}, U., {Braine}, J., \& {Wielebinski}, R. 1995, \aaps, 114,
  21

\bibitem[{{Niklas} {et~al.}(1997){Niklas}, {Klein}, \&
  {Wielebinski}}]{NIKLAS97}
{Niklas}, S., {Klein}, U., \& {Wielebinski}, R. 1997, \aap, 322, 19

\bibitem[{{Onodera} {et~al.}(2010){Onodera}, {Kuno}, {Tosaki}, {Kohno},
  {Nakanishi}, {Sawada}, {Muraoka}, {Komugi}, {Miura}, {Kaneko}, {Hirota}, \&
  {Kawabe}}]{ONODERA10}
{Onodera}, S., {Kuno}, N., {Tosaki}, T., {Kohno}, K., {Nakanishi}, K.,
  {Sawada}, T., {Muraoka}, K., {Komugi}, S., {Miura}, R., {Kaneko}, H.,
  {Hirota}, A., \& {Kawabe}, R. 2010, \apjl, 722, L127

\bibitem[{{Osterbrock} \& {Ferland}(2006)}]{OSTERBROCK06}
{Osterbrock}, D.~E., \& {Ferland}, G.~J. 2006, {Astrophysics of gaseous nebulae
  and active galactic nuclei}, ed. {Osterbrock, D.~E.~\& Ferland, G.~J.}

\bibitem[{{P{\'e}rez-Gonz{\'a}lez} {et~al.}(2006){P{\'e}rez-Gonz{\'a}lez},
  {Kennicutt}, {Gordon}, {Misselt}, {Gil de Paz}, {Engelbracht}, {Rieke},
  {Bendo}, {Bianchi}, {Boissier}, {Calzetti}, {Dale}, {Draine}, {Jarrett},
  {Hollenbach}, \& {Prescott}}]{PEREZGONZALEZ06}
{P{\'e}rez-Gonz{\'a}lez}, P.~G., {Kennicutt}, Jr., R.~C., {Gordon}, K.~D.,
  {Misselt}, K.~A., {Gil de Paz}, A., {Engelbracht}, C.~W., {Rieke}, G.~H.,
  {Bendo}, G.~J., {Bianchi}, L., {Boissier}, S., {Calzetti}, D., {Dale}, D.~A.,
  {Draine}, B.~T., {Jarrett}, T.~H., {Hollenbach}, D., \& {Prescott}, M.~K.~M.
  2006, \apj, 648, 987

\bibitem[{{Prescott} {et~al.}(2007){Prescott}, {Kennicutt}, {Bendo},
  {Buckalew}, {Calzetti}, {Engelbracht}, {Gordon}, {Hollenbach}, {Lee},
  {Moustakas}, {Dale}, {Helou}, {Jarrett}, {Murphy}, {Smith}, {Akiyama}, \&
  {Sosey}}]{PRESCOTT07}
{Prescott}, M.~K.~M., {Kennicutt}, Jr., R.~C., {Bendo}, G.~J., {Buckalew},
  B.~A., {Calzetti}, D., {Engelbracht}, C.~W., {Gordon}, K.~D., {Hollenbach},
  D.~J., {Lee}, J.~C., {Moustakas}, J., {Dale}, D.~A., {Helou}, G., {Jarrett},
  T.~H., {Murphy}, E.~J., {Smith}, J.-D.~T., {Akiyama}, S., \& {Sosey}, M.~L.
  2007, \apj, 668, 182

\bibitem[{{Prugniel} \& {Heraudeau}(1998)}]{PRUGNIEL98}
{Prugniel}, P., \& {Heraudeau}, P. 1998, \aaps, 128, 299

\bibitem[{{Rahman} {et~al.}(2010){Rahman}, {Bolatto}, {Wong}, {Leroy},
  {Walter}, {Rosolowsky}, {West}, {Bigiel}, {Ott}, {Xue}, {Herrera-Camus},
  {Jameson}, {Blitz}, \& {Vogel}}]{RAHMAN10}
{Rahman}, N., {Bolatto}, A.~D., {Wong}, T., {Leroy}, A.~K., {Walter}, F.,
  {Rosolowsky}, E., {West}, A.~A., {Bigiel}, F., {Ott}, J., {Xue}, R.,
  {Herrera-Camus}, R., {Jameson}, K., {Blitz}, L., \& {Vogel}, S.~N. 2010,
  ArXiv e-prints

\bibitem[{{Reipurth}(2008{\natexlab{a}})}]{REIPURTH08A}
{Reipurth}, B. 2008{\natexlab{a}}, {Handbook of Star Forming Regions, Volume I:
  The Northern Sky}, ed. {Reipurth, B.}

\bibitem[{{Reipurth}(2008{\natexlab{b}})}]{REIPURTH08B}
---. 2008{\natexlab{b}}, {Handbook of Star Forming Regions, Volume II: The
  Southern Sky}, ed. {Reipurth, B.}

\bibitem[{{Rosolowsky} {et~al.}(2008){Rosolowsky}, {Pineda}, {Kauffmann}, \&
  {Goodman}}]{ROSOLOWSKY08}
{Rosolowsky}, E.~W., {Pineda}, J.~E., {Kauffmann}, J., \& {Goodman}, A.~A.
  2008, \apj, 679, 1338

\bibitem[{{Roussel} {et~al.}(2005){Roussel}, {Gil de Paz}, {Seibert}, {Helou},
  {Madore}, \& {Martin}}]{ROUSSEL05}
{Roussel}, H., {Gil de Paz}, A., {Seibert}, M., {Helou}, G., {Madore}, B.~F.,
  \& {Martin}, C. 2005, \apj, 632, 227

\bibitem[{{Salim} {et~al.}(2007){Salim}, {Rich}, {Charlot}, {Brinchmann},
  {Johnson}, {Schiminovich}, {Seibert}, {Mallery}, {Heckman}, {Forster},
  {Friedman}, {Martin}, {Morrissey}, {Neff}, {Small}, {Wyder}, {Bianchi},
  {Donas}, {Lee}, {Madore}, {Milliard}, {Szalay}, {Welsh}, \& {Yi}}]{SALIM07}
{Salim}, S., {Rich}, R.~M., {Charlot}, S., {Brinchmann}, J., {Johnson}, B.~D.,
  {Schiminovich}, D., {Seibert}, M., {Mallery}, R., {Heckman}, T.~M.,
  {Forster}, K., {Friedman}, P.~G., {Martin}, D.~C., {Morrissey}, P., {Neff},
  S.~G., {Small}, T., {Wyder}, T.~K., {Bianchi}, L., {Donas}, J., {Lee}, Y.,
  {Madore}, B.~F., {Milliard}, B., {Szalay}, A.~S., {Welsh}, B.~Y., \& {Yi},
  S.~K. 2007, \apjs, 173, 267

\bibitem[{{Schlegel} {et~al.}(1998){Schlegel}, {Finkbeiner}, \&
  {Davis}}]{SCHLEGEL98}
{Schlegel}, D.~J., {Finkbeiner}, D.~P., \& {Davis}, M. 1998, \apj, 500, 525

\bibitem[{{Schnee} {et~al.}(2008){Schnee}, {Li}, {Goodman}, \&
  {Sargent}}]{SCHNEE08}
{Schnee}, S., {Li}, J., {Goodman}, A.~A., \& {Sargent}, A.~I. 2008, \apj, 684,
  1228

\bibitem[{{Schruba} {et~al.}(2010){Schruba}, {Leroy}, {Walter}, {Sandstrom}, \&
  {Rosolowsky}}]{SCHRUBA10}
{Schruba}, A., {Leroy}, A.~K., {Walter}, F., {Sandstrom}, K., \& {Rosolowsky},
  E. 2010, \apj, 722, 1699

\bibitem[{{Schruba} {et~al.}(2011){Schruba}, {Leroy}, {Walter}, {Sandstrom}, \&
  {Rosolowsky}}]{SCHRUBA11}
---. 2011, \apj, 722, 1699

\bibitem[{{Schuster} {et~al.}(2004){Schuster}, {Boucher}, {Brunswig}, {Carter},
  {Chenu}, {Foullieux}, {Greve}, {John}, {Lazareff}, {Navarro}, {Perrigouard},
  {Pollet}, {Sievers}, {Thum}, \& {Wiesemeyer}}]{SCHUSTER04}
{Schuster}, K.-F., {Boucher}, C., {Brunswig}, W., {Carter}, M., {Chenu}, J.-Y.,
  {Foullieux}, B., {Greve}, A., {John}, D., {Lazareff}, B., {Navarro}, S.,
  {Perrigouard}, A., {Pollet}, J.-L., {Sievers}, A., {Thum}, C., \&
  {Wiesemeyer}, H. 2004, \aap, 423, 1171

\bibitem[{{Storey} \& {Hummer}(1995)}]{STOREY95}
{Storey}, P.~J., \& {Hummer}, D.~G. 1995, \mnras, 272, 41

\bibitem[{{Strong} \& {Mattox}(1996)}]{STRONG96}
{Strong}, A.~W., \& {Mattox}, J.~R. 1996, \aap, 308, L21

\bibitem[{{Thilker} {et~al.}(2007){Thilker}, {Boissier}, {Bianchi}, {Calzetti},
  {Boselli}, {Dale}, {Seibert}, {Braun}, {Burgarella}, {Gil de Paz}, {Helou},
  {Walter}, {Kennicutt}, {Madore}, {Martin}, {Barlow}, {Forster}, {Friedman},
  {Morrissey}, {Neff}, {Schiminovich}, {Small}, {Wyder}, {Donas}, {Heckman},
  {Lee}, {Milliard}, {Rich}, {Szalay}, {Welsh}, \& {Yi}}]{THILKER07}
{Thilker}, D.~A., {Boissier}, S., {Bianchi}, L., {Calzetti}, D., {Boselli}, A.,
  {Dale}, D.~A., {Seibert}, M., {Braun}, R., {Burgarella}, D., {Gil de Paz},
  A., {Helou}, G., {Walter}, F., {Kennicutt}, Jr., R.~C., {Madore}, B.~F.,
  {Martin}, D.~C., {Barlow}, T.~A., {Forster}, K., {Friedman}, P.~G.,
  {Morrissey}, P., {Neff}, S.~G., {Schiminovich}, D., {Small}, T., {Wyder},
  T.~K., {Donas}, J., {Heckman}, T.~M., {Lee}, Y., {Milliard}, B., {Rich},
  R.~M., {Szalay}, A.~S., {Welsh}, B.~Y., \& {Yi}, S.~K. 2007, \apjs, 173, 572

\bibitem[{{Thilker} {et~al.}(2000){Thilker}, {Braun}, \&
  {Walterbos}}]{THILKER00}
{Thilker}, D.~A., {Braun}, R., \& {Walterbos}, R.~A.~M. 2000, \aj, 120, 3070

\bibitem[{{Vacca} {et~al.}(1996){Vacca}, {Garmany}, \& {Shull}}]{VACCA96}
{Vacca}, W.~D., {Garmany}, C.~D., \& {Shull}, J.~M. 1996, \apj, 460, 914

\bibitem[{{Walter} {et~al.}(2008){Walter}, {Brinks}, {de Blok}, {Bigiel},
  {Kennicutt}, {Thornley}, \& {Leroy}}]{WALTER08}
{Walter}, F., {Brinks}, E., {de Blok}, W.~J.~G., {Bigiel}, F., {Kennicutt},
  R.~C., {Thornley}, M.~D., \& {Leroy}, A. 2008, \aj, 136, 2563

\bibitem[{{Wilson} {et~al.}(2009){Wilson}, {Warren}, {Israel}, {Serjeant},
  {Bendo}, {Brinks}, {Clements}, {Courteau}, {Irwin}, {Knapen}, {Leech},
  {Matthews}, {M{\"u}hle}, {Mortier}, {Petitpas}, {Sinukoff}, {Spekkens},
  {Tan}, {Tilanus}, {Usero}, {van der Werf}, {Wiegert}, \& {Zhu}}]{WILSON09}
{Wilson}, C.~D., {Warren}, B.~E., {Israel}, F.~P., {Serjeant}, S., {Bendo}, G.,
  {Brinks}, E., {Clements}, D., {Courteau}, S., {Irwin}, J., {Knapen}, J.~H.,
  {Leech}, J., {Matthews}, H.~E., {M{\"u}hle}, S., {Mortier}, A.~M.~J.,
  {Petitpas}, G., {Sinukoff}, E., {Spekkens}, K., {Tan}, B.~K., {Tilanus},
  R.~P.~J., {Usero}, A., {van der Werf}, P., {Wiegert}, T., \& {Zhu}, M. 2009,
  \apj, 693, 1736

\bibitem[{{Wong} \& {Blitz}(2002)}]{WONG02}
{Wong}, T., \& {Blitz}, L. 2002, \apj, 569, 157

\bibitem[{{Wyder} {et~al.}(2007){Wyder}, {Martin}, {Schiminovich}, {Seibert},
  {Budav{\'a}ri}, {Treyer}, {Barlow}, {Forster}, {Friedman}, {Morrissey},
  {Neff}, {Small}, {Bianchi}, {Donas}, {Heckman}, {Lee}, {Madore}, {Milliard},
  {Rich}, {Szalay}, {Welsh}, \& {Yi}}]{WYDER07}
{Wyder}, T.~K., {Martin}, D.~C., {Schiminovich}, D., {Seibert}, M.,
  {Budav{\'a}ri}, T., {Treyer}, M.~A., {Barlow}, T.~A., {Forster}, K.,
  {Friedman}, P.~G., {Morrissey}, P., {Neff}, S.~G., {Small}, T., {Bianchi},
  L., {Donas}, J., {Heckman}, T.~M., {Lee}, Y., {Madore}, B.~F., {Milliard},
  B., {Rich}, R.~M., {Szalay}, A.~S., {Welsh}, B.~Y., \& {Yi}, S.~K. 2007,
  \apjs, 173, 293

\bibitem[{{Young} {et~al.}(1995){Young}, {Xie}, {Tacconi}, {Knezek}, {Viscuso},
  {Tacconi-Garman}, {Scoville}, {Schneider}, {Schloerb}, {Lord}, {Lesser},
  {Kenney}, {Huang}, {Devereux}, {Claussen}, {Case}, {Carpenter}, {Berry}, \&
  {Allen}}]{YOUNG95}
{Young}, J.~S., {Xie}, S., {Tacconi}, L., {Knezek}, P., {Viscuso}, P.,
  {Tacconi-Garman}, L., {Scoville}, N., {Schneider}, S., {Schloerb}, F.~P.,
  {Lord}, S., {Lesser}, A., {Kenney}, J., {Huang}, Y.-L., {Devereux}, N.,
  {Claussen}, M., {Case}, J., {Carpenter}, J., {Berry}, M., \& {Allen}, L.
  1995, \apjs, 98, 219

\bibitem[{{Zibetti} {et~al.}(2009){Zibetti}, {Charlot}, \& {Rix}}]{ZIBETTI09}
{Zibetti}, S., {Charlot}, S., \& {Rix}, H. 2009, \mnras, 400, 1181

\end{thebibliography}

\begin{appendix}

\section{Estimating the Radiation Field Associated With the Cirrus}
\label{sec:ufield}

\begin{figure*}
\epsscale{0.75}
\plotone{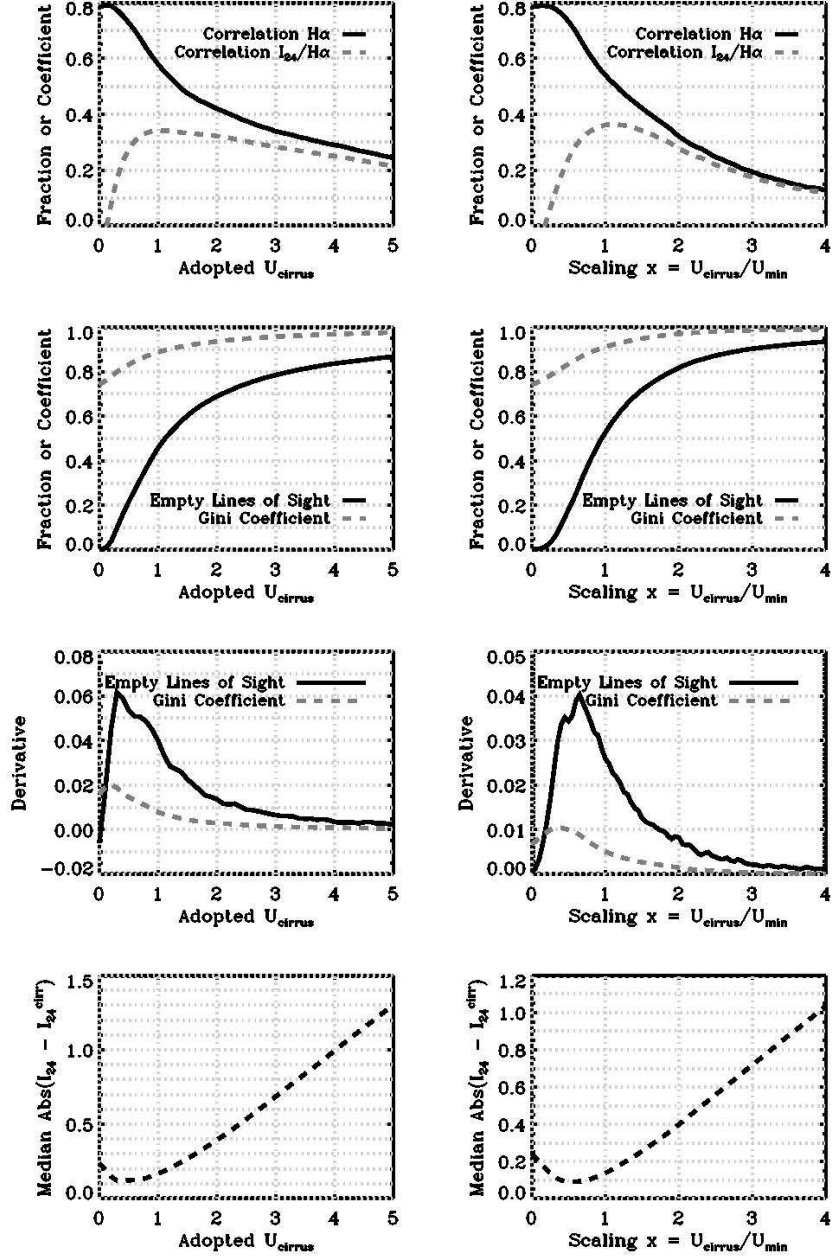}
\caption{The structure of 24$\mu$m emission after cirrus subtraction as a function of ({\em left column}) adopted cirrus
    radiation field, $U_{\rm cirrus}$ and ({\em right column}) adopted scaling factor $x$ when $U_{\rm cirrus} = x~U_{\rm min}$. The top row shows the correlation of 24$\mu$m emission with H$\alpha$ emission or $A_{\rm H\alpha}$ inferred from the ratio $I_{24}/I_{\rm H\alpha}$. The second row shows the
  fraction of lines of sight where the cirrus removes all emission,
  and the Gini coefficient, a non-parametric measure of the
  unevenness of the distribution. The third row shows the derivatives of the fraction
  of pixels set to zero by the cirrus subtraction and the Gini
  coefficient as a function of the radiation field. The bottom row shows the median absolute difference
  between the cirrus map calculated from the indicated radiation field
  and the observed ``low-lying'' emission, with low-lying defined as
  all data less than the median.}
\label{fig:cirrus_struct}
\end{figure*}

Our fits to the dust properties yield the radiation field powering dust emission in each ring, including the best-fit single radiation field illuminating the bulk of the dust mass, $U_{\rm min}$. The fits do not distinguish how much of this $U_{\rm min}$ is associated with recently formed stars and how much results from an older stellar population. Because of this uncertainty, we cannot simply identify all 24$\mu$m emission associated with $U_{\rm min}$ as cirrus. In this appendix, we attempt an independent estimate of the radiation field, $U_{\rm cirrus}$, not directly associated with star formation. In the regime that we consider the intensity of the cirrus emission at 24$\mu$m, $I_{24}$, depends linearly on the incident radiation field (Section \ref{sec:cirrus} and Figure \ref{fig:emis24}), so that $I_{24}^{\rm cirrus} \propto U_{\rm cirrus}$.

Estimating $U_{\rm cirrus}$ requires that we define a ``successful" cirrus subtraction, or equivalently a goodness-of-fit metric for a cirrus estimate. After considerable experimentation we settled on the following: {\em a successful cirrus estimate removes most emission from faint regions but does not overcorrect the map or delete emission actually associated with star formation.} This means that we look for $U_{\rm cirrus}$ that significantly affects low-lying emission but does not systematically oversubtract. In practice we use five measurements of structure in the cirrus-subtracted 24$\mu$m map to assess the quality of our $U_{\rm cirrus}$ determination: 

\begin{enumerate}
\item The fraction of lines of sight for which the 24$\mu$m is set to zero, that is, completely removed, by the cirrus subtraction (this will include lines of sight where the 24$\mu$m has been oversubtracted). 
\item The median absolute difference between the cirrus estimate and the initial 24$\mu$m map for faint emission. We define ``faint" emission as emission below the
  median $I_{\rm 24}$ in the map. This measures the degree to which the cirrus matches the 24$\mu$m map over the faint lines of sight. 
\item The Gini coefficient \citep[][]{GINI1912}, a non-parametric measure of the unevenness of the distribution of emission \citep[see discussion in][]{MUNOZMATEOS09A}. This measures the degree to which the IR emission is concentrated into a few bright regions after the cirrus subtraction.
\item The rank correlation coefficient between H$\alpha$ emission and cirrus-subtracted 24$\mu$m emission. In principle, this should measure the degree to which the morphology of the cirrus-subtracted 24$\mu$m map matches that of the H$\alpha$, a relatively unambiguous signature of high-mass star formation (but see complications below).
\item The rank correlation coefficient relating H$\alpha$ emission and the ratio of 24$\mu$m-to-H$\alpha$ emission, or equivalently $A_{\rm H\alpha}$. This measures the degree to which the cirrus-subtracted map obeys the expectation that more actively star-forming regions exhibit more extinction (but see below).
\end{enumerate}

{\bf Calculations:} For each galaxy, we carry out a cirrus subtraction for a range of test radiation fields and measure each of these five statistics for each test radiation field. We consider: (1) a single fixed radiation field, $U_{\rm cirrus}$, for the whole sample; (2) a single fixed radiation field, $U_{\rm cirrus}$, per galaxy; and (3) the case where the cirrus is powered by a scaled version of the best-fit $U_{\rm min}$ from the dust SED fits (i.e., $U_{\rm cirrus} = x U_{\rm min}$; recall $U_{\rm min}$ is derived from a fit for each radial ring). We vary these test fields from $U_{\rm cirrus} = 0$--$5$ and try scaling $U_{\rm cirrus} = x U_{\rm min}$ with $x$ from 0--4. We consider $x > 1$, which may seem counterintuitive, to allow the possibility that the cirrus is driven by a radiation field locally more intense than that indicated by the $U_{\rm min}$ parameter fit at coarser ($40\arcsec$) resolution or because of contribution from fields $> U_{\rm min}$. In the end we do find $x < 1$ to yield the best results.

Figure \ref{fig:cirrus_struct} shows these measurements for all of our data, plotting each as a function of the radiation field $U_{\rm cirrus}$ assumed to power the cirrus (left column) or the factor $x$ by which we scale $U_{\rm min}$, $U_{\rm cirrus} = x U_{\rm min}$  (right column). A single $U_{\rm cirrus}$ for the whole sample is certainly too simplistic, but the plots still give a good overview of the calculations. We include all data with  $\left< \Sigma_{\rm SFR} \right> > 10^{-4}$~M$_\odot$~yr$^{-1}$~kpc$^{-2}$. We will find $\Sigma_{\rm SFR}$ to be unreliable at this low level but including these low-lying data in the cirrus calculations is important. A major goal of the cirrus subtraction will be to correct for faint but pervasive emission that extends down to this level.

The left column of Figure \ref{fig:cirrus_struct} shows that over the range $U_{\rm cirrus} \sim 0.3$--$1.5$, the cirrus subtraction varies between creating no empty lines of sight at all to removing all emission associated with more than half of the lines of sight. Over this same range, the correlation of extinction with H$\alpha$ surface brightness improves. The Gini coefficient also increases, implying a stronger concentration of IR emission into a few bright regions. The deviation between the calculated cirrus and the faint parts of the sample is minimized for $U_{\rm cirrus} \sim 0.4$ (bottom left panel). The right column shows similar behavior scaling the fit radiation field ($U_{\rm min}$) by $\sim 0.3$--$1$, with the match between the calculated cirrus and faint emission best for $\sim 0.6~U_{\rm min}$.

{\bf Identification of $U_{\rm cirrus}$:} Figure \ref{fig:cirrus_struct} suggests a rough magnitude of the cirrus. Low values, $U_{cirrus} \lesssim 0.3$ will have little or no effect on the 24$\mu$m map and not produce a meaningful correction. High values $U_{cirrus} \gtrsim 1$ will suppress a very large part of the map and oversubtract emission from low intensity regions. We want to pick a cirrus correction in the intermediate regime, near the regions of steep positive slope in the fraction of blank lines of sight or Gini coefficient where the cirrus subtraction has the maximum effect without oversubtraction. The derivative of the fraction of lines of blank lines of sight and the Gini coefficient appear in the middle row of Figure \ref{fig:cirrus_struct} and indeed peak in the desired range. Even more basically, we argue that the most appropriate $U_{\rm cirrus}$ will be the one that generates the minimum scatter between the calculated cirrus and the real map for low-level emission. The bottom row in Figure \ref{fig:cirrus_struct} plots the median absolute difference between the cirrus and low-lying emission, defined as $I_{\rm 24}$ below the median.

Thus, we have several candidates for $U_{\rm cirrus}$: 

\begin{enumerate}
\item $U_{\rm cirrus}$ that yields the maximum derivative of the fraction of blank lines of sight vs. $U$.
\item $U_{\rm cirrus}$ that yields the maximum of derivative of the Gini coefficient vs. $U$.
\item $U_{\rm cirrus}$ that yields the minimum absolute difference between the calculated cirrus and the observed 24$\mu$m intensity for faint lines of sight.
\item $U$ at which the fraction of blank pixels passes some fiducial threshold; $0.2$ works well for our data based on the tests in the next section.
\item $U$ that yields the maximum correlation of H$\alpha$ and 24$\mu$m.
\item $U$ that yields the maximum correlation of H$\alpha$ with $A_{\rm H\alpha}$, traced by $I_{\rm 24} / I_{\rm H\alpha}$ after cirrus subtraction.
\end{enumerate}

{\bf Testing of Our $U$-Fitting Metrics:} We evaluate these metrics by testing their ability to recover a known radiation field. We generate a series of simulated 24$\mu$m data sets with cirrus powered by a range of $U_{\rm cirrus}$ and a similar set powered by scaled versions of $U_{\rm min}$. To do so we:

\begin{enumerate}
\item Scale the H$\alpha$ data to correspond to the 24$\mu$m emission expected for $A_{\rm H\alpha} = 1$~mag. This represents a plausible core of IR associated with star formation.
\item Rescale the calculated cirrus for each line of sight to have a known $U_{\rm cirrus}$ between 0 and 5 and then add it to this ``star forming" 24$\mu$m emission to produce a realistic mix of star formation and cirrus. We run a parallel set of calculations with $U_{\rm cirrus} = x U_{\rm min}$ and $x$ ranging from 0 to 4. 
\item Add lognormal (multiplicative) scatter with magnitude $0.15$~dex  to the cirrus map to reflect uncertainty in the cirrus estimate.
\end{enumerate}

\begin{figure}
\epsscale{1.0}
\plottwo{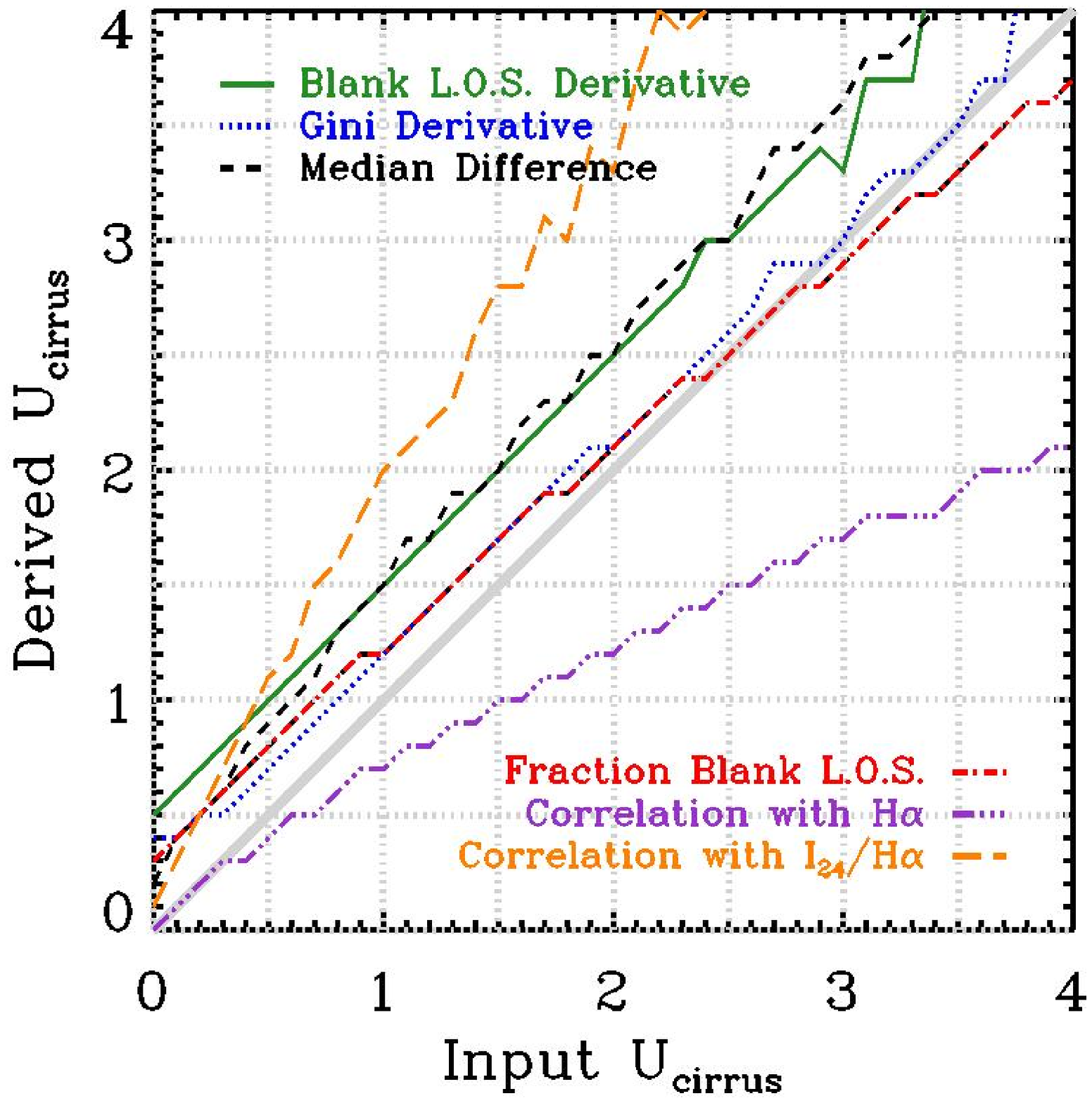}{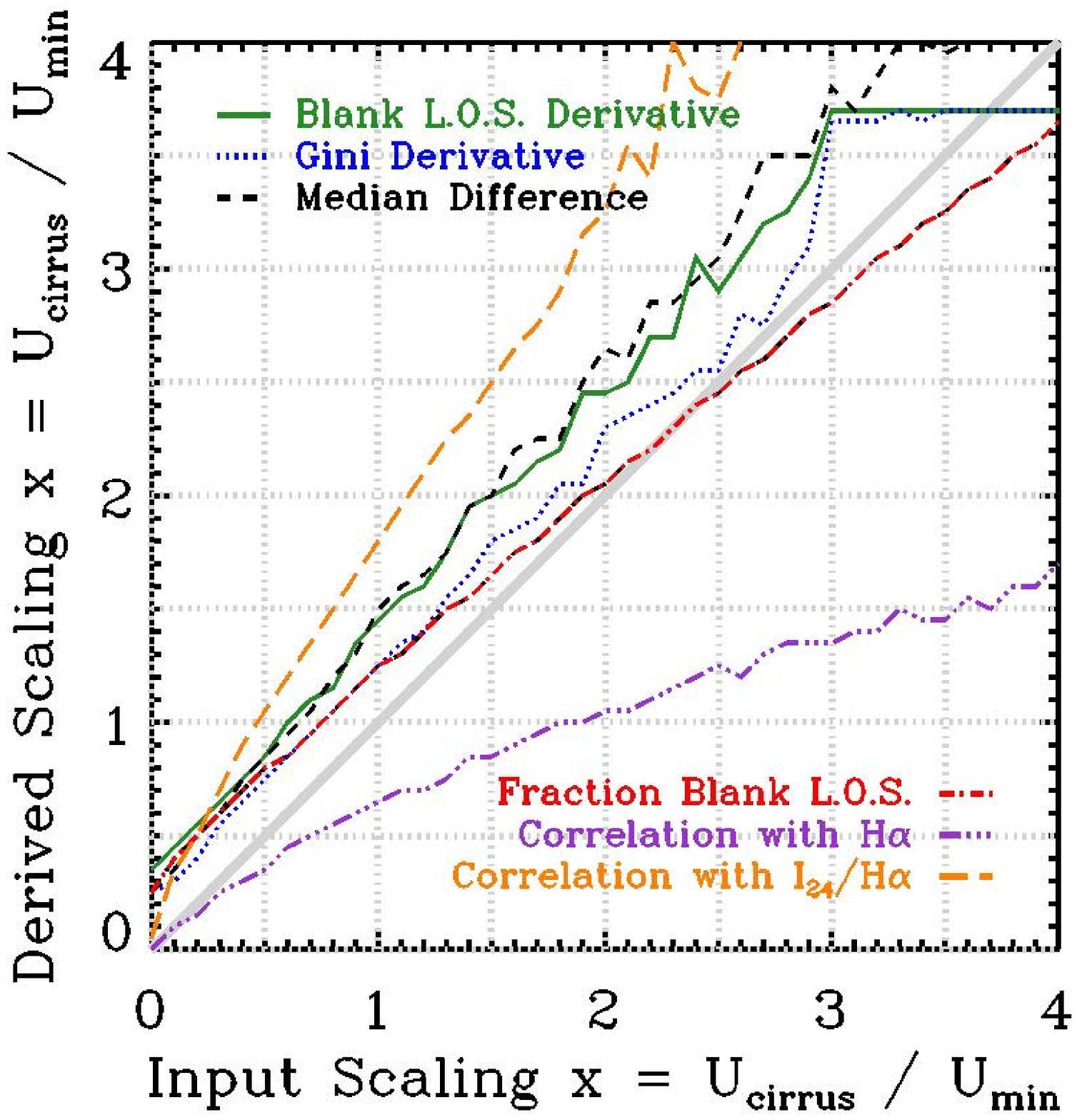}
\caption{{\em Test of $U$-fitting metrics.} Ability of our $U_{\rm cirrus}$ goodness-of-fit metrics to recover the radiation field powering the cirrus in model case where the true $U_{\rm cirrus}$ is known. The lines show $U$ derived from maximizing the derivative of the fraction of blanked lines of sight; maximizing the derivative of the Gini coefficient; minimizing the scatter between the cirrus and low-lying emission; matching a fixed fraction (here 20\%) of blanked lines of sight; maximizing the rank correlation between H$\alpha$ and $I_{24}$; and maximizing the rank correlation between H$\alpha$ and $I_{24} / I_{\rm H\alpha}$ (a quantity closely related to the H$\alpha$ extinction, $A_{\rm H\alpha}$). The {\em left} panel shows the results of Monte Carlo calculations aimed at recovering a fixed $U_{\rm cirrus}$, the {\em right} panel explores the ability of our metrics to recover the scaling factor $x$ when $U_{\rm cirrus} = x U_{\rm min}$, with $U_{\rm min}$ fit from the IR SED. A solid gray line in each panel shows equality for reference.
}
\label{fig:metric_test}
\end{figure}

We apply each metric to each simulated 24$\mu$m map and to estimate either $U_{\rm cirrus}$ or $x$. Figure \ref{fig:metric_test} plots the ability of the methods to recover the input cirrus radiation field. The left panel plots results for a fixed $U_{\rm cirrus}$, the right panel for a scaled version of $U_{\rm min}$, so that $U_{\rm cirrus} = x U_{\rm min}$.

The tests show that the derivative-based and minimum scatter metrics recover the input $U_{\rm cirrus}$ well with a small bias to recover higher values than the input. This small high bias reflects the fact that low lying H$\alpha$ pervades much of the map so that forcing many lines of sight to zero actually moves some ``star formation'' emission into ``cirrus.'' The effect is mild. Picking the $U_{\rm cirrus}$ that recovers 20\% completely blank lines of sight also works well, but this particular value of 20\% had to be tuned for our data.

The rank-correlation based metrics, H$\alpha$ with either $I_{24}$ or $I_{24}/I_{\rm H\alpha}$, show strong biases, especially toward large $U_{\rm cirrus}$ or $x$. The correlation of $I_{24}$ with $I_{\rm H\alpha}$ is biased toward to recover lower $U_{\rm cirrus}$ than the input because the cirrus correction is noisy. As we apply progressively more cirrus correction to $I_{\rm 24}$, this scatter artificially suppresses the correlation. The extinction correlation is biased to recover higher $U_{\rm cirrus}$ than input because the two quantities, $I_{\rm H\alpha}$ and $I_{\rm 24}/I_{\rm H\alpha}$, are intrinsically anti-correlated for random $I_{\rm 24}$. 

We conclude from this test that the derivative of the blank-pixel fraction, the derivative of the Gini coefficient, a tuned blank-pixel fraction threshold, and the minimum absolute deviation between calculated cirrus and observed intensity all represent useful tools to identify the radiation field driving the cirrus (though they have a mild bias to recover too much cirrus). We expect that the rank correlation coefficients, which fail in our data, will emerge as very useful for higher resolution studies but the resolution of our current data set is too coarse to take full advantage of them.

{\bf Galaxy-by-Galaxy Estimates:} So far we have examined the ensemble of our data, but the radiation field certainly varies among and within galaxies. Our metrics require a population of points to identify a best-fit $U_{\rm cirrus}$. Therefore they cannot be applied point-by-point, but our individual galaxies do have enough data to derive $U_{\rm cirrus}$.

\begin{figure}
\epsscale{1.0}
\plottwo{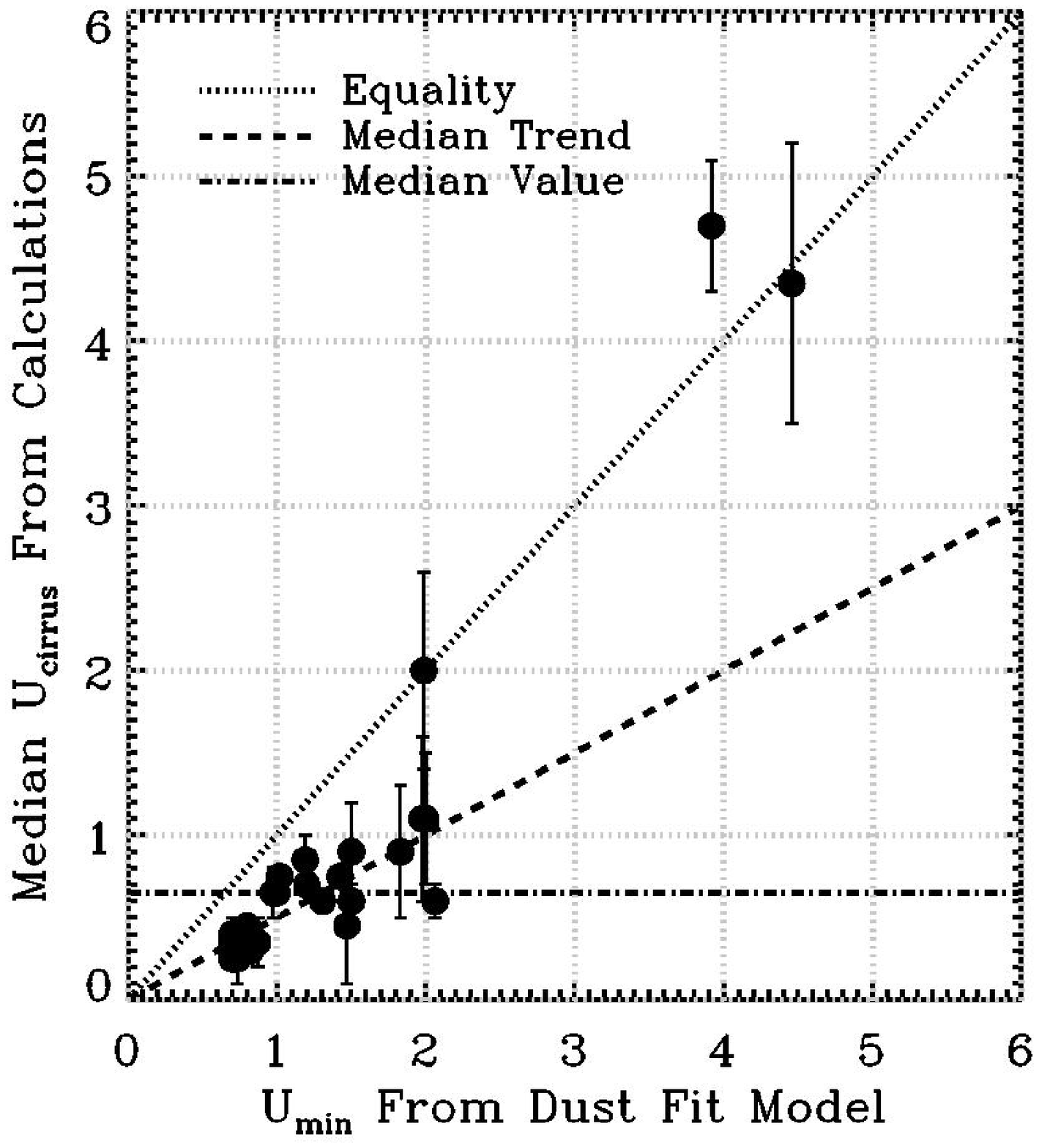}{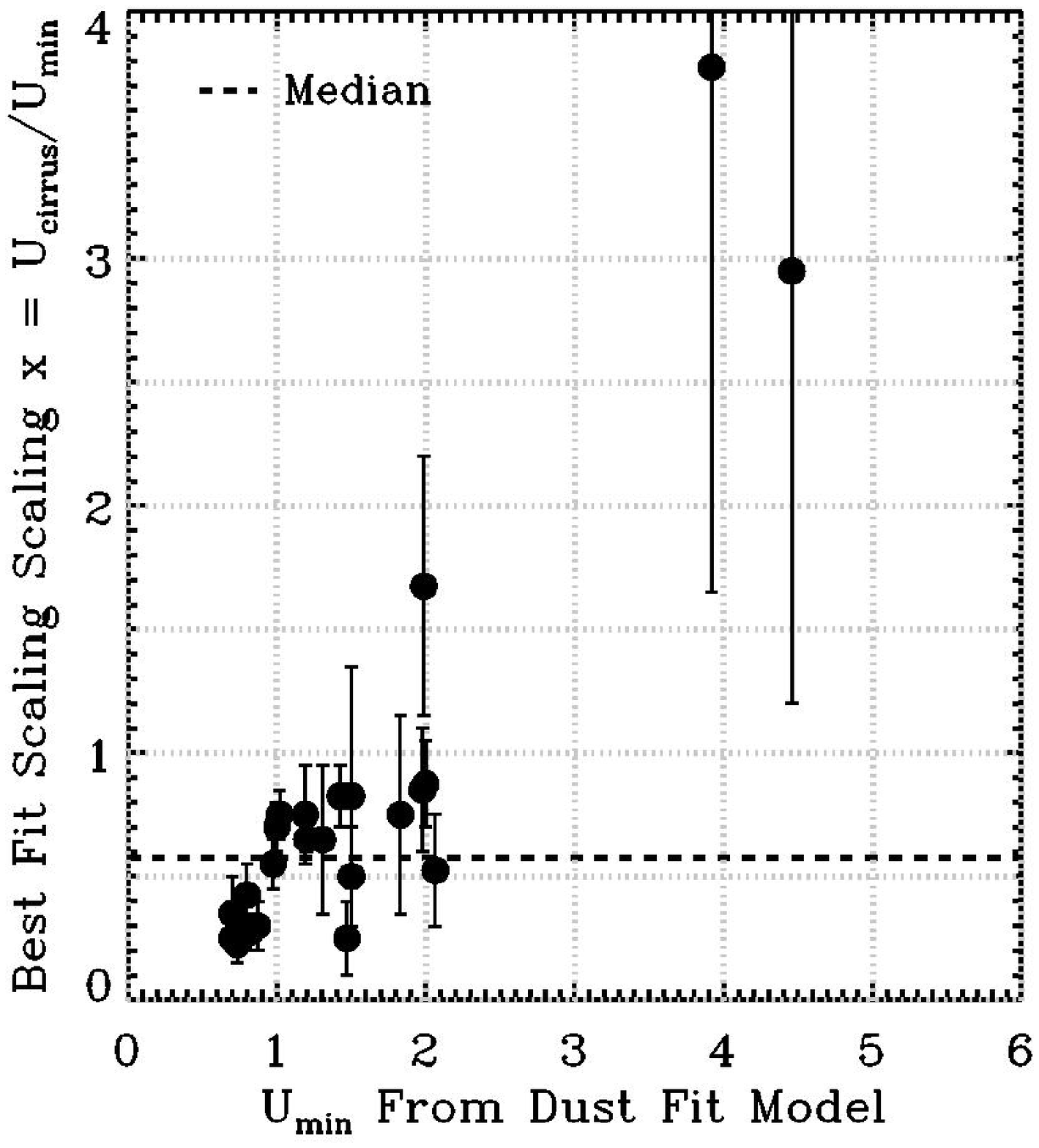}
\caption{({\em left}) Calculated $U_{\rm cirrus}$ vs. $U_{\rm min}$ from fitting 
    dust models to the IR SED. We show the median $U_{\rm cirrus}$ ($y$-axis) estimated for individual galaxies (points) using our four reliable methods. We plot $U_{\rm cirrus}$ as a function of each galaxy's average $U_{\rm min}$, the fixed radiation yielded from fitting the \citet{DRAINE07A} models to the IR SED. $U_{\rm cirrus} $ tracks $U_{\rm min}$ well, but tends to be lower than $U_{\rm min}$ by a factor of $\approx 2$. The dashed line shows $U_{\rm cirrus} = 0.5~U_{\rm min}$, which is a good description of the data. The dash-dotted line show $U_{\rm cirrus} \sim 0.6$, the median value. The {\em right} panel shows an analogous determination of the best-fit factor $x$ considering $U_{\rm cirrus} = x~U_{\rm min}$. The dashed line shows the median $x \approx 0.6$. NGC~337 and NGC~7331 appear as outliers in both panels.}
\label{fig:ufit_vs_umin}
\end{figure}

Figure \ref{fig:ufit_vs_umin} shows the median $U_{\rm cirrus}$ (left) and the best-fit scaling of $U_{\rm min}$ (right) derived from our four useful goodness-of-fit  statistics: maximizing the derivative of the blank line of sight, maximizing the derivative of the Gini coefficient, finding $U_{\rm cirrus}$ that minimizes the difference
between 24$\mu$m and cirrus for faint lines of sight, and finding $U_{\rm cirrus}$ that blanks a fixed fraction of lines of sight. We plot both $U_{\rm cirrus}$ and the best-fit scaling $x$ in $U_{\rm cirrus} = x U_{\rm min}$ as a function of the median fit radiation field $U_{\rm min}$ for that galaxy from the \citet{DRAINE07A} models. 

The left panel of Figure \ref{fig:ufit_vs_umin} reveals a median $U_{\rm cirrus}$ of $0.6$ and a good correspondence between our $U_{\rm cirrus}$ and fit $U_{\rm min}$. The rank correlation relating the two is $\sim 0.9$, but the derived $U_{\rm cirrus}$ are smaller than the fit $U_{\rm min}$ by a factor of two on average (dashed line). The exceptions are the two bright systems NGC~337 and NGC~7331 where both the dust fitting and our metrics suggest a very high diffuse field $U \sim 4$--$5$. These targets are interesting points of follow-up but appear to be special cases, as we expect the very high field to be partially due to resolution and geometric effects. We do not follow up further here, but note these galaxies as excellent targets for {\em Herschel} study. 

The right panel of Figure \ref{fig:ufit_vs_umin} shows the best-fit scaling $x$ for the case where $U_{\rm cirrus} = x U_{\rm min}$, again as a function of the fit field $U_{\rm min}$. The median scaling is $x \approx 0.6$ with about a factor of two scatter. There is certainly some correlation between host galaxy and best scaling and again NGC~337 and NGC~7331 appear as outliers, though here with very large error bars, meaning large scatter in $x$ determined by different methods. This median scaling in the right panel and the typical ratio of $U_{\rm cirrus}$ to to $U_{\rm min}$ in the left panel agree with one another fairly well, both suggesting a cirrus radiation field about half of the fit $U_{\rm min}$.

The good correspondence between our calculated $U_{\rm cirrus}$ and the fit $U_{\rm min}$ suggests that we can use a scaled version of $U_{\rm min}$ for local
estimates of the cirrus. This is desirable, as it allows us to capture local variations in the radiation field driving the cirrus. The correspondence of $U_{\rm cirrus}$ and $U_{\rm min}$ also reinforces that our calculations have returned a meaningful result, inasmuch as they return information similar to an SED fit.

{\bf Conclusions From Radiation Field Calculations:} We define and test a series of goodness-of-fit metrics that attempt to identify the radiation field powering the cirrus at 24$\mu$m. We require that subtracting a cirrus driven by this radiation field significantly affect the faint emission in the map but not dramatically oversubtract emission.

For our data set, the cirrus correction should employ $U_{\rm cirrus} \sim 0.5 U_{\rm min}$, a median $U_{\rm cirrus} \sim 0.6$--$0.7$. There is a factor of $\sim 2$ scatter in these estimates from galaxy to galaxy with two notable outliers.  In practice this lowers the median $U$ used in our calculation from $\sim 1.4$ times
the Solar Neighborhood ISRF on average to $\sim 0.6$. This appears reasonable; as noted, the Solar Neighbornood ISRF likely includes significant contributions from nearby young stars. When we refer to ``cirrus'' throughout the rest of the paper we will adopt $U_{\rm cirrus} = 0.5 \times U_{\rm min}$. To test the impact of this conclusion, we will also occasionally refer to ``twice" or ``double" cirrus, which indicates that we have taken $U_{\rm cirrus} = 1.0 \times U_{\rm min}$, i.e., double our best estimate of the appropriate $U_{\rm cirrus}$ value.

The critical points are that the rough magnitude of the cirrus is slightly below the Solar Neighborhood ISRF and does appear to track the radiation field estimated from dust fits. The specific scaling $U_{\rm cirrus} \sim 0.5~U_{\rm min}$ almost certainly results from the interaction of resolutions for which we carry out our dust fits ($40\arcsec$) and cirrus calculations ($1$~kpc$\sim 13\arcsec$) and we do not expect it to be a general result . For a recent demonstration of the shortcomings of simple power law distributions of $U$ to yield scale-independent dust-fitting results see the work by \citet{GALLIANO11} on the Large Magellanic Cloud. The magnitude, $U_{\rm cirrus} \sim 0.6$, on the other hand, seems plausible and may represent a reasonable conservative starting point (as might $U\sim1$). In any case, our understanding of IR cirrus emission will benefit greatly from achieving good IR SED coverage at very high spatial resolution \citep[e.g.,][]{LAWTON10}. This will allow the kind of dust-SED based analysis that we apply here to be undertaken on maps with the ability to cleanly distinguish quiescent and star-forming parts of a galaxy. In this regime, we expect the rank correlation statistics that we test and reject ($I_{\rm 24}$ with H$\alpha$ morphology) and simple estimates of the total $U$ in quiescent regions to become powerful tools.

\section{$\Sigma_{\rm SFR}$ Estimates for HERACLES Disk Galaxies}

\begin{figure*}
\plotone{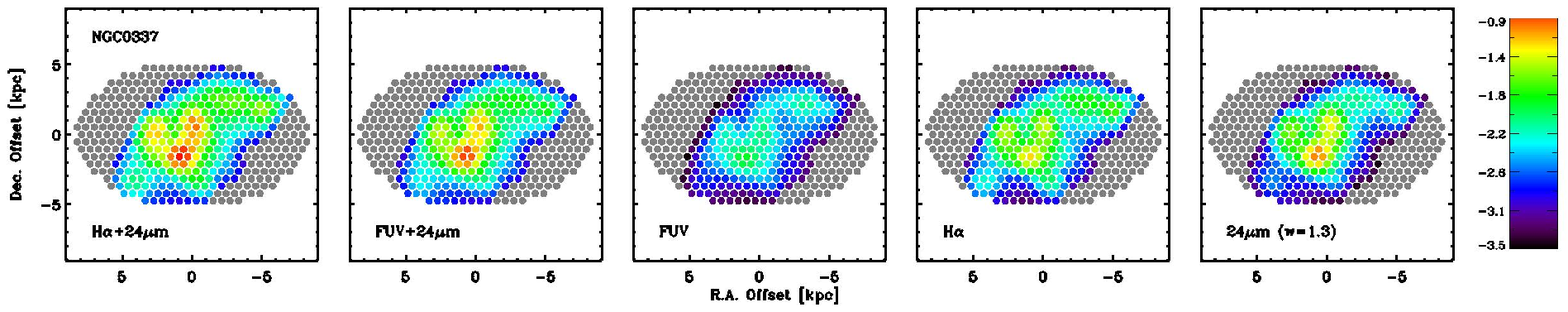}
\plotone{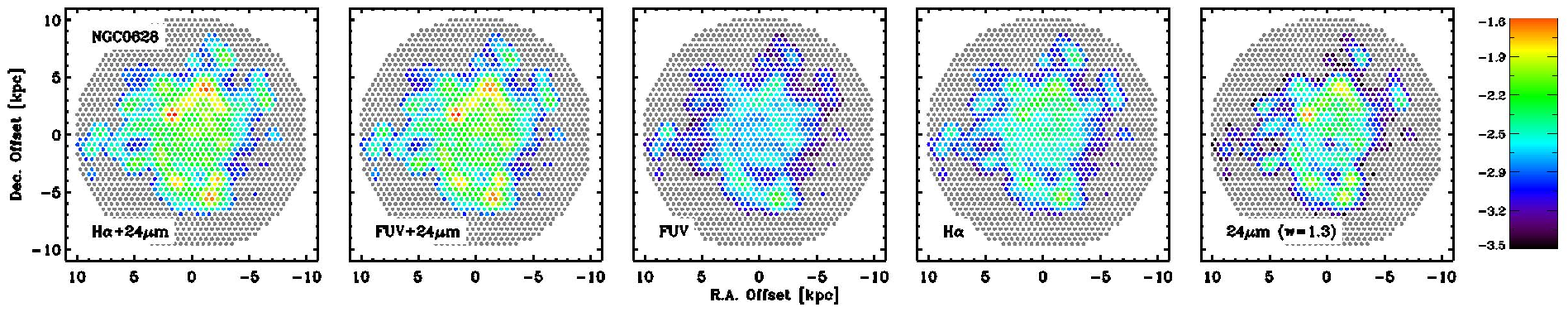}
\plotone{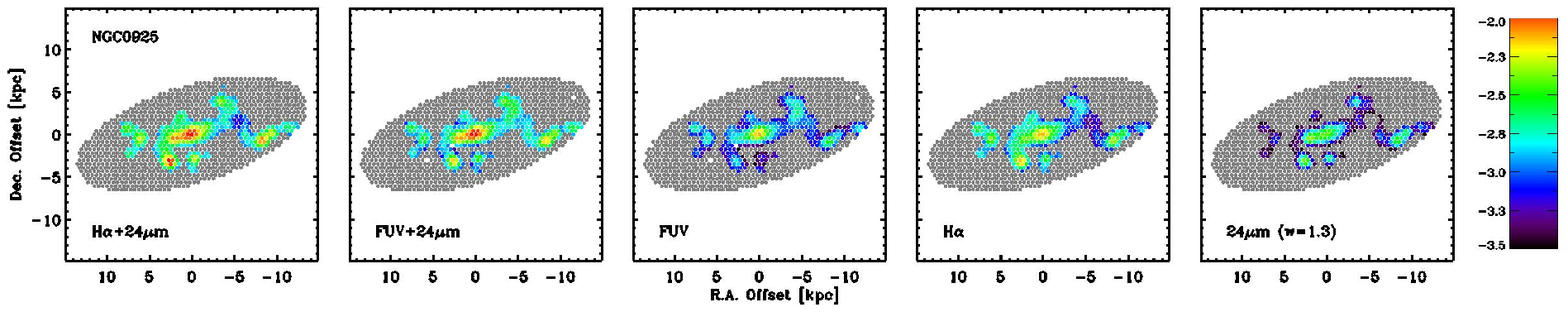}
\plotone{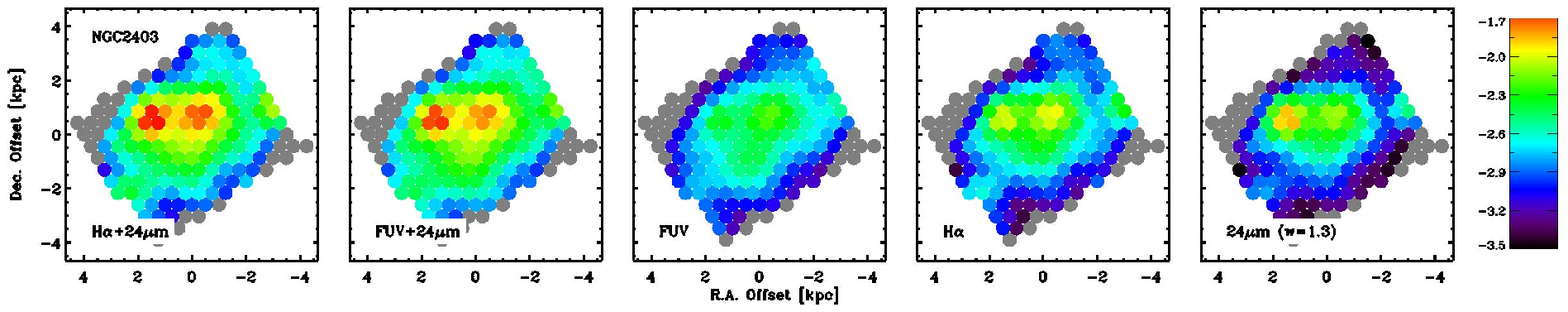}
\plotone{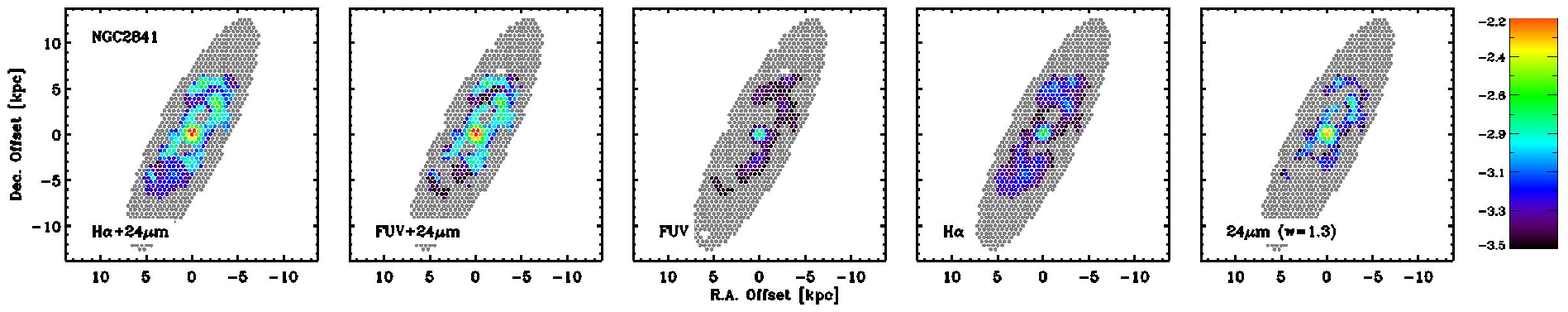}
\plotone{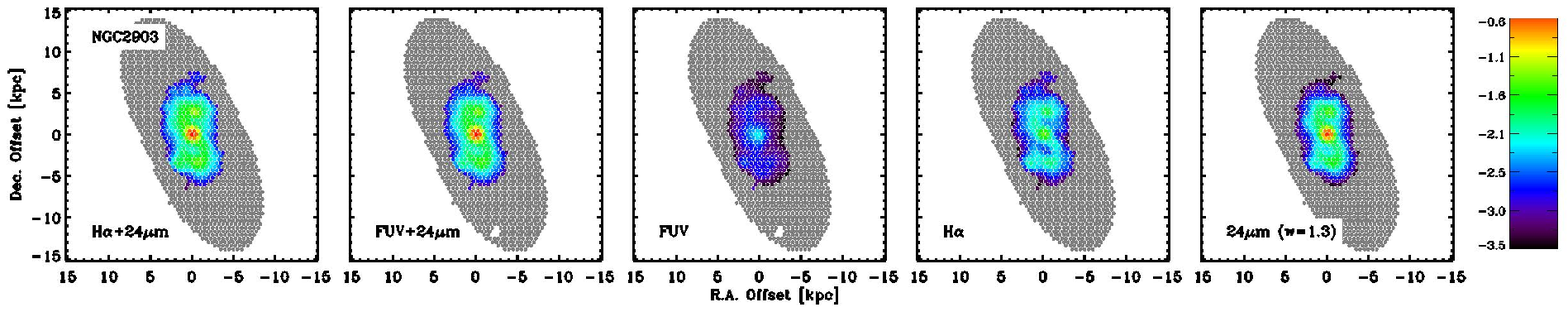}
\caption{$\Sigma_{\rm SFR}$ estimates for the individual lines of
  sight studied throughout this paper. The axes show displacement from
  the galaxy center in kpc. We color code points by
  $\log_{10}~\Sigma_{\rm SFR}$ estimated following the recommendations
  in the main text. We fix the scale for each galaxy to allow
  comparison of different components but the color scale varies from
  target to target according to the dynamic range in the maps. Points
  where $\Sigma_{\rm SFR}$ estimated from H$\alpha$+24$\mu$m $<
  10^{-3}$~M$_\odot$~yr$^{-1}$~kpc$^{-2}$ are color coded gray in all
  panels.}
\label{fig:sfrmaps_1}
\end{figure*}

\begin{figure*}
\plotone{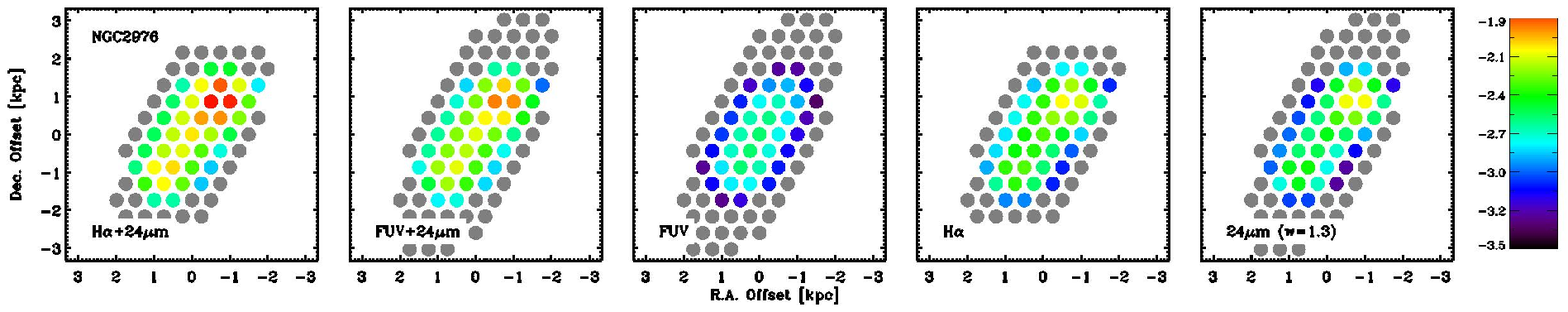}
\plotone{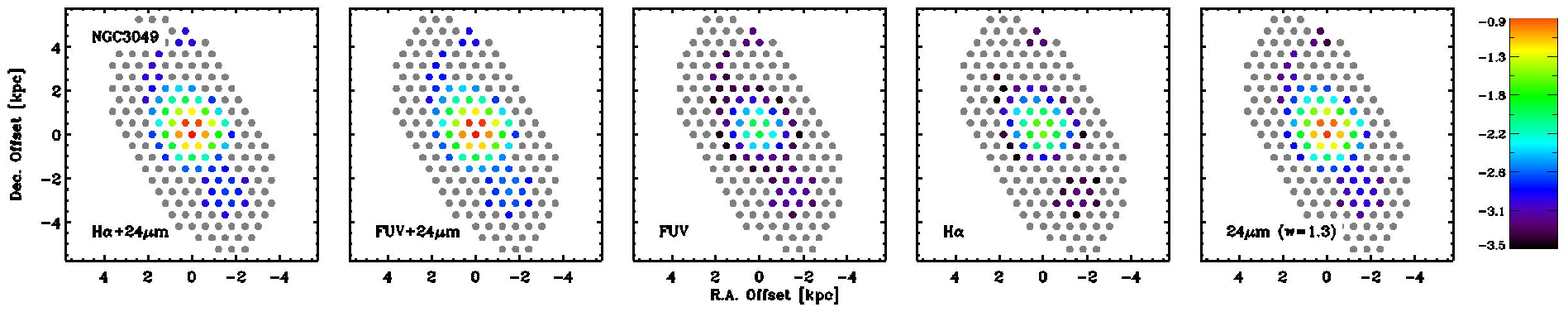}
\plotone{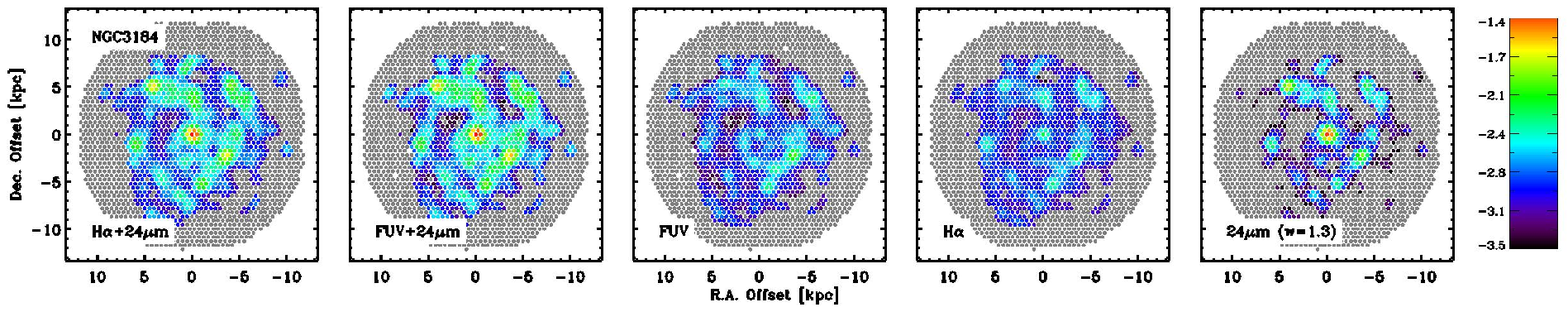}
\plotone{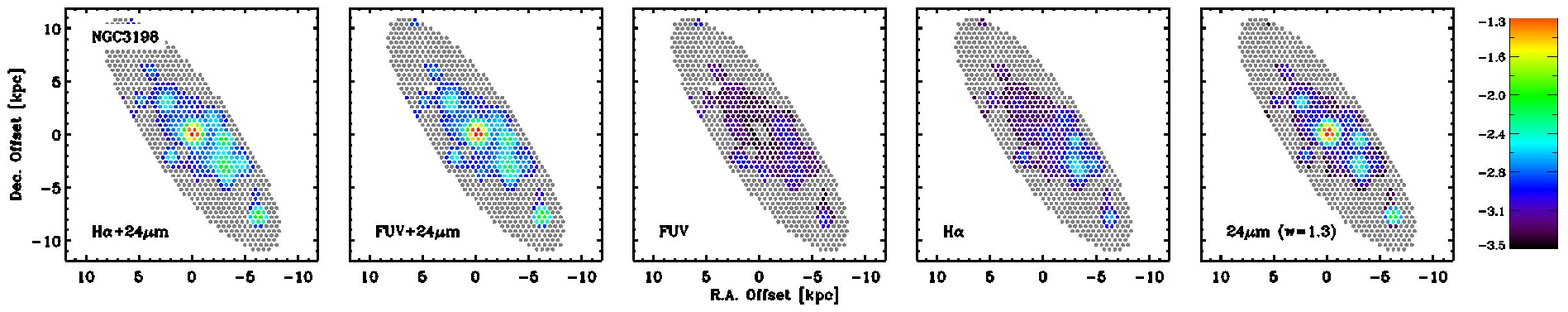}
\plotone{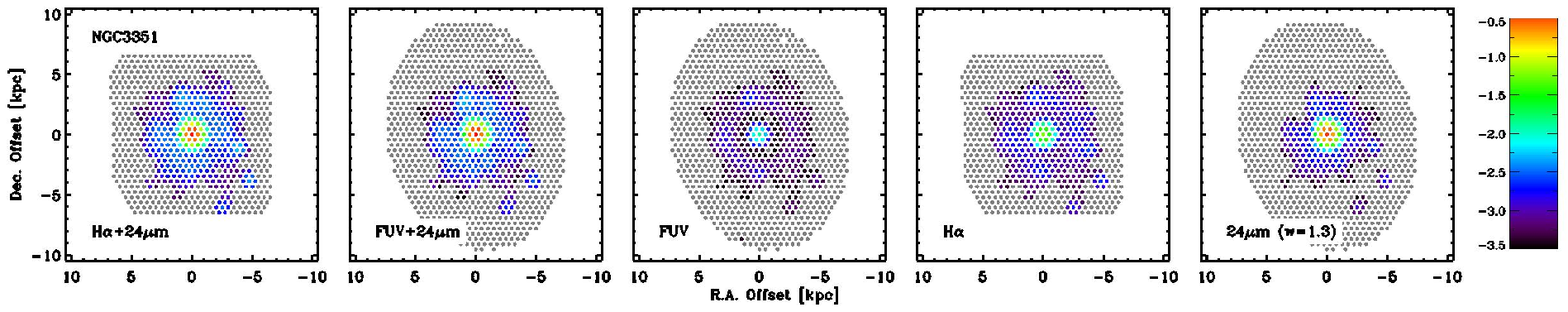}
\plotone{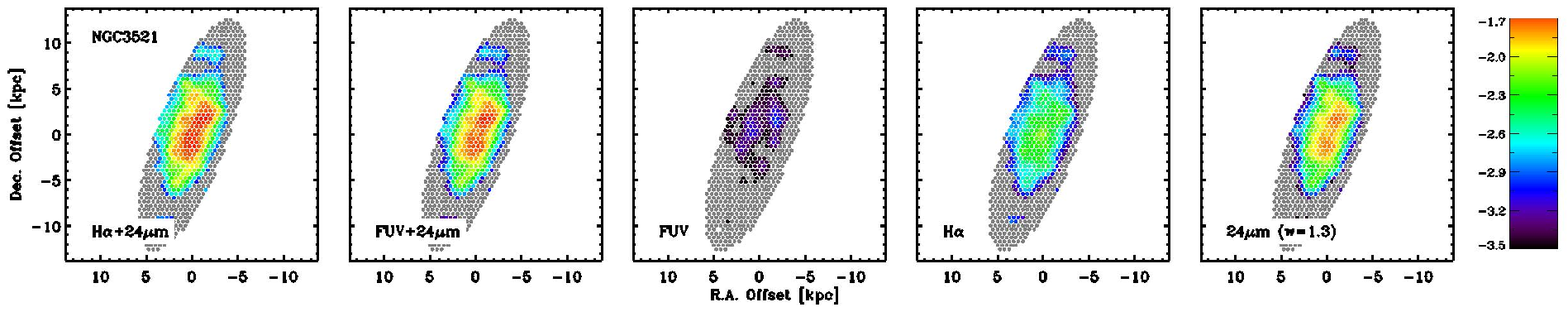}
\caption{As Figure \ref{fig:sfrmaps_1}.}
\label{fig:sfrmaps_2}
\end{figure*}

\begin{figure*}
\plotone{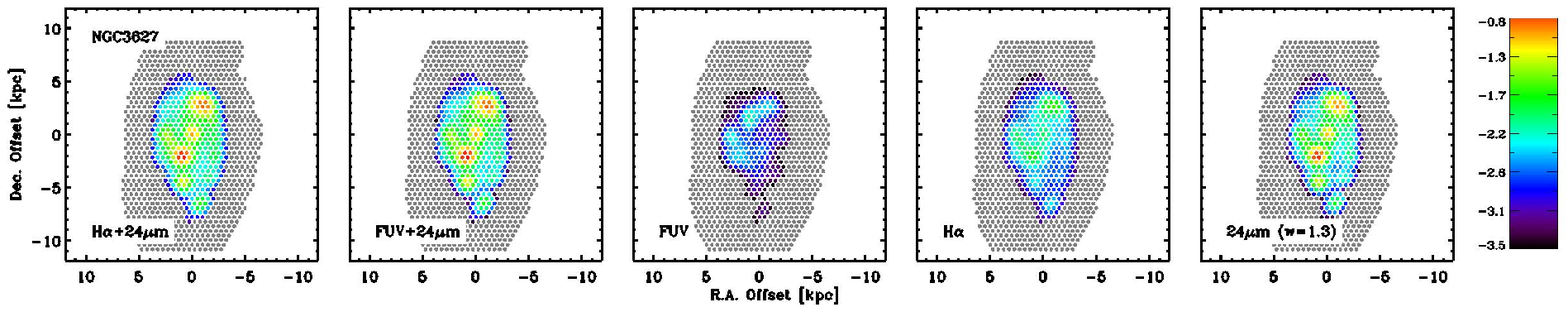}
\plotone{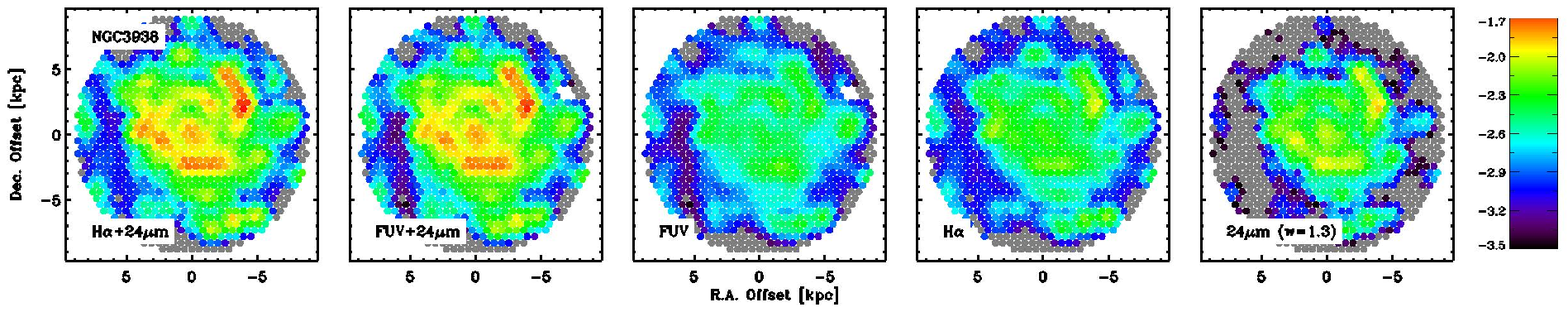}
\plotone{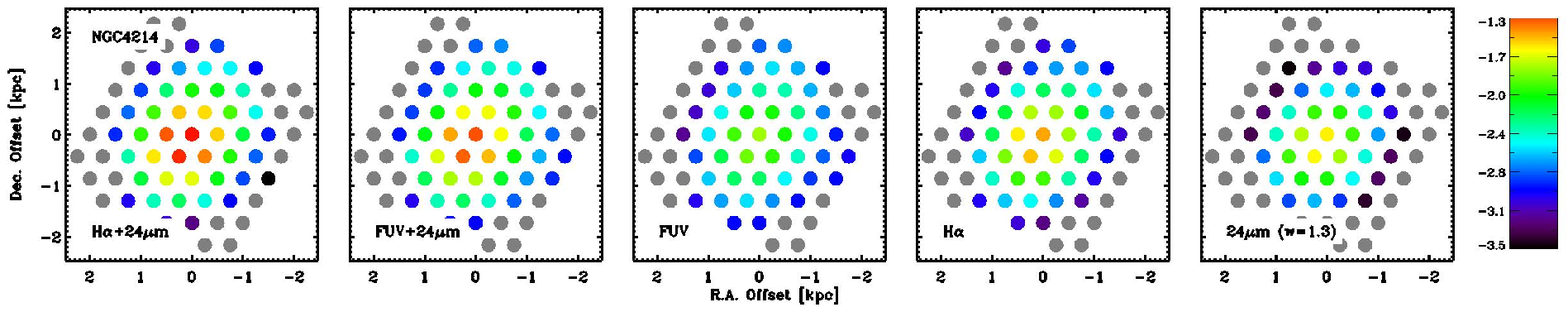}
\plotone{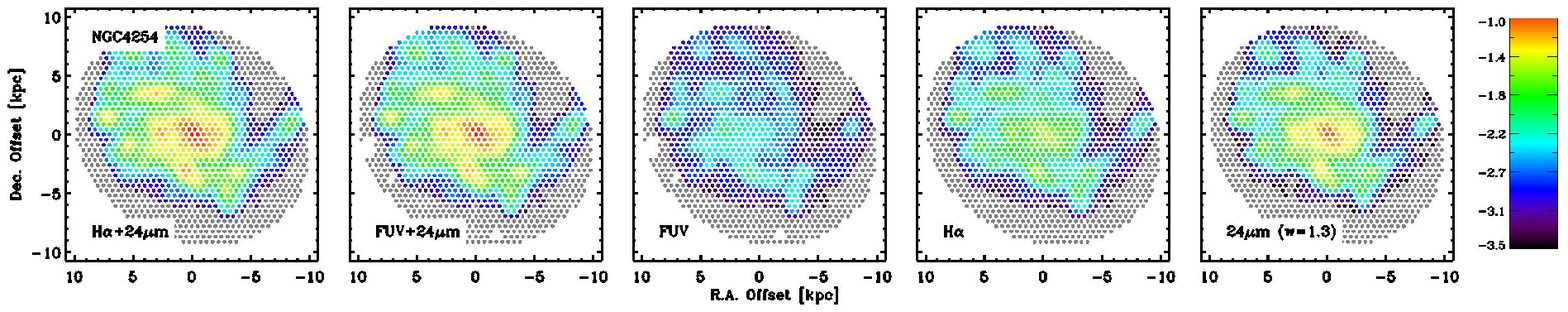}
\plotone{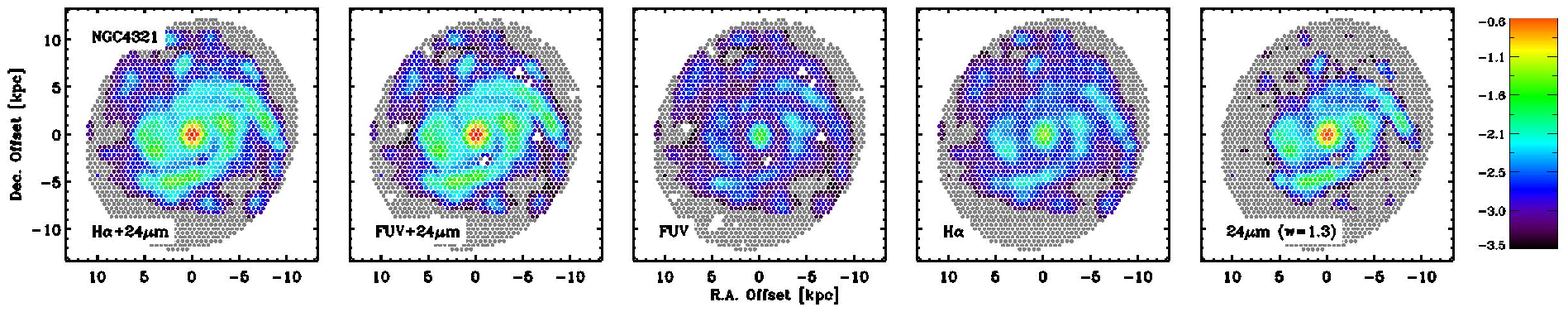}
\plotone{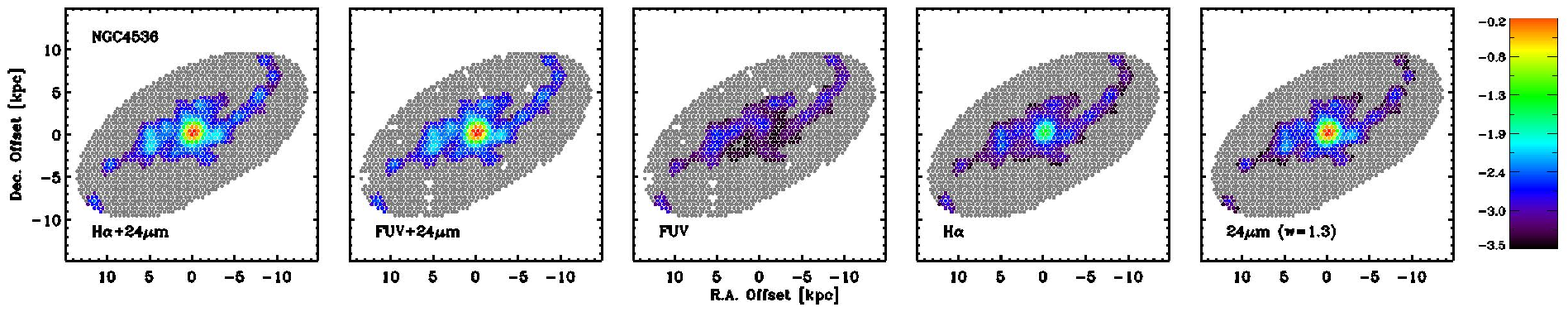}
\caption{As Figure \ref{fig:sfrmaps_1}.}
\label{fig:sfrmaps_3}
\end{figure*}

\begin{figure*}
\plotone{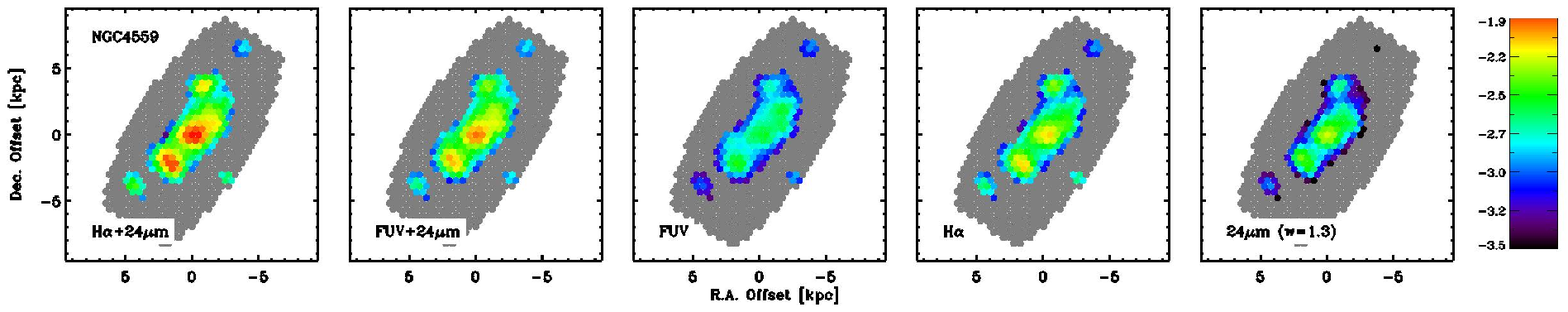}
\plotone{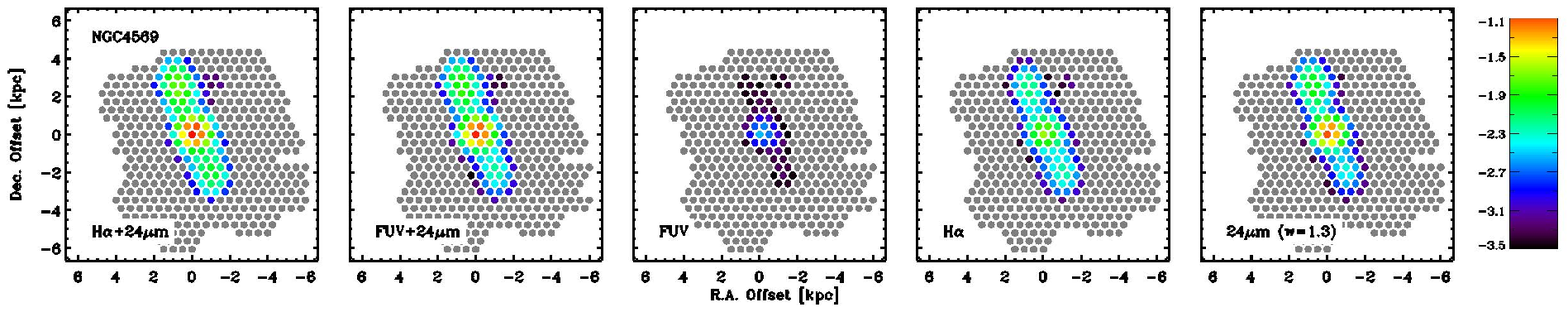}
\plotone{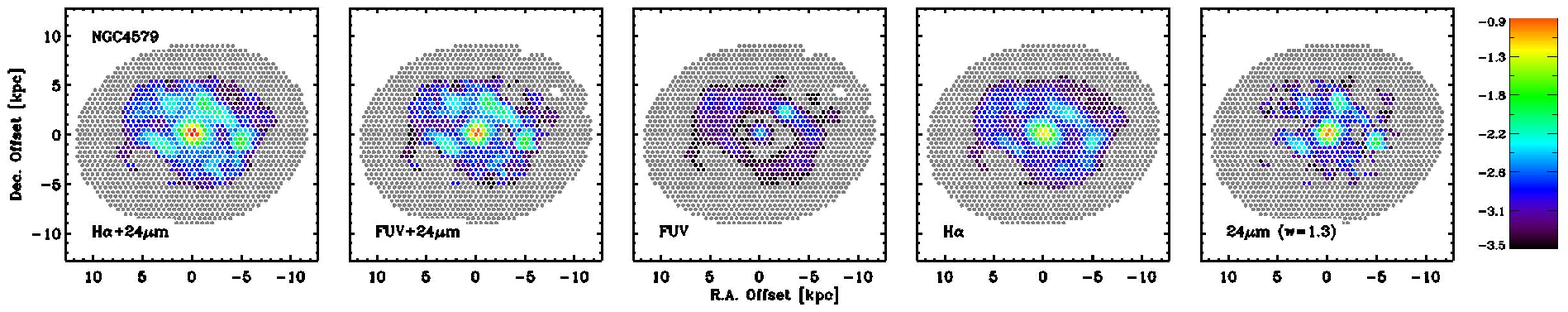}
\plotone{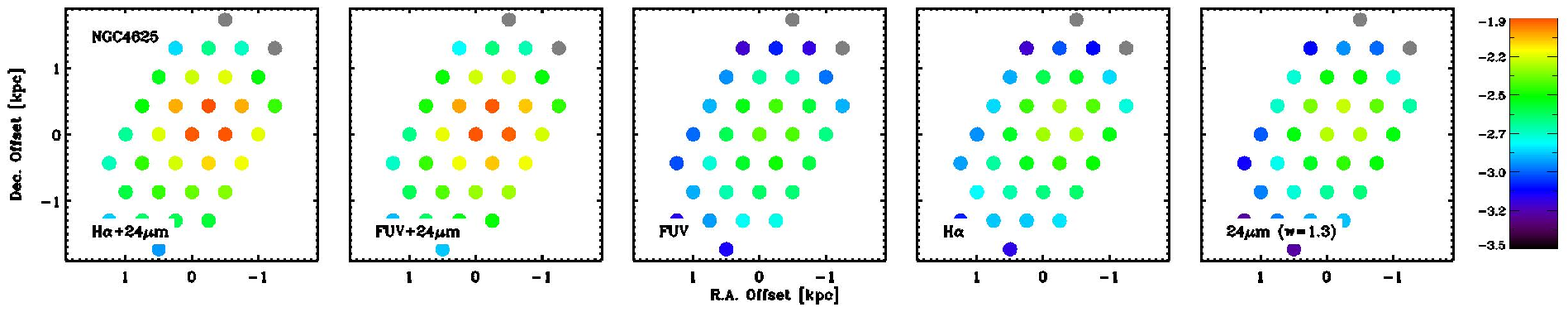}
\plotone{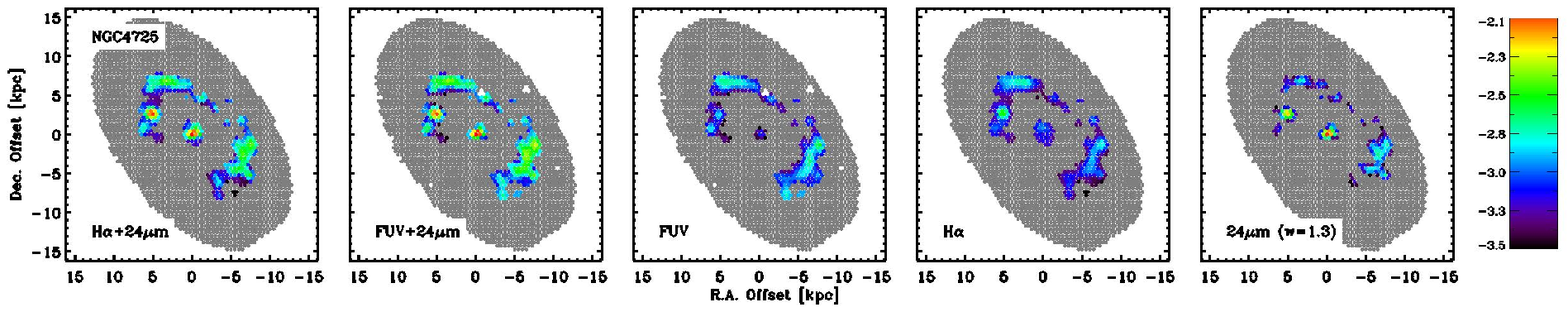}
\plotone{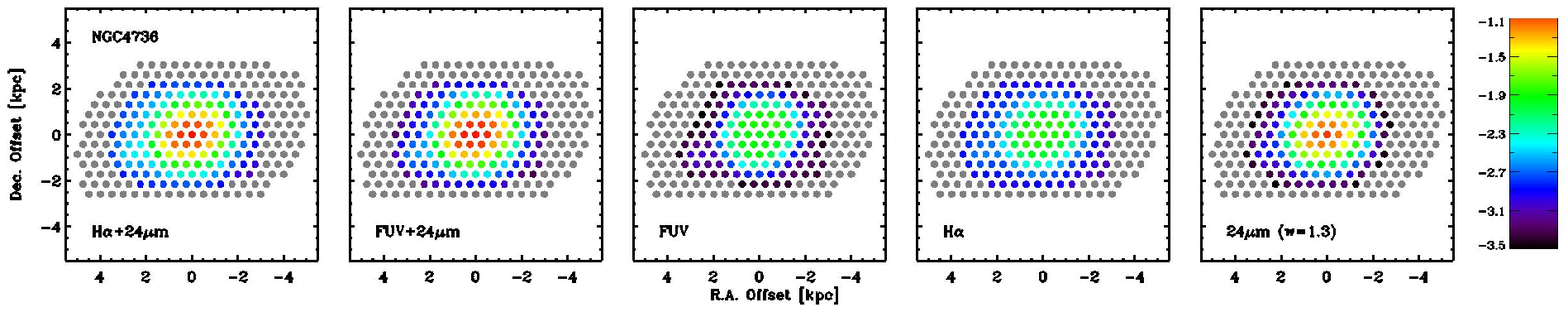}
\caption{As Figure \ref{fig:sfrmaps_1}.}
\label{fig:sfrmaps_4}
\end{figure*}

\begin{figure*}
\plotone{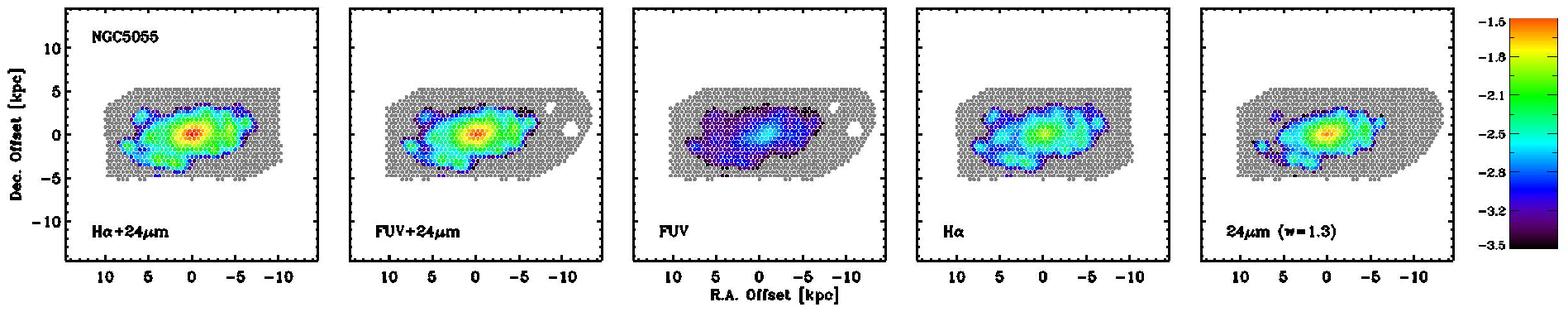}
\plotone{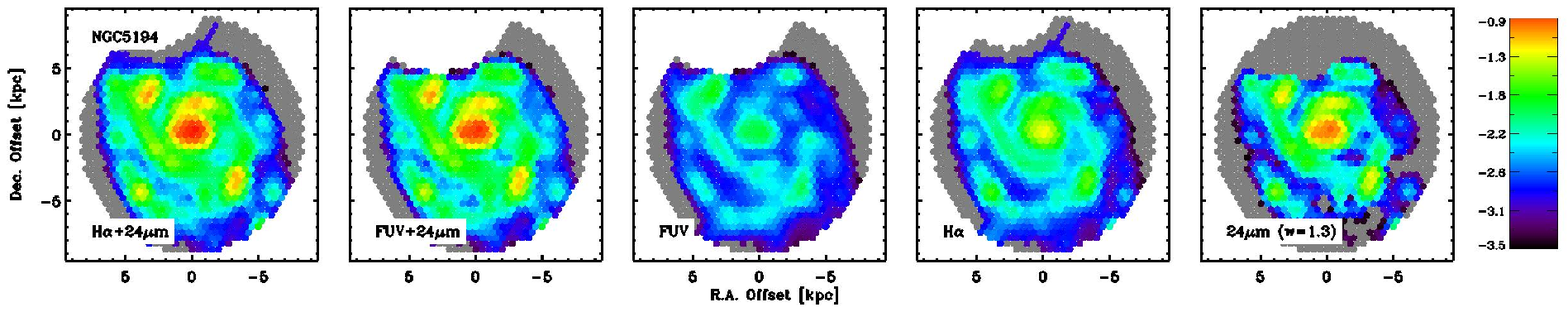}
\plotone{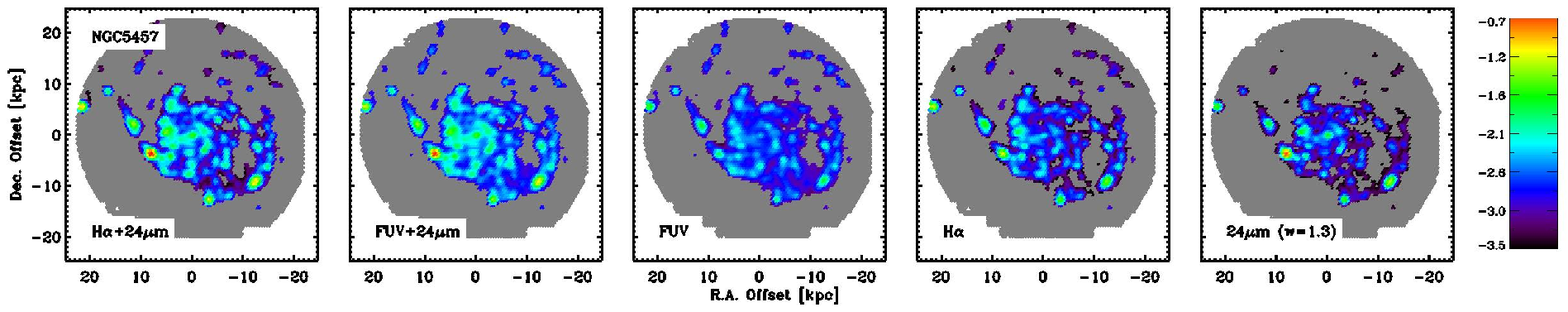}
\plotone{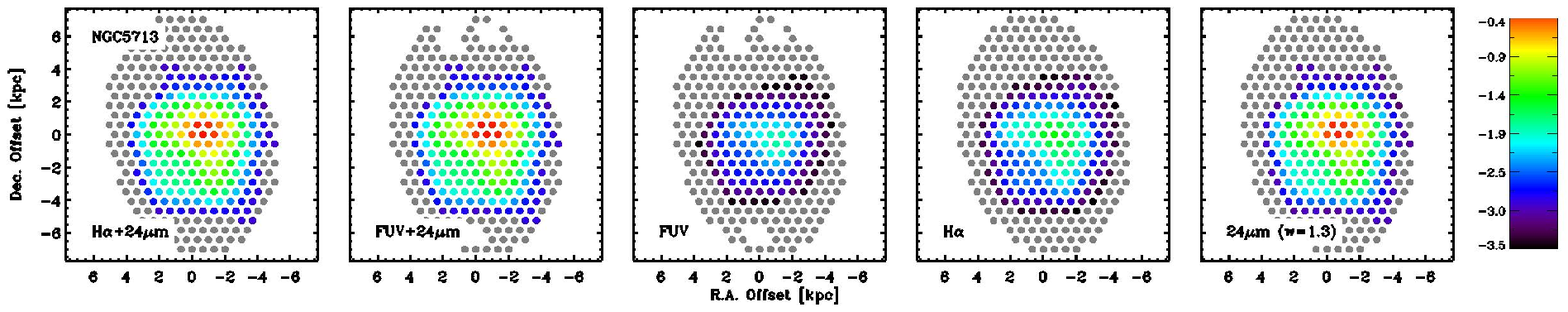}
\plotone{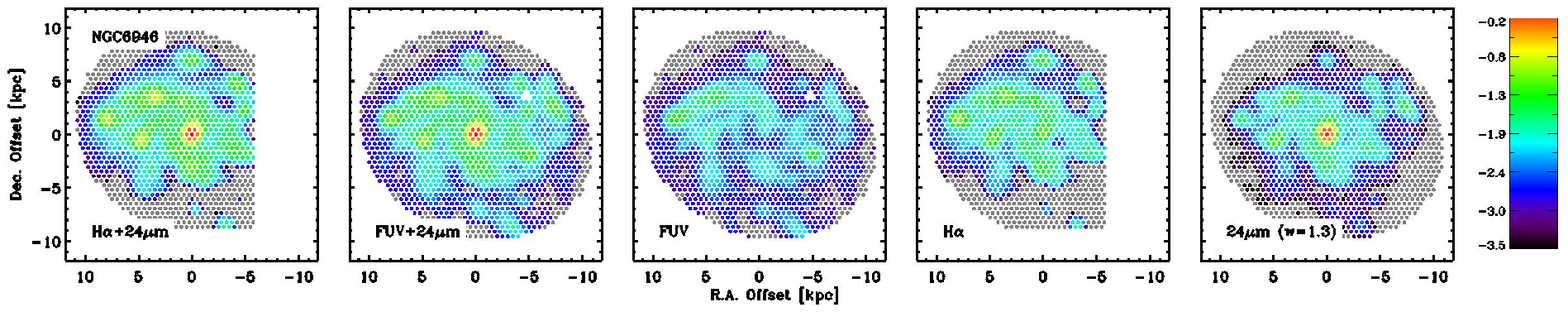}
\plotone{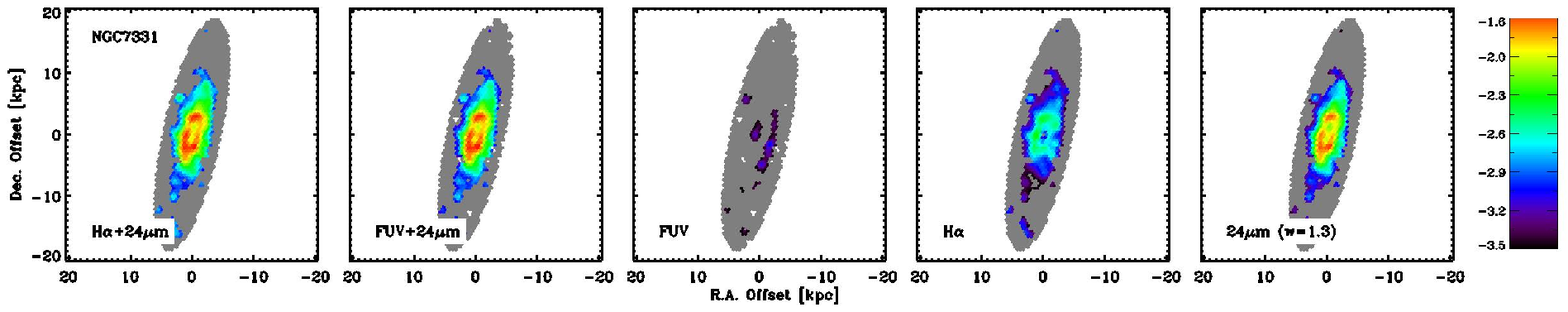}
\caption{As Figure \ref{fig:sfrmaps_1}.}
\label{fig:sfrmaps_5}
\end{figure*}

In this appendix we plot $\Sigma_{\rm SFR}$ estimates based on our
data. Figures \ref{fig:sfrmaps_1} -- \ref{fig:sfrmaps_5} show each
line of sight considered in this paper as a point color-coded by
$\Sigma_{\rm SFR}$. As described in Section \ref{sec:data}, these data
are sampled on a hexagonal grid with spacing 0.5~kpc except for a few
distant cases noted Table \ref{tab:sample}, which are sampled with
6.5$\arcsec$ (half-beam) spacing. Gray points indicate where the
H$\alpha$+24$\mu$m based $\Sigma_{\rm SFR} <
10^{-3}$~M$_\odot$~yr$^{-1}$~kpc$^{-2}$, our limit for a robust
measurement (Section \ref{sec:compare}). From left to right panels show
$\Sigma_{\rm SFR}$ estimated from H$\alpha$+24$\mu$m, FUV+24$\mu$m,
FUV only, H$\alpha$ only, and 24$\mu$m only. In all panels we have
applied our recommended cirrus correction (Section \ref{sec:cirrus}) to the
24$\mu$m data and adopt our recommended $w_{\rm H\alpha} = 1.3$ and
$w_{\rm FUV} = 1.7$. The 24$\mu$m-only panel uses $w = 1.3$.

\clearpage

\clearpage

\clearpage

\clearpage

\clearpage

\clearpage

\end{appendix}

\end{document}